\documentclass[useAMS,usenatbib,usegraphicx]{mn2e}

\usepackage{amsmath}
\usepackage{aas_macros}
\usepackage[british]{babel}
\usepackage{multirow}
\usepackage{url}
\bibpunct{(}{)}{,}{a}{}{,} 
\newcommand{\lt}{<}
\newcommand{\gt}{>}

\newcommand{\pms}{pre-MS}
\newcommand{\zams}{ZAMS}

\newcommand{\eos}{EOS}
\newcommand{\ml}{$\alpha_\mathrm{ML}$}

\newcommand{\bcs}{BCs}

\newcommand{\lsun}{L$_{\sun}$}
\newcommand{\msun}{M$_{\sun}$}

\title[Uncertainties on low and very-low mass stars radii]{Theoretical uncertainties on the radius of low- and very-low mass stars.}
\author[E. Tognelli, P. G. Prada Moroni, \& S. Degl'Innocenti]
{E. Tognelli$^{1,2}$\thanks{e-mail: ema.tog$@$gmail.com}, P.G. Prada Moroni$^{2,3}$\thanks{e-mail: pier.giorgio.prada.moroni$@$unipi.it}, 
S. Degl'Innocenti$^{2,3}$\\
$^{1}$Osservatorio Astronomico di Teramo, via M. Maggini, I-64100, Teramo, Italy\\
$^{2}$INFN, Sezione di Pisa, Largo Bruno Pontecorvo 3, I-56127, Pisa, Italy\\
$^{3}$Dipartimento di Fisica `E.Fermi', Universit\'a of Pisa, Largo Bruno Pontecorvo 3, I-56127, Pisa, Italy\\
}
\begin{document}
\date{Accepted 2018 January 19. Received 2017 November 30; in original form 2017 July 19}
\pagerange{\pageref{firstpage}--\pageref{lastpage}} \pubyear{2018}
\maketitle
\label{firstpage}
%
\begin{abstract}
We performed an analysis of the main theoretical uncertainties that affect the radius of low- and very-low mass-stars predicted by current stellar models. We focused on stars in the mass range 0.1-1~\msun{}, on both the zero-age main-sequence (ZAMS) and on 1, 2 and 5~Gyr isochrones. First, we quantified the impact on the radius of the uncertainty of several quantities, namely the equation of state, radiative opacity, atmospheric models, convection efficiency and initial chemical composition. Then, we computed the cumulative radius error stripe obtained by adding the radius variation due to all the analysed quantities. 

As a general trend, the radius uncertainty increases with the stellar mass. For ZAMS structures the cumulative error stripe of very-low mass stars is about $\pm 2$ and $\pm 3$~percent, while at larger masses it increases up to $\pm 4$ and $\pm 5$~percent. The radius uncertainty gets larger and age dependent if isochrones are considered, reaching for $M\sim 1$~\msun{} about $+12(-15)$ percent at an age of 5~Gyr.   

We also investigated the radius uncertainty at a fixed luminosity. In this case, the cumulative error stripe is the same for both ZAMS and isochrone models and it ranges from about $\pm 4$~percent to $+7$ and $+9$($-5$)~percent. 

We also showed that the sole uncertainty on the chemical composition plays an important role in determining the radius error stripe, producing a radius variation that ranges between about $\pm 1$ and $\pm 2$~percent on ZAMS models with fixed mass and about $\pm 3$ and $\pm 5$~percent at a fixed luminosity.
\end{abstract}
\maketitle
%
\section{Introduction}
\label{sec:intro}
In recent years, accurate measurements of stellar radii for low and very-low mass stars have become available both for single stars, through interferometric measurements, and for eclipsing binary systems \citep[see e.g.][]{segransan2003,berger2006,mathieu07,demory2009,torres2010,boyajian2012a,boyajian2012b}. 

The continuously growing sample of stars for which accurate mass and radius measurements are available suggests that stellar models tend to systematically underestimate the radius of low-mass (LM) and very low-mass (VLM) stars by about 3-20 percent, depending on the stellar mass \citep[see e.g.][]{torres2002,chabrier2005,berger2006,morales2009,boyajian2012b,spada2013,torres2013}. Such a sizeable disagreement prompted a renewed interest on theoretical models of low-mass stars and on their evolution.

In spite of the significant improvement of stellar evolution computations in the last decades, models are still affected by uncertainties coming from the adopted input physics [i.e. equation of state (EOS), radiative opacity, boundary conditions (BCs)], initial chemical composition and from the still oversimplified treatment of the convection in super-adiabatic regimes \citep[i.e. the largely adopted mixing length theory, MLT,][]{bohm58}. This last point is particularly important for stars with deeper super-adiabatic envelopes (i.e. M$\ga 0.3$--0.4~\msun, for solar chemical composition). The discrepancy between the expected and observed mass-radius relation leaded also to investigate possible non-standard mechanisms that might produce the observed radius inflation. 

From the observational point of view, \citet{kraus2011} claimed  a possible correlation between the radius inflation and the orbital period of binary stars. They found that short period ($\lt 1.5$ days)  tidally locked stars seem to exhibit larger radius than longer period counterparts. However, the correlation is still debated \citep{feiden2012a}. Moreover, \citet{spada2013} found similar levels of radius inflation in both single and binary stars. 

Systems with inflated radii tend to show also an high level of magnetic activity and/or spot coverage \citep{torres2002,chabrier2007,lopez2007,ribas2008,stassun2012,feiden13,somers2015}, although a clear correlation has not yet been found \citep{mann2015}. Moreover, the presence of stellar spots alters also the stellar properties derived from the light curve analysis of binary systems \citep[see e.g.][]{morales10,windmiller2010}.

The effect of a relatively strong magnetic fields in stellar models has been recently investigated in several papers \citep[][and references therein]{chabrier2005,feiden2012,feiden13,feiden2014}. Feiden and collaborators showed that the introduction of a magnetic field acts to reduce the convection efficiency in the external super-adiabatic region. The magnetic field strength can be tuned to improve the agreement between data and models, at least in stars with radiative cores and convective envelopes \citep{feiden13}. On the other hand, in the case of fully convective stars, extremely large and unrealistic magnetic fields are required. As such, magnetic field is unlikely to be the sole mechanism acting to inflate the radius  \citep{feiden2014b,feiden2014,feiden2014c}. Interestingly, \citet{feiden13} showed also that the main effect of magnetic convection inhibition can be reproduced, in standard models with no magnetic fields, by simply changing the mixing length parameter (\ml). In particular they showed that a value of \ml{} much lower than the solar calibrated one allows to mimic the main effects of the presence of an internal magnetic field. Models of main sequence (MS) and pre-MS low-mass stars with reduced external convection efficiency (\ml$\le 1$) show, in some cases, a better agreement with observational data \citep[i.e. radius or surface lithium abundance, see e.g.][]{chabrier2007,tognelli12}.

Another aspect related to the presence of strong magnetic fields is surface stellar spots phenomenon. The presence of spots with long enough lifetime, which depends on thermal time-scales of the convective envelope \citep{spruit1986}, might lead to a reduction of surface energy flux, due to the presence of spotted cooler regions, thus inducing a radius inflation and a change in the expected colours [spectral energy distribution (SED)]. Depending on the assumptions made on the spots (i.e. their depth, temperature and coverage fraction) a radius variation up to about 10~percent can be obtained \citep[depending on stellar mass and age, see e.g.][]{somers2015}. \citet{chabrier2007} showed that models that account for both spots and magnetic convection inhibition lead to a better agreement with observed low-mass stars radii.

Recently \citet{chen2014} were able to get a good agreement between models and data for low-mass stars by modifying the temperature profile in the atmospheric models used as BC for the computation of stellar models. They suggested that a possible reason for part of the disagreement between the observed and predicted mass-radius relation might be found in the still not satisfactory synthetic atmospheric structures used to obtain outer BCs for stellar models. However, such models systematically underestimate the effective temperature of VLM and LM stars in clusters, as shown by \citet{randich2017}.
 
The aim of this paper is to focus on the prediction of standard models (i.e. without the presence of magnetic fields, spots, or rotation) and to give a quantitative estimation of the actual uncertainty in the predicted stellar radius, due to the uncertainty in the adopted input physics, initial chemical composition and convection efficiency. As a result we obtained a cumulative error stripe in the predicted radius, when all the uncertainty sources are accounted for at the same time. 

The paper is structured as it follows. In Section \ref{sec:models} we introduce the standard set of models. Then, we analyse the contribution to the radius uncertainty due to the errors of several adopted input physics in Section \ref{sec:fis} and due to the uncertainty in the adopted initial chemical composition in Sections \ref{sec:chm}. In Section \ref{sec:total} the cumulative error stripe for the stellar radius computed, by adding all the contribution to the radius uncertainty analysed in the paper, is presented. In Section \ref{sec:conclusions} we summarize the main results of this work.

\section{The models}
\label{sec:models}
\begin{figure}
	\centering
	\includegraphics[width=\columnwidth]{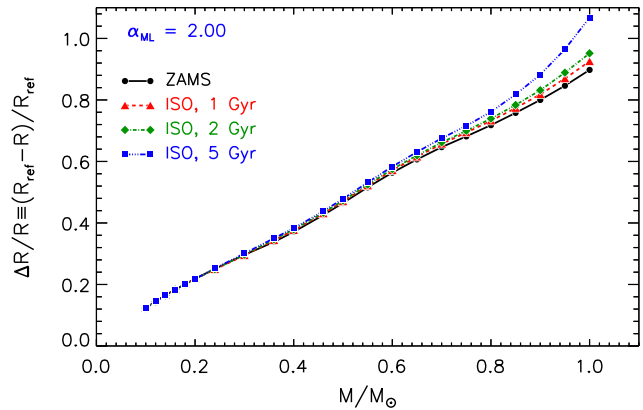}
	\caption{Stellar radius (in solar units) as a function of the mass for our reference set of models located on the ZAMS (black-solid line) and on 1, 2, and 5~Gyr isochrones (coloured-lines).}
	\label{fig:ref}
\end{figure}
We computed LM and VLM stellar models using the Pisa version of the \textsc{franec} code \citep[\textsc{prosecco}; ][]{tognelli11,dellomodarme12}, described in details in \citet{tognelli15a} and \citet{tognelli15b}.  Here we briefly recall the main input physics relevant for the present analysis. We adopted \textsc{opal} 2005 radiative opacities in the interior for $\log T$(K)$\ge 4.5$ extended with the \citet[][hereafter F05]{ferguson05} at lower temperatures. Both the \textsc{opal} and the F05 opacities are computed for the \citet[][hereafter AS09]{asplund09} solar mixture. The EOS has been obtained using the \textsc{opal} \eos{} 2006 \citep{rogers02} as reference, extended by the \textsc{scvh95} \eos{} \citep{saumon95} in the low temperature - high  pressure regime. The outer BCs have been extracted from the non-grey \citet[][hereafter AHF11]{allard11}\footnote{The AHF11 tables are available at the URL:\\ \url{https://phoenix.ens-lyon.fr/Grids/BT-Settl/AGSS2009/}.} atmospheric structures. The atmosphere adopts the same heavy elements mixture (i.e. AS09) and mixing length parameter (i.e. \ml =2.0) used in the interiors. To be noted that \ml=2.0 is the solar calibrated value we obtained with our models.

Besides the reference  value \ml=2.00, we also computed models for a much lower mixing length parameter value, namely \ml=1.00, corresponding to a much less efficient super-adiabatic convection. To this regard, \citet{chabrier2007} showed that the reduction of super-adiabatic convection efficiency might partially reduce the disagreement level between the predicted and observed radii of low-mass stars. The availability of two sets of stellar tracks with significantly different \ml{} values allows us to investigate whether the theoretical uncertainty on the stellar radius depends on such a poorly constrained parameter.

For the standard set of stellar tracks we chose [Fe/H]=$+0.0$, which corresponds to an initial helium abundance $Y=0.274$ and metallicity $Z=0.013$ by assuming the AS09 solar heavy-element mixture. 

The stellar models cover the mass range [0.1, 1.0]~\msun, with a variable mass spacing of 0.02~\msun{} for $M\in [0.1,\,0.5]$~\msun{} and 0.05~\msun{} for $M> 0.5~$\msun. From the stellar tracks we then built the zero-age main sequence (\zams), identified as the sequence of models of different mass whose central hydrogen abundance has been reduced by 0.02~percent with respect to the initial one. Although this is not the rigorous definition of \zams, we checked that it is a good approximation, with the advantage of being easily and consistently implementable in all the mass range we analysed\footnote{A different definition/approximation of \zams{} might be adopted, however the impact on the final results is completely negligible if compared to the variation produced by the perturbed quantities discussed in this work.}. We recall that the models on \zams{} do not share the same age. Indeed, stars with a different mass ignite the central hydrogen burning at different ages, and the lower is the stellar mass, the larger is the ZAMS age. In particular, in the selected mass range, the \zams{} sequence is populated by models with ages from 50~Myr ($\sim 1.0~$\msun) up to about 5~Gyr ($\sim 0.1~$\msun). To investigate the uncertainty in the stellar radius at fixed age, isochrones for three ages, namely 1, 2 and 5~Gyr have been calculated.

We show in Fig.~\ref{fig:ref} a comparison between the theoretical stellar radius as a function of the mass obtained from our reference  models on the ZAMS and on the 1, 2 and 5~Gyr isochrone sequences. 

\section{Uncertainty in the adopted input physics}
\label{sec:fis}
As a first step of our analysis, we quantified the uncertainty in the stellar radius caused by the uncertainty in the adopted input physics by varying, separately, each of the main quantities relevant for the stellar evolution up to the \zams /MS phase. 

We followed two approaches when analysing the effect of one perturbed quantity on the stellar radius. In one case, we simply take the relative difference between the radius prediction -- at a given mass -- of the reference set and of the set with one perturbed quantity. The latter have been computed by changing/perturbing a single input physics and keeping the other quantities to their reference values. In the second approach, we performed a new solar calibration of the mixing length parameter in the models with the perturbed input physics. This second case deserves a brief explanation. 

As previously mentioned, the reference set of models adopted a solar calibrated mixing length parameter (\ml$_{\sun}$). Such a value has been obtained by requiring that a 1~\msun{} stellar model, at the age of the Sun, must reproduce simultaneously the radius, luminosity and surface $(Z/X)$ of the Sun. The solar calibration is performed by tuning three initial parameters, namely \ml$_{\sun}$, the initial helium $Y_{\sun}$ and the metal $Z_{\sun}$ content, which affect the solar luminosity, radius and surface $(Z/X)$. However, in the present analysis we did not fully use the solar calibrated values, in the sense that the initial chemical composition adopted for the computation is fixed and derived assuming [Fe/H]=0 (see Section \ref{sec:chm}) and it is not that resulting from the solar calibration. Thus, here with solar calibration we mean models computed with only the solar calibrated \ml.

In this second approach we thus performed the solar calibration for models with the perturbed input physics, to assure that the 1~\msun{} model reproduces the solar characteristics. This approach is different from the first one where no calibration is done. In some cases and for masses close to the 1~\msun, it might happen that the differences between the reference set of models and that with the perturbed input physics can be partially counterbalanced by the recalibrated \ml$_{\sun}$ value.

In the following, we will discuss the results of adopting these two approaches in each analysed case.

\subsection{Outer boundary conditions: atmospheric models}
\label{sec:bc}
The outer BCs necessary to integrate the stellar structure equations consist of the pressure $P_\rmn{bc}\equiv P(\tau_\rmn{bc})$ and temperature $T_\rmn{bc}\equiv T(\tau_\rmn{bc})$ -- obtained from an atmospheric model -- at a given optical depth $\tau_\rmn{bc}$, corresponding to the point where the atmosphere matches the interior of the star (defined as the region where $\tau \gt \tau_\rmn{bc}$). When considering the uncertainty due to the adopted \bcs{}, two aspects have to be analysed: (1) the atmospheric model used to obtain $P_\rmn{bc}$ and $T_\rmn{bc}$; and (2) the adopted $\tau_\rmn{bc}$ value. 

Concerning the first point, at the moment no estimation of the uncertainty on the results of atmospheric model computations is available. As such, a firm evaluation of the uncertainty in the adopted $P_\rmn{bc}$ and $T_\rmn{bc}$ values cannot be consistently obtained. A possible way to estimate the effect of the \bcs{} on the models is to compare the results obtained adopting different atmospheric structures in the stellar evolutionary code. To do this we used the \bcs{} provided by two detailed (non-grey) sets of atmospheric models, namely the \citet[][hereafter BH05]{brott05} and the \citet[][hereafter CK03]{castelli03}. We also computed a set of models using the still largely used \citet[][hereafter KS66] {krishna66} atmosphere that adopts an empirical $T=T(T_\rmn{eff},\,\tau)$ relation calibrated on the Sun. In the latter case the atmospheric structure is calculated directly inside the stellar evolutionary code, with exactly the same input physics used for the computation of the interior.

\begin{figure}
	\centering
	\includegraphics[width=\columnwidth]{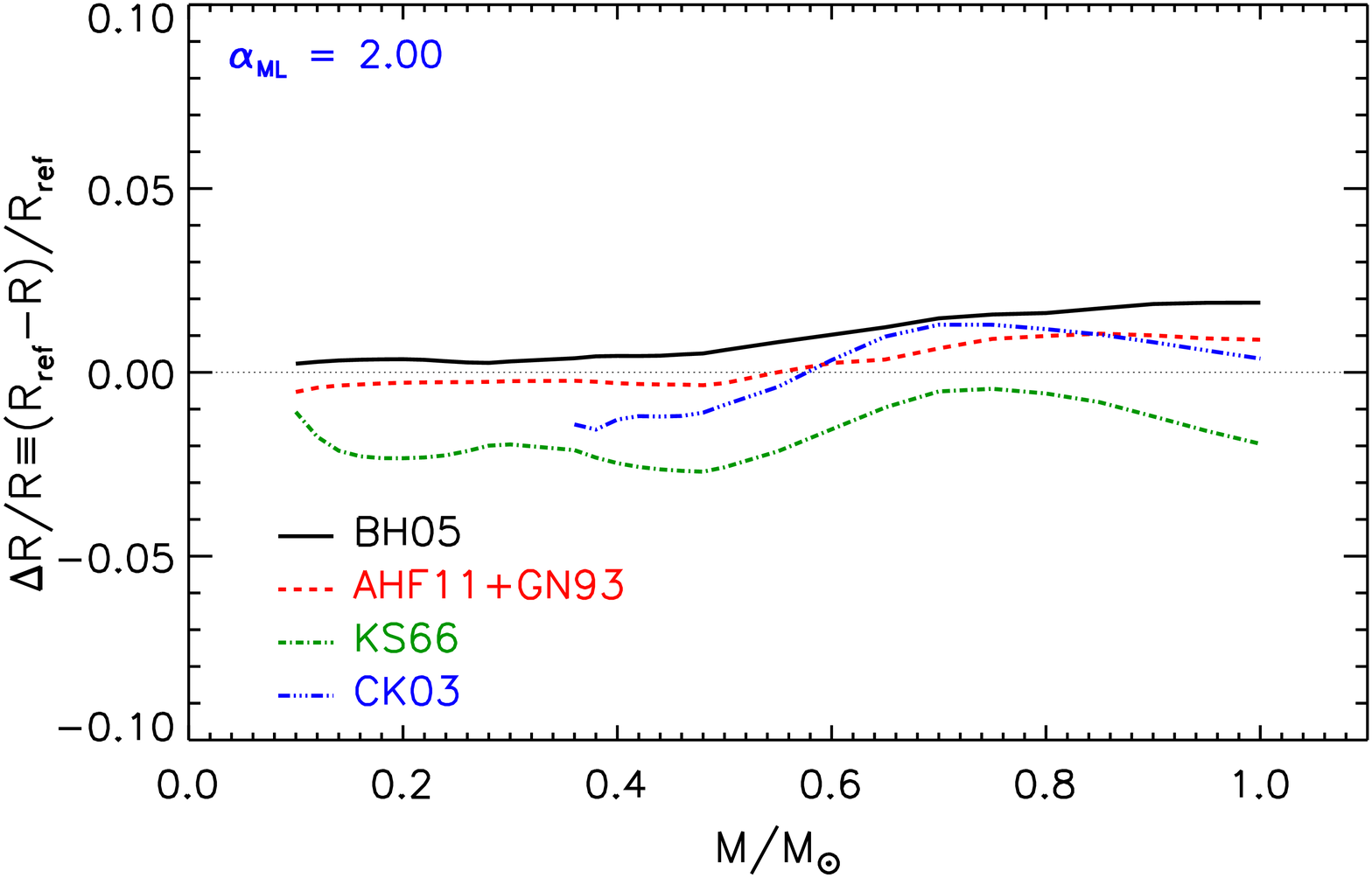}\\
    \includegraphics[width=\columnwidth]{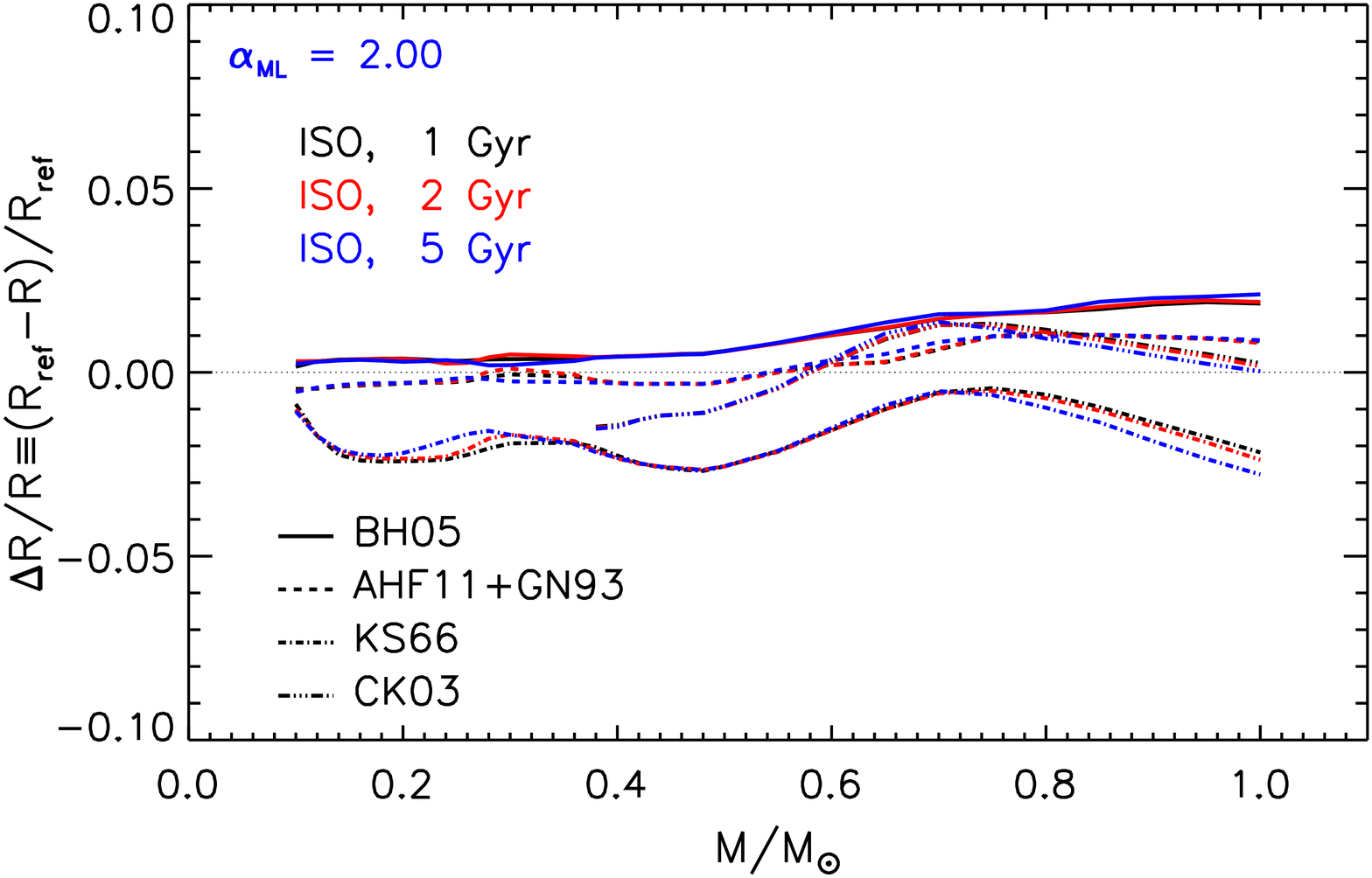}	
	\caption{Relative radius variation as a function of the stellar mass due to the adoption of different atmospheric models (i.e. BH05, CK03, AHF11+GN93 and KS66) with respect to the reference one (AHF11). Top panel: models on the ZAMS. Bottom panel: models on the 1, 2, and 5 Gyr isochrones.}
	\label{fig:bc}
\end{figure}

The top panel of Fig.~\ref{fig:bc} shows the relative radius variation due to the adoption of the quoted \bcs{} instead of the reference AHF11 for \zams{} models, at a fixed value of the stellar mass. To better understand the radius variation due to the \bcs, more details are required. The BH05 have been computed by means of the same atmospheric code \citep[\textsc{phoenix}; see e.g.][]{hauschildt99} used to compute our reference set of atmospheres (AHF11), but the latter has been recently upgraded to better describe the atmospheric structure of LM and VLM stars. The two sets of \textsc{phoenix} atmosphere models differ in several aspects: (1) updated solar mixture (metal abundances), which has a crucial role in determining the opacity in the atmosphere (molecules/dust); (2) updated opacity lines (especially for H$_2$O, CH$_4$, NH$_3$ and CO$_2$); and (3) updated treatment of cloud/dust formation/diffusion (important for cool atmospheres, below 2600--2700~K). The global effect of all the differences between the two sets of atmospheric structures is to produce a radius variation that depends on the stellar mass. For $M~\la~0.5$~\msun{} the radius variation is very small and less than 1 percent, while it progressively increases reaching about $+2$ percent at 1~\msun. 

As previously stated, one of the differences between the AHF11 and the BH05 atmosphere is the adopted solar mixture. The recent AHF11 atmospheric structures are computed using the AS09 -- which is consistent with the interior calculation -- while the BH05 used the \citet[][hereafter GN93]{grevesse93}. It is evident that the adoption of the BH05 \bcs{} produces an inconsistency between the interior and atmospheric structures, at least for what concerns the metals abundances. Unfortunately, the BH05 are not available for the AS09 mixture, so we cannot analyse the effect of the mixture in this case. However, the AHF11 have been made available also for the GN93 mixture (AHF11+GN93 set of models in Fig.~\ref{fig:bc}). This allows us to estimate the effect of a mixture change in the atmosphere, keeping fixed that in the interior (AS09). As clearly visible in figure, the radius variation due to the adoption of a different solar mixture in the atmosphere is very small in VLM stars, below 0.6--0.7~\msun, and it reaches a plateau at about 1 percent for larger masses. It is interesting to compare the result found for the BH05 and the AHF11+GN93 models. The two cases show a very similar behaviour, but the two sets of models are shifted by an almost constant offset, which indicates that part of the effect on the radius is caused by the different mixture used in the interior and in the atmosphere, while the remaining part has to be found in the other differences between the two adopted atmospheric structures, as discussed above. 

It is interesting to analyse also the effect of adopting a totally different atmospheric models. To do this we showed also the comparisons with the \textsc{atlas} code atmosphere \citep{kurucz1970}, in the CK03 configuration. To be noted that the CK03 are available only for $T_\mathrm{eff} \ge 3500$~K, which in our case corresponds to about 0.36~\msun, in ZAMS. From Fig.~\ref{fig:bc} it is evident that the maximum radius variation strongly depends on the stellar mass, reaching about $-1.5$ percent for $M\la 0.4$~\msun{} and $+1.5$ percent at about 0.7~\msun. The \textsc{atlas} code is different from the \textsc{phoenix} one in many aspects, so it is difficult to clearly identify the discrepancy sources. Some of possible relevant differences might be found in (1) the opacity sampling/calculation, (2) opacity lines list, (3) solar mixture \citep[i.e.][ in CK03]{grevesse98}, and (4) \ml{} (1.25 in CK03). All these quantities have a role in determining the final BCs, but the detailed analysis of their individual impact in the models is beyond the scope of the present work.  

We also showed, the effect on the radius of adopting the semi-empirical solar calibrated KS66 atmosphere. We emphasize that such an atmospheric model is a too rough approximation for VLM stars, being calibrated on the Sun. However, grey $T(\tau)$ hydrostatic atmosphere had been adopted in the past in the regime of LM and VLM mas stars to compute still used stellar evolutionary libraries \citep[e.g.][]{dantona94,dantona97}. We thus used the KS66 as representative of this class of stellar models. The KS66 \bcs{} produces the largest radius variation in almost the whole analysed mass range, which reaches about $-3$ percent. To be noted that in the KS66 models the radius is systematically larger than the reference one and the resulting models are cooler.

The bottom left panel of Fig.~\ref{fig:bc} shows the relative radius variation obtained by adopting the same \bcs{} discussed above, but for models on the 1, 2, and 5~Gyr isochrones. In almost all the cases, the radius change is similar to that obtained for the \zams{} in the whole selected mass range, with no evident dependence on the age. Only in the case of the KS66 \bcs, the radius variation slightly increases with the age for masses larger than about 0.8~\msun.  

These results have been obtained adopting \ml{}~=~2.00. We performed the same analysis using \ml{}~=~1.00, to evaluate the dependence on \ml. The adoption of the non-grey \bcs{} (i.e. BH05, CK03 or AHF11+GN93) produces a radius variation almost identical to that obtained in the models with \ml~=~2.00. On the contrary, the use of the KS66 results in a larger surface radius variation that reaches values about 1.5 times larger than those obtained in the models with \ml~=~2.00 in both the \zams{} and the isochrone case, but only for $M \ga 0.3$~\msun. We want to precise that, when changing the \ml{} value, we actually change it in the interiors ($\tau \gt \tau_\rmn{bc}$), as in the non-grey atmosphere \ml{} is fixed. Thus, we do not properly account for the effect of \ml{} in the whole structure. Moreover, the dependence of the KS66 models on the adopted \ml{} probably resides in the fact that KS66 adopts $\tau_\rmn{bc}=2/3$, as suggested in the original paper, instead of $\tau_\rmn{bc}=10$ (our reference in non-grey models). This means that when the KS66 BCs are used, the interior calculations extends to lower values of $\tau$, thus covering a portion of the star that in non-grey BCs is incorporated in the atmosphere and thus independent of the value of \ml{} used for the interiors. Such most external regions can be super-adiabatic, especially at progressively larger values of the stellar mass analysed here, and sensitive to the adopted \ml. Such an effect is almost inconsequential in  VLM, which have adiabatic external layers that extends to the bottom of the atmosphere. This qualitatively explains the dependence of the KS66 models to the adopted \ml, as the mass increases.
\begin{figure}
	\centering
	\includegraphics[width=\columnwidth]{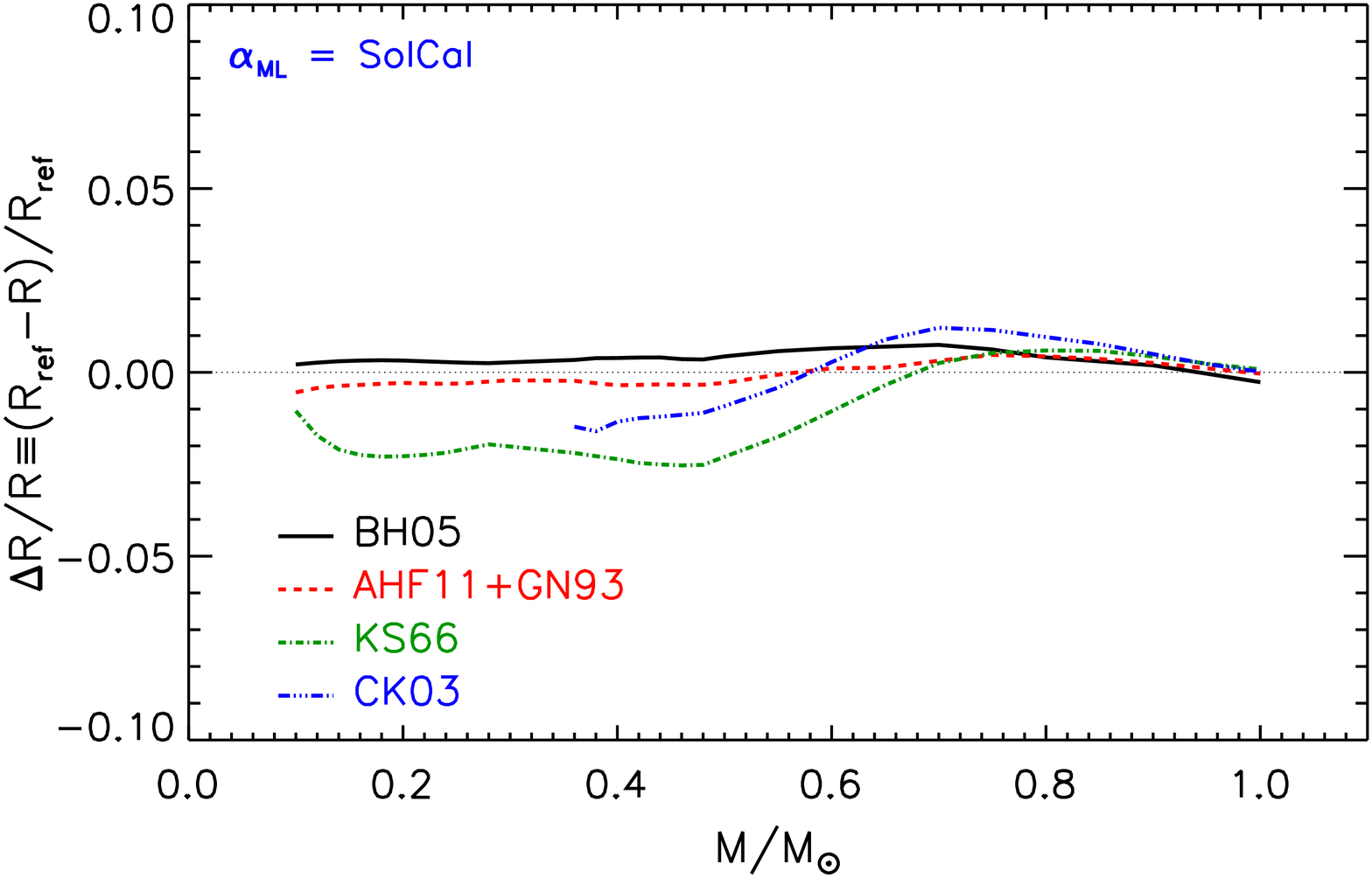}
	\caption{As in Fig.~\ref{fig:bc} but for solar calibrated \ml{} values.}
	\label{fig:bc_solcal}
\end{figure}

We also analysed the radius change due to the adoption of different \bcs{} when the solar calibration is performed for each set of  perturbed models. To be noted that the outer \bcs{} mainly affect the radius of the star. As a consequence, the solar calibration results in the same initial $Y_{\sun}$ and $Z_{\sun}$ values but in different \ml$_{\sun}$. In particular we obtained the following values: \ml$_{\sun}=1.74$ for BH05, \ml$_{\sun}=1.88$ for AHF11+GN93, \ml$_{\sun}=1.95$ for CK03 and \ml$_{\sun}=2.25$ for KS66. The constancy of the initial chemical composition means that all the radius change due to the use of different \bcs{} is counterbalanced in 1~\msun{} model by the adoption of a properly tuned \ml$_{\sun}$. Thus, around 1~\msun{} the radius change is cancelled if the solar calibrated \ml$_{\sun}$ is adopted. However, star of different masses are sensitive in a different way to the \ml, so we do not expect that the solar calibrated \ml$_{\sun}$ is able to counterbalance the radius change in all the selected mass range.

Fig.~\ref{fig:bc_solcal} shows the radius variation in ZAMS induced by the adopted \bcs{} for solar calibrated \ml$_{\sun}$ values. It is clearly visible that the radius change is not affected by the adopted \ml$_{\sun}$ value for $M\la 0.6$~\msun, while for larger masses, as expected, the adoption of a solar value of \ml{} progressively reduces the radius variation.

\subsection{Outer boundary conditions: $\tau_\mathrm{bc}$}
\label{sec:bc_tau}
\begin{figure}
	\centering
	\includegraphics[width=\columnwidth]{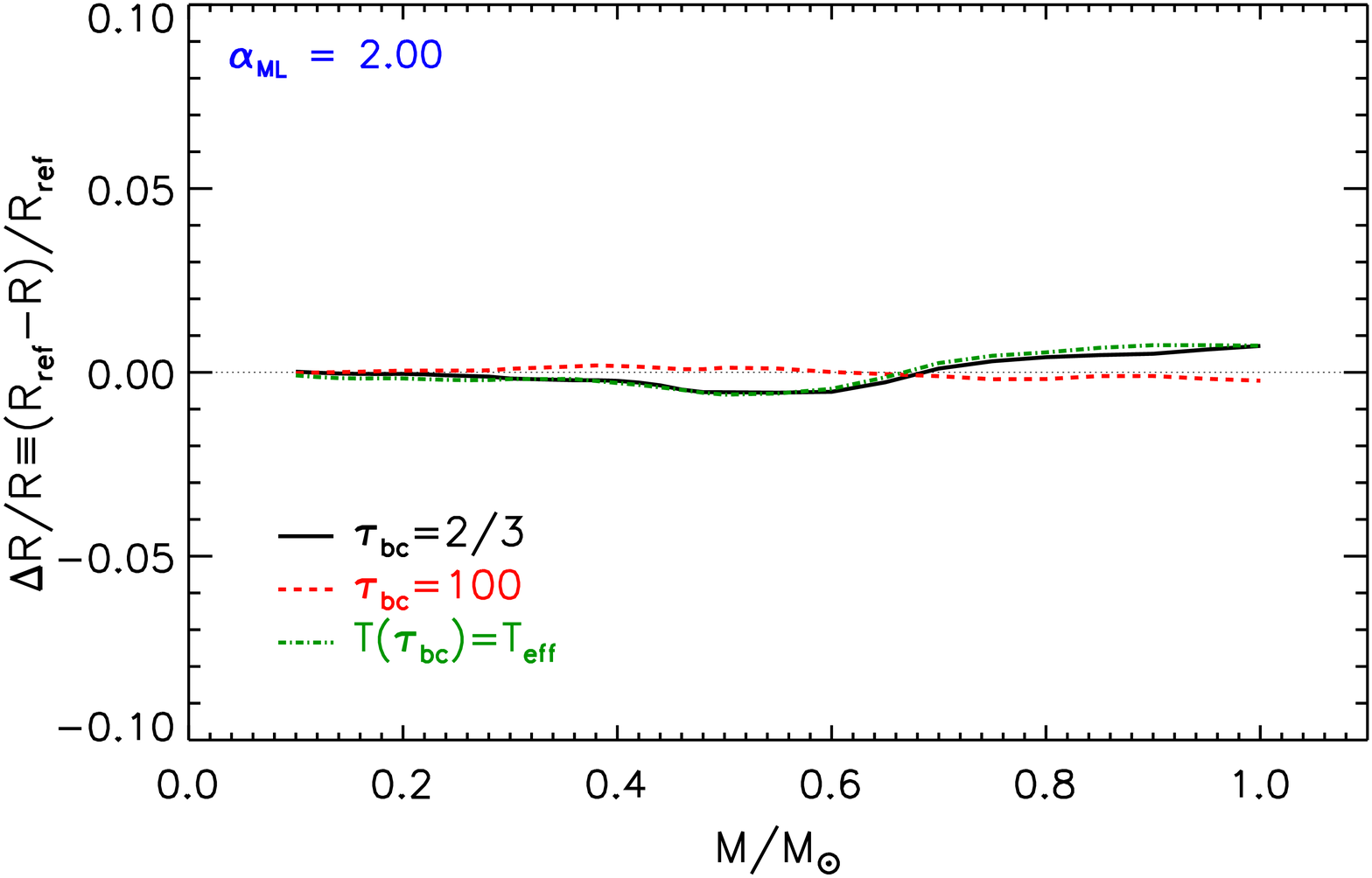}\\
	\includegraphics[width=\columnwidth]{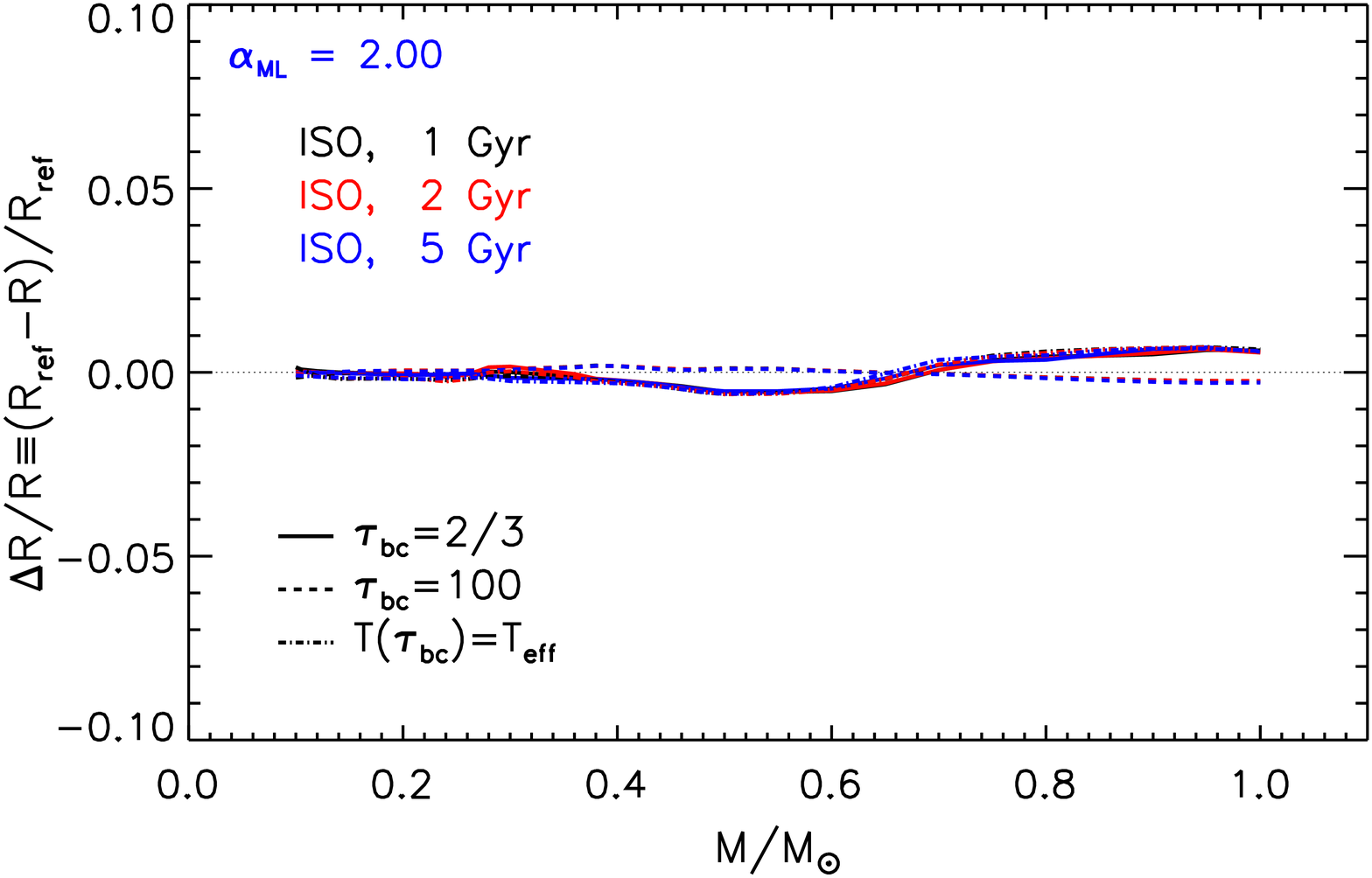}
	\caption{Relative radius variation as a function of the stellar mass due to the adoption of different $\tau_\rmn{bc}$ values (i.e. $\tau_\rmn{bc}=2/3$, 100 and $T(\tau_\rmn{bc})~=~T_\rmn{eff}$) with respect to the reference one ($\tau_\rmn{bc}=10$). Top panel: models on the ZAMS. Bottom panel: models on the 1, 2, and 5 Gyr isochrones.}
	\label{fig:tau}
\end{figure}
As anticipated in the previous section, the \bcs{} are also affected by the choice of $\tau_\rmn{bc}$. This quantity is freely chosen and generally it assumes values in the interval $2/3~\la~\tau_\rmn{bc}~\la~100$ \citep[see e.g. Table 2 in][]{tognelli11}. The choice of $\tau_\rmn{bc}$ is not trivial. Here we limit to recall that in the computations of the stellar interior (i.e. for $\tau~\gt~\tau_\rmn{bc}$) the diffusive radiative approximation has to be fulfilled. To this regard, \citet{morel94} showed that a value of $\tau_\rmn{bc}\ge 10$ should be adopted \citep[see also][]{trampedach14}. To test the dependency of the predicted radius on $\tau_\rmn{bc}$, we computed models for three values of $\tau_\rmn{bc}$, namely 2/3, 10 (our reference) and 100 (with the reference AHF11 atmosphere). We also computed a set of models where the atmosphere and the interior are matched at the point where the temperature in the atmosphere is equal to the star effective temperature, i.e. where $T_\rmn{bc}\equiv T(\tau_\rmn{bc})~=~T_\rmn{eff}$. This condition -- adopted by some authors -- does not correspond to a unique value of $\tau_\rmn{bc}$ during the whole stellar evolution, but to a range of values that generally are lower than 1 and close to 2/3. 

Fig.~\ref{fig:tau} shows the relative radius variation induced by the use of the quoted $\tau_\rmn{bc}$ in the case of \zams{} models. The adopted value of $\tau_\rmn{bc}$ has a small impact on the radius, being smaller than 1 percent. To be noted that the largest radius variation (about 1~percent) occurs when small values of $\tau_\rmn{bc}$ are adopted, i.e. $\tau_\rmn{bc}=2/3$ or $T(\tau_\rmn{bc})=T_\rmn{eff}$.  

We performed the same analysis using the isochrones, as shown in bottom panel of Fig.~\ref{fig:tau}. We found that the effect on the radius of the adopted $\tau_\rmn{bc}$ is independent of the age and it is the same shown for the \zams.
\begin{figure}
	\centering
	\includegraphics[width=\columnwidth]{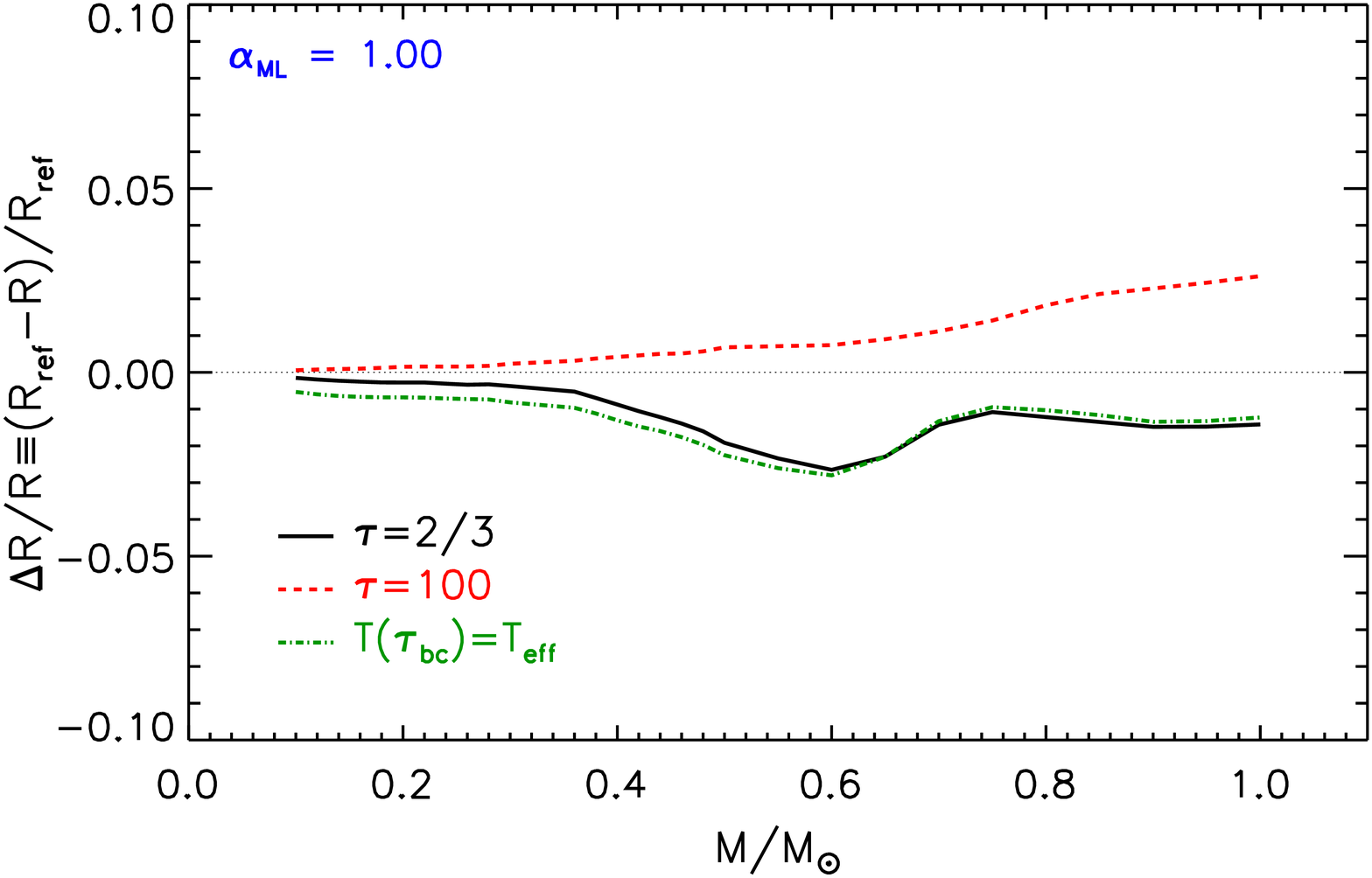}\\
    \includegraphics[width=\columnwidth]{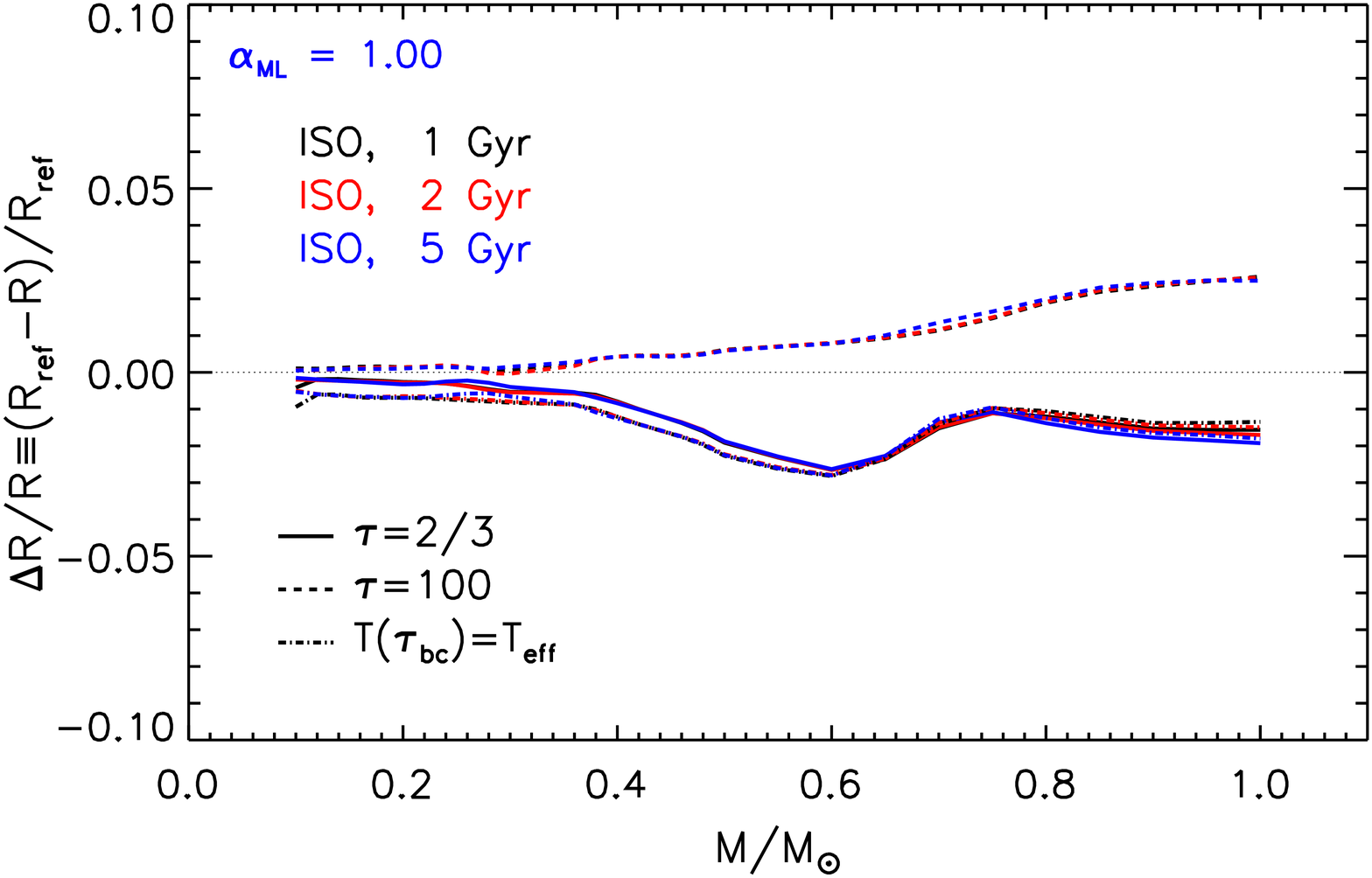}	
	\caption{The same as in Fig.~\ref{fig:tau} but for models with \ml=1.0.}
	\label{fig:tau_ml1}
\end{figure}

If \ml~=~1.00 is adopted, the same variations of $\tau_\rmn{bc}$ discussed above produce an effect much larger on the radius, but only for $M\ga 0.5$~\msun, as shown in Fig.~\ref{fig:tau_ml1}. For $\tau_\rmn{bc}=100$, the effect on the radius progressively increases from about $+1$ percent at 0.5~\msun{} to about $+2.5$ percent at 1~\msun. If $\tau_\rmn{bc}=2/3$ (or $T(\tau_\rmn{bc})=T_\rmn{eff}$) is adopted, the maximum of radius change occurs between 0.4~\msun{} ($-1$ percent) and 0.7~\msun{} (about $-1$ percent), with a peak at about 0.6~\msun{} (about $-2.5$ percent). For larger masses there is a constant variation of about $-1.5$ percent. The dependence of the effect of a $\tau_\rmn{bc}$ change on the \ml{} value is probably caused by the fact that we cannot consistently modify the \ml{} in the atmosphere, which in all the cases, is computed for \ml=2.00. This introduces a possible discontinuity in the temperature gradient at the matching point, i.e. at $\tau_\rmn{bc}$. However, depending on the stellar mass, if $\tau_\rmn{bc}$ is large enough the \ml{} step change occurs in a region where the super-adiabaticity is not large. In this case the \ml{} variation (from interior to atmosphere) has a small impact, leading to a very small discontinuity in the temperature gradient. This happens, for example, when one adopts large values of $\tau_\rmn{bc}$ in VLM stars. On the other hand, if small value of $\tau_\rmn{bc}$ are used, the match between interior and atmosphere occurs  where the super-adiabaticity is larger, thus more sensitive to the adopted \ml. In this case, the temperature gradient discontinuity can be sizeable, depending on the stellar mass, leading to an appreciable effect on the stellar structure and on its radius.  
\begin{figure}
	\centering
	\includegraphics[width=\columnwidth]{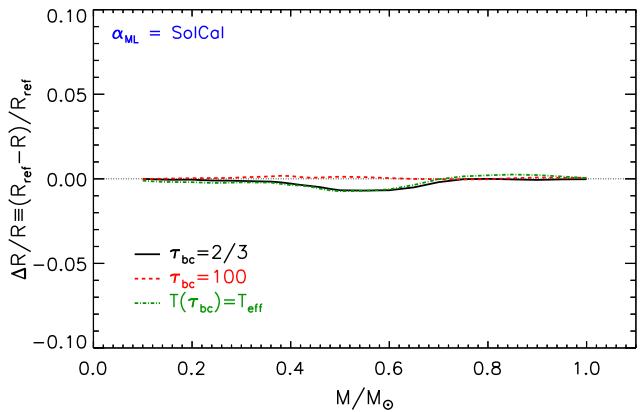}
	\caption{As in Fig~\ref{fig:tau} but for solar calibrated \ml{} values.}
	\label{fig:tau_solcal}
\end{figure}

As done for the atmospheric models, we analysed the effect on the radius of adopting a solar calibrated \ml{} in the models with perturbed $\tau_\rmn{bc}$. Fig.~\ref{fig:tau_solcal} shows the radius variation in ZAMS induced by the quoted $\tau_\rmn{bc}$ with the corresponding solar calibrated \ml, namely \ml$_{\sun}=1.92$ ($\tau_\rmn{bc}=2/3$), \ml$_{\sun}=1.93$ ($T(\tau_\rmn{bc})=T_\rmn{eff}$) and \ml$_{\sun}=2.05$ ($\tau_\rmn{bc}=100$). The adoption of different values of $\tau_\rmn{bc}$ in the solar calibration does not modify the initial chemical composition but only the adopted \ml. The radius variation due to the adopted $\tau_\rmn{bc}$ is progressively counterbalanced in models larger than about $0.7$~\msun{} by the use of a calibrated \ml$_{\sun}$, as shown in figure. For lower masses, the solar calibration has no effect on the stellar radius.

\subsection{Radiative opacity}
\label{sec:opacity}
\begin{figure}
	\centering
	\includegraphics[width=\columnwidth]{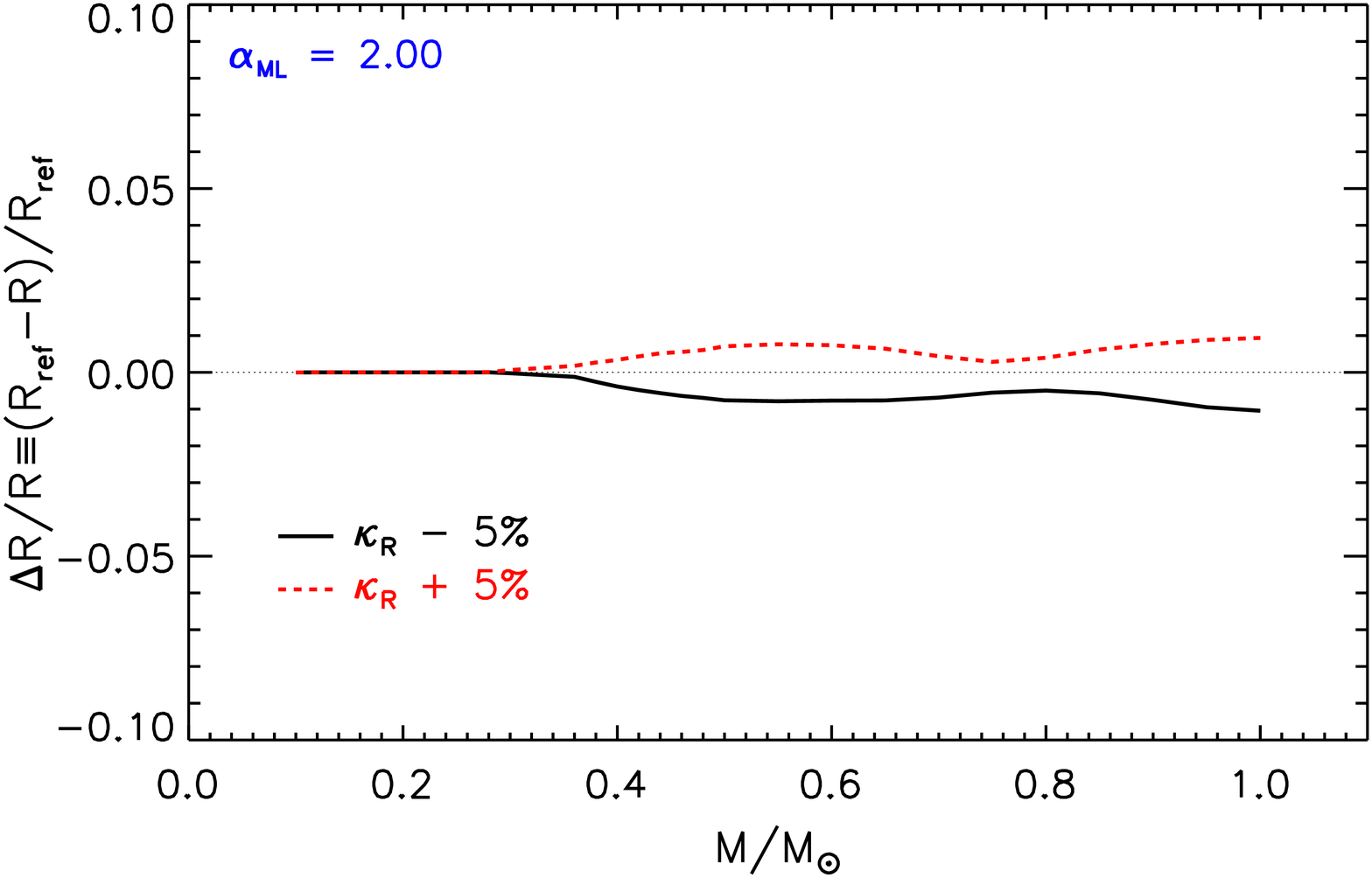}\\
	\includegraphics[width=\columnwidth]{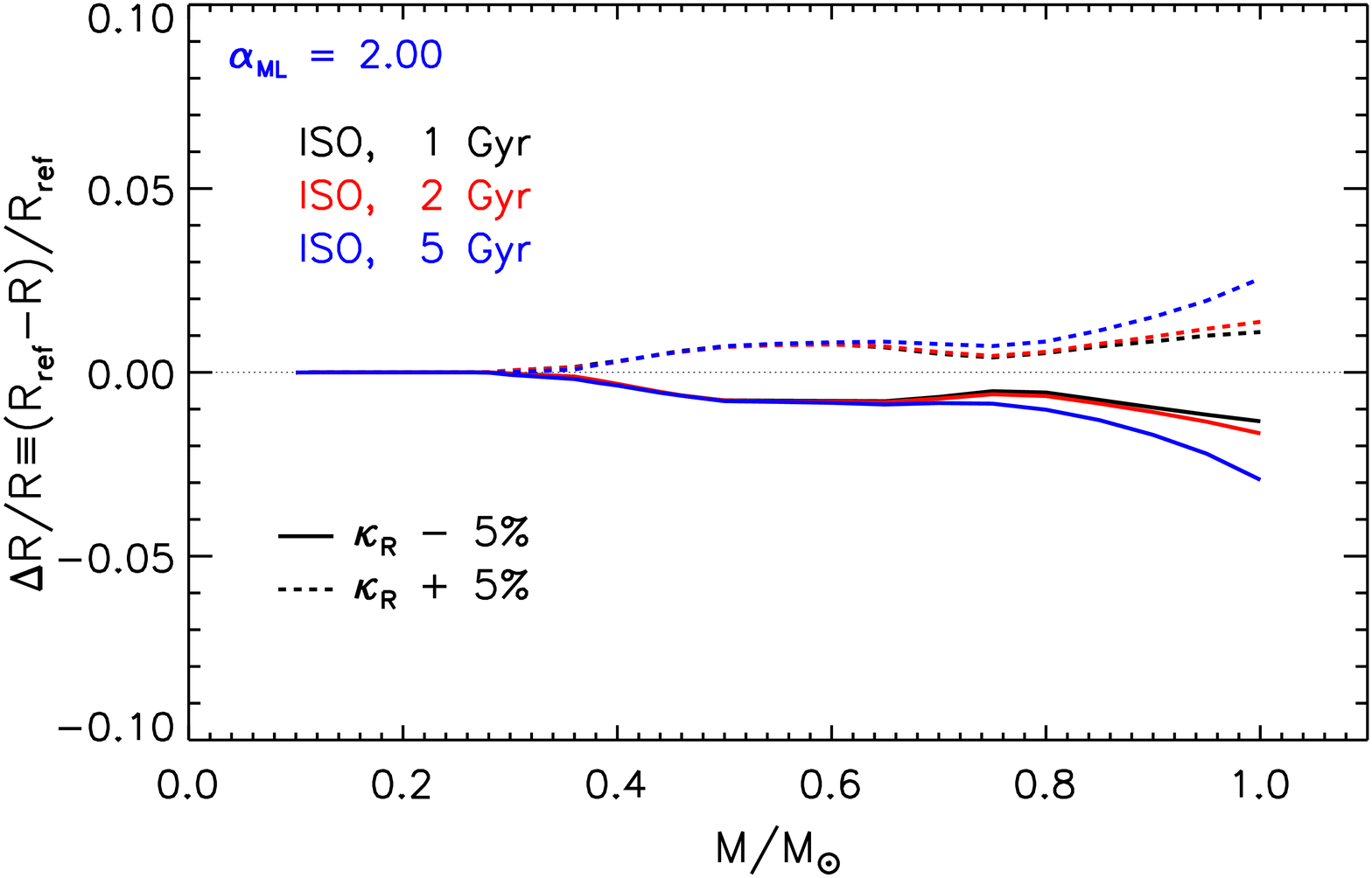}
	\caption{Relative radius variation as a function of the stellar mass due to the adoption of an uncertainty of $\pm 5 \%$ on the radiative opacity coefficients (only in the stellar interior) with respect to the reference one. Top panel: models on the ZAMS. Bottom panel: models on the 1, 2, and 5 Gyr isochrones.}
	\label{fig:kappa}
\end{figure}

The uncertainty in the Rosseland radiative opacity $\overline{\kappa}_\rmn{R}$ is not provided in the opacity tables. However, it is possible to obtain an estimate of the uncertainty by comparing our reference tables (\textsc{opal05}) with other tables used in the literature, i.e. the OP \citep{badnell05}. \citet[][see also \citet{tognelli12,tognelli15b}]{valle13a}  showed that such a comparison suggests an average uncertainty on $\overline{\kappa}_\rmn{R}$ of about $\pm 5$~percent in the temperature-density regime we are interested in. Such an uncertainty is approximatively the same found by \citet{lepennec2015} when comparing the \textsc{opal} and the \textsc{opas} \citep{blancard2012,mondet2015} new set of Rosseland mean opacity tables, in solar conditions (about 6~percent). If a wider range of temperature-pressure is considered, \citet{mondet2015} showed that the differences between the \textsc{opas} and the \textsc{opal} tables are a bit larger reaching about 12--13~percent, but mainly in region of low-temperature and low-density. We also note that recently \citet{bailey2015} found that, in conditions similar to those at the bottom of the solar convective envelope, the predicted monochromatic iron opacity in specific wavelength ranges is drastically underestimated when compared to measured values. They also suggested that the effect of using the measured iron opacity would increase the global Rosseland mean opacity by about 4--10~percent, at least for conditions similar to those at the bottom of solar convective envelope.

For the purpose of our investigation, we preferred to adopt an average value for the uncertainty on the radiative opacity instead of a temperature-density dependent error. To be consistent with the uncertainty values quoted above, we adopted a rigid uncertainty of $\pm 5$~percent in the whole temperature-density domain covered by our calculations.

The top panel of Fig.~\ref{fig:kappa} shows the effect on the radius of a radiative opacity variation of $\pm5$ percent with respect to the reference value for the \zams{} sequence. Notice that such an opacity variation is limited to the internal region of the star, i.e. $\tau~\gt~\tau_\rmn{bc}$ since the models shown in figure are computed adopting the AHF11 BCs. In this case we cannot vary the opacity in the atmosphere, since the atmospheric structure is provided as a pre-computed table. 

For $M~\la~0.3$~\msun{} the opacity variation is inconsequential. This is expected as the internal structure of a star with $M\la 0.3$~\msun{} is almost adiabatic in \zams /MS. Under this conditions the temperature gradient is mainly determined by the EOS (through the adiabatic gradient) and it is completely independent of the radiative opacity. On the other hand as the stellar mass increases, the temperature gradient in the external convective region becomes progressively more and more super adiabatic, thus more  sensitive to the adopted $\overline{\kappa}_\rmn{R}$. In addition, as the stellar mass increases above 0.4--0.5~\msun, the \zams /MS model is no more fully convective, as it has developed a radiative core during the \pms{} evolution. In this cases, the opacity plays a role in determining the temperature profile also inside the internal radiative region, thus also this part of the star is sensitive to $\overline{\kappa}_\rmn{R}$ variation. The effect of varying $\overline{\kappa}_\rmn{R}$ of $\pm 5$ percent is about 1 percent for $M~\ga~0.4$~\msun. 

The bottom panel of Fig.~\ref{fig:kappa} shows the effect of the opacity perturbation on the radius for models on the isochrones of 1, 2 and 5~Gyr. The relative radius variation is similar to that discussed for the \zams{} for $M~\la~0.7$--0.8~\msun{} while it becomes progressively more and more sensitive to the stellar age as the mass increases. In the worst case, i.e. 5~Gyr and 1~\msun, the perturbed radiative opacity leads to a radius variation of about 3 percent.
\begin{figure}
	\centering
	\includegraphics[width=\columnwidth]{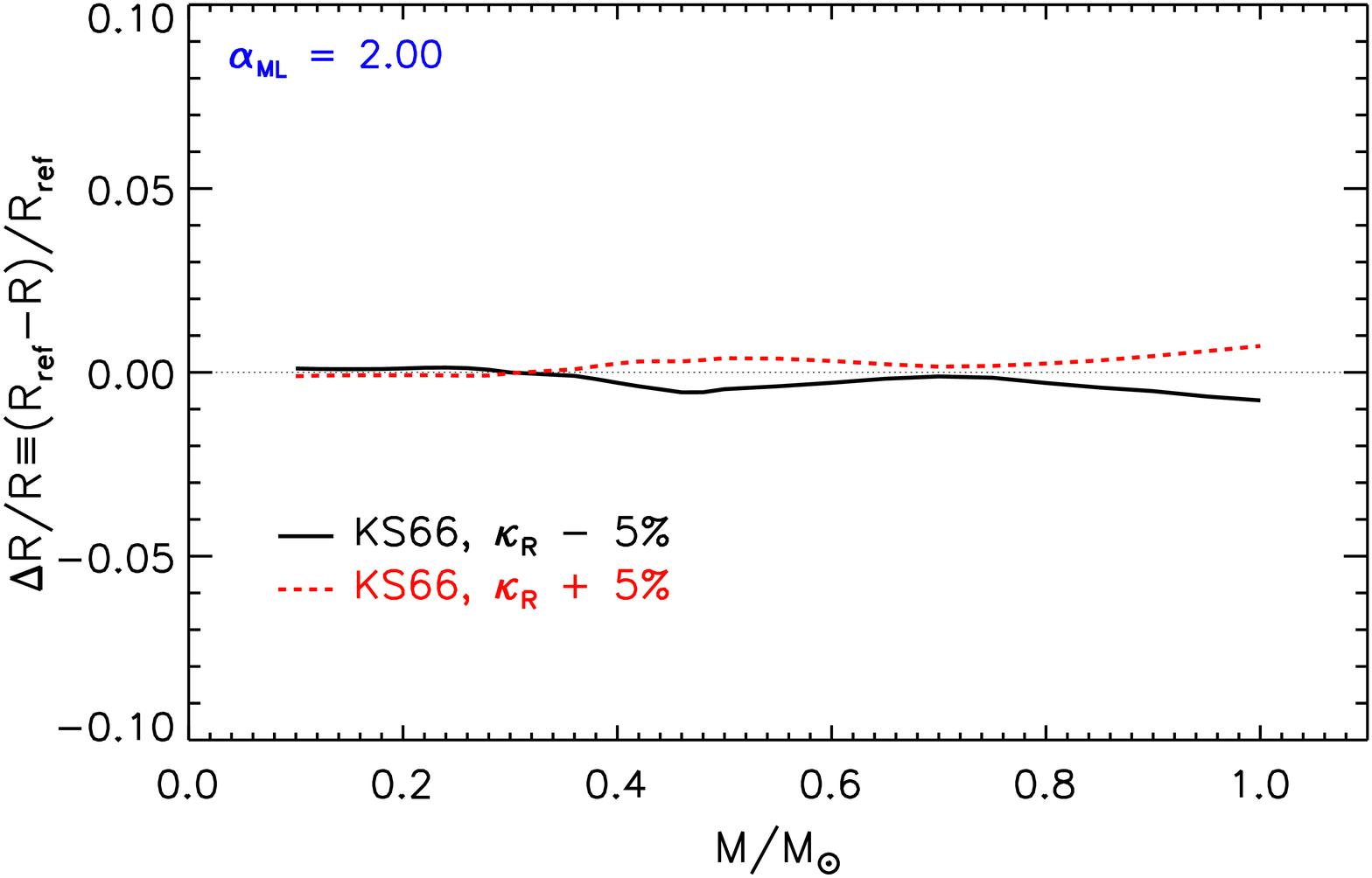}\\
	\includegraphics[width=\columnwidth]{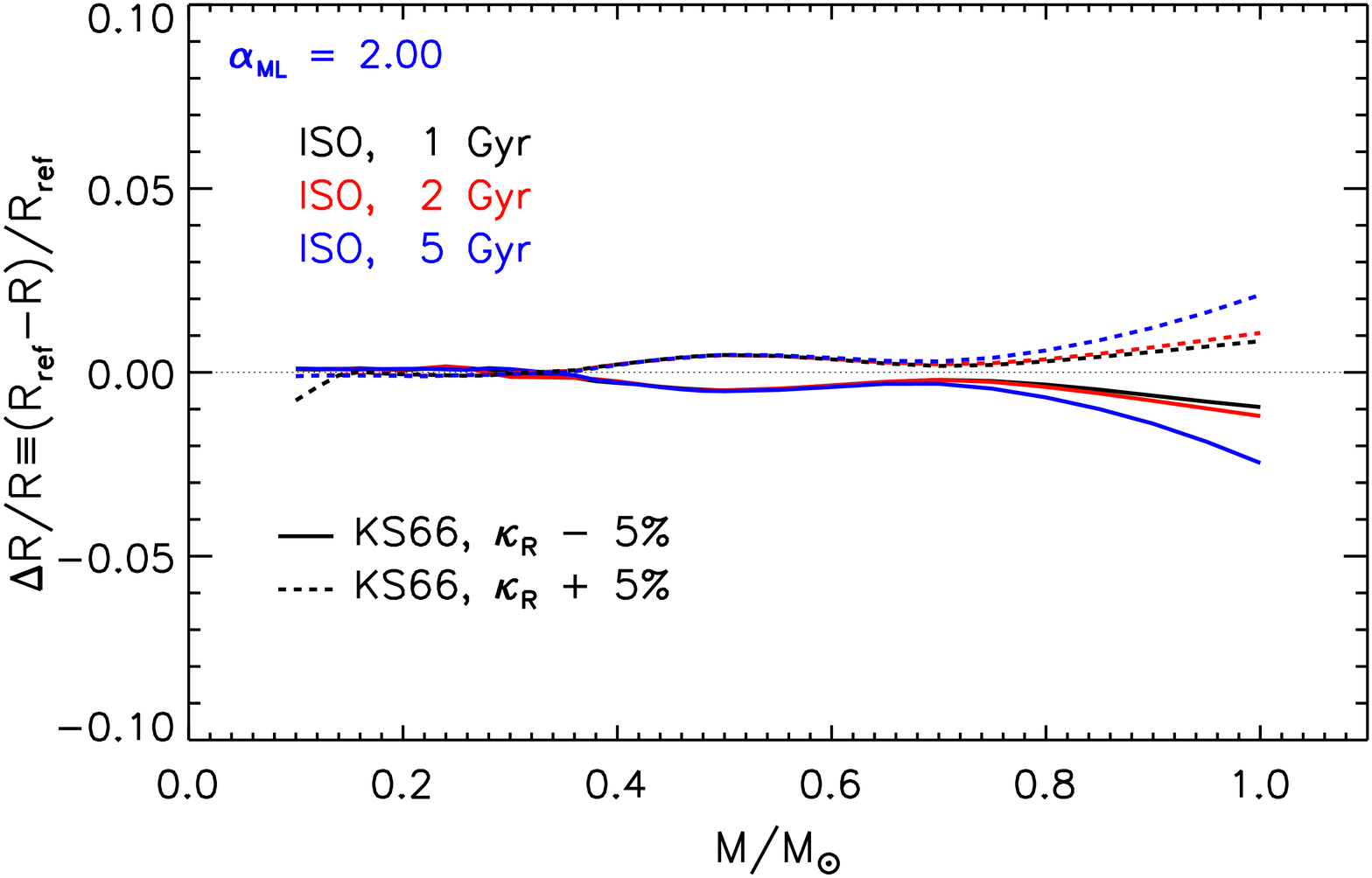}	
	\caption{Relative radius variation as a function of the stellar mass due to the adoption of an uncertainty of $\pm 5 \%$ on the radiative opacity coefficients in the whole structure with respect to the reference one. The perturbed and reference models have been computed adopting the KS66 atmospheric structure. Top panel: models on the ZAMS. Bottom panel: models on the 1, 2, and 5 Gyr isochrones.}
	\label{fig:kappa_ks66}
\end{figure}

The radius variation shown in Fig.~\ref{fig:kappa} does not account for the opacity change in the atmosphere. Thus, to estimate the effect of an opacity variation of $\pm 5$ percent in the whole structure, atmosphere included, we computed sets of perturbed and reference models with the KS66 \bcs. In these cases, the atmospheric structure is actually computed inside our stellar code together with the interiors,  allowing us to modify the opacity in the atmosphere too\footnote{We are aware that the KS66 is not a good choice for low-mass stars, but here we are interested in a differential analysis, which is only marginally affected by the use of such atmosphere.}. 

The top panel of Fig.~\ref{fig:kappa_ks66} shows the effect of an opacity variation of $\pm 5$ percent in the whole structure when the KS66 \bcs{} are adopted for the \zams{} sequence. For $M~\la~0.3$~\msun{} the models are almost unaffected by the opacity variation even in the atmosphere, but the dependency on $\overline{\kappa}_\rmn{R}$ increases with the stellar mass. For $M~\ga~0.4$~\msun{} the relative differences in radius are less than 1 percent, slightly smaller than those obtained using the non-grey \bcs{}. Comparing the results shown in Figs~\ref{fig:kappa} and \ref{fig:kappa_ks66} it emerges that the effect due to the opacity variation in the atmosphere partially counterbalances that caused by the opacity change in the interiors, thus slightly reducing the total effect on the stellar radius.

The bottom panel of Fig.~\ref{fig:kappa_ks66} shows the effect on the isochrones of the same opacity perturbation. The effect on the radius of the perturbed opacity depends on the age. The largest variation occurs for the 1~\msun{} model at an age of 5~Gyr, with a relative radius change of about 2--3 percent.

Another point that deserves to be discussed is the dependence of $\overline{\kappa}_\rmn{R}$ on the adopted heavy elements abundance (i.e. elements heavier than boron). Indeed, it is well known that metals strongly contribute to the radiative opacity coefficients, especially in the external regions of a star \citep[see e.g.][]{sestito06}. In the present models we assumed that the metals relative abundances are equal to those of the Sun (solar scaled mixture). However, the solar metal abundances issue is still under debate, as witnessed by the several solar abundance revisions released in recent years \citep[e.g.][AS09]{asplund05,caffau08,caffau11}. Thus, it is worth to evaluate the effect of adopting different solar metal abundances. To do this, we compared the results obtained using the AS09  compilation (reference) and the still largely adopted \citet[][hereafter GS98]{grevesse98} mixture, when computing the $\overline{\kappa}_\rmn{R}$ coefficients. 
\begin{figure}
	\centering
	\includegraphics[width=\columnwidth]{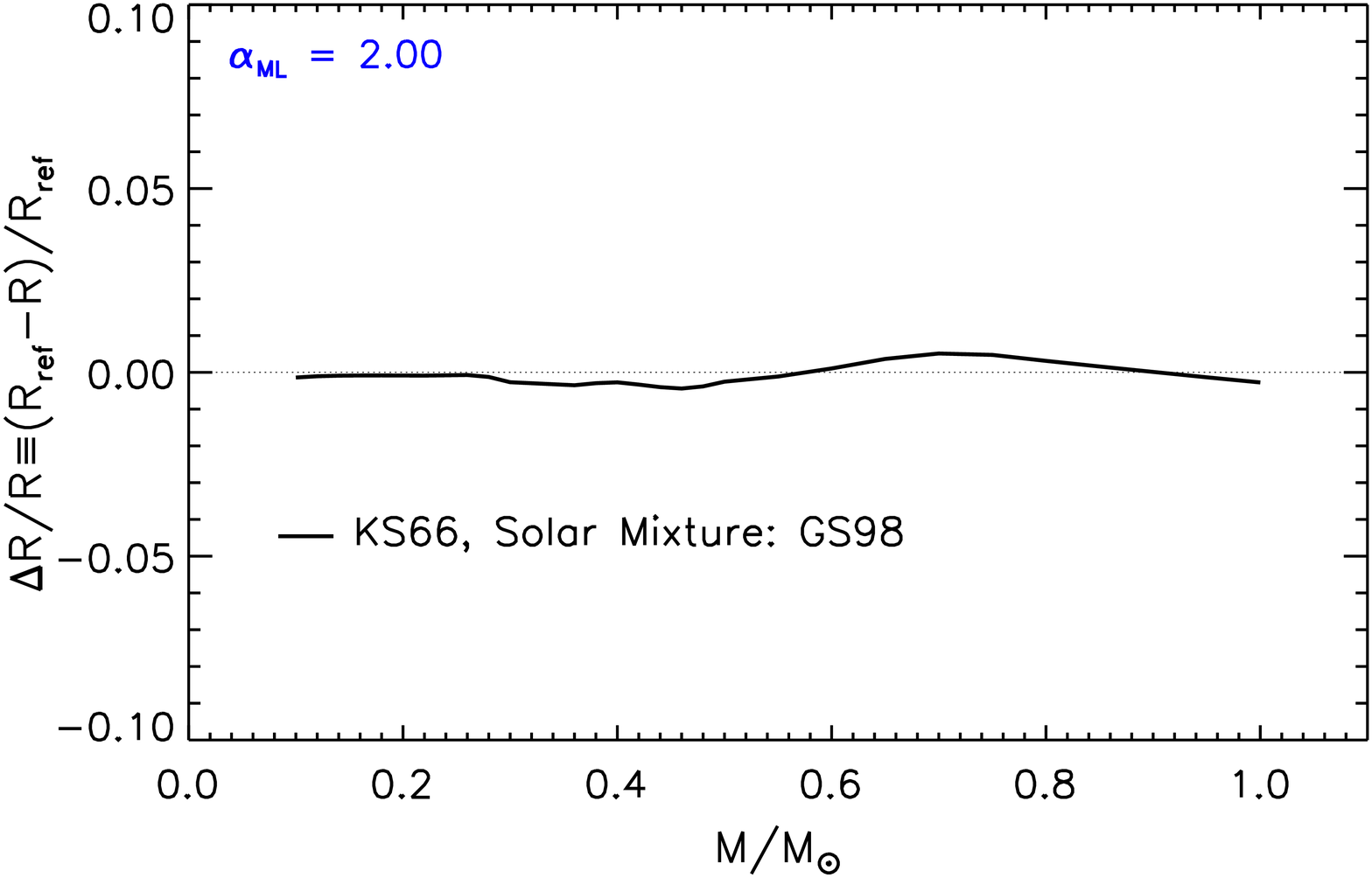}\\
	\includegraphics[width=\columnwidth]{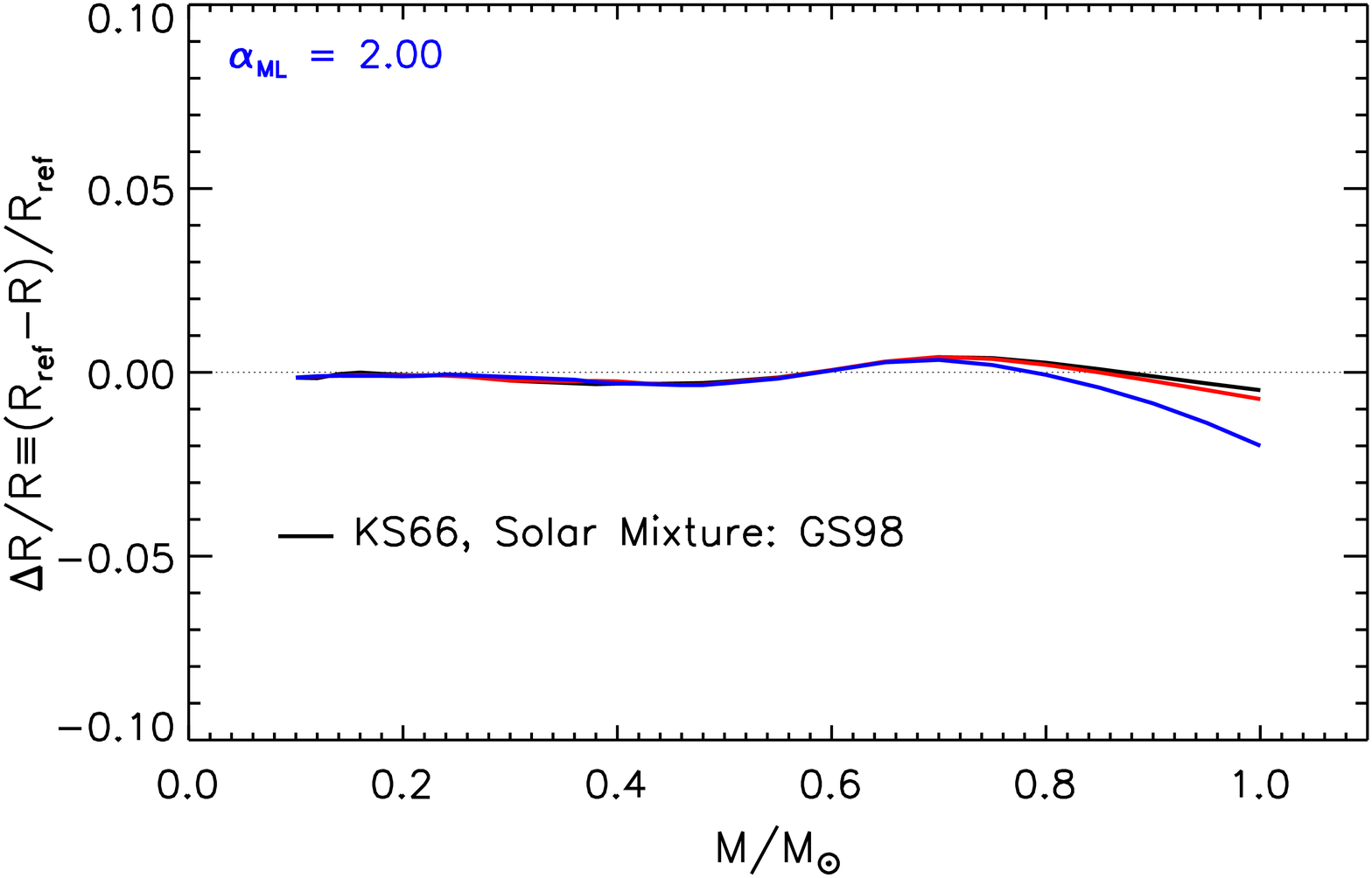}
	\caption{Relative radius variation as a function of the stellar mass due to the adoption of the GS98 solar mixture in the radiative opacity coefficients with respect to the reference one (AS09). Top panel: models on the ZAMS. Bottom panel: models on the 1, 2, and 5 Gyr isochrones.}
	\label{fig:kappa_gs98}
\end{figure}

Top panel of Fig.~\ref{fig:kappa_gs98} shows the relative variation of the surface radius in ZAMS at a given mass if the GS98 mixture is adopted (instead of the AS09) in the $\overline{\kappa}_\rmn{R}$ coefficient computations. We used exactly the same opacity tables used for the standard models, i.e. the F05 (for low temperatures) and the \textsc{opal} (for high temperatures), but with the GS98 solar heavy elements abundances. The KS66 BCs have been used to check the effect of the opacity variation also in the atmosphere. The effect of the mixture on the stellar radius is almost negligible (if compared to the other uncertainty sources), being generally smaller than 1 percent over the whole selected mass range. 

Bottom panel of Fig.~\ref{fig:kappa_gs98} shows the effect of the opacity variation due to the heavy element abundances on the radius for models on the isochrones. As in the previous cases, only for $M~\ga~0.7$~\msun{} the radius change gets sensitive to the age. The maximum effect reaches 2 percent at 1~\msun. We obtained similar results (for ZAMS and isochrones) using the non-grey AHF11 BCs.

For both the \zams{} and isochrones models the radius variation due to the radiative opacity change (i.e. $\pm 5$ percent perturbation or heavy element abundance modification) is slightly dependent on the adopted \ml. In particular we found that the peculiar `waving' around 0.70--0.85~\msun{} tends to disappear if \ml=1.00 models are used in both ZAMS and isochrones models. The other masses are not affected by the adopted \ml{} value.

We also checked the impact of the solar calibration on the models with the perturbed radiative opacity. A perturbation of $\overline{\kappa}_\rmn{R}$ affects not only the stellar radius but also its luminosity and temporal evolution. As a consequence, the solar calibration procedure yields values of \ml$_{\sun}$ and of the initial chemical composition ($Y_{\sun}$ and $Z_{\sun}$) different from those obtained for the reference case. In particular, by varying $\overline{\kappa}_\rmn{R}$ of $+5$ and $-5$ percent we obtained, respectively, ($Y_{\sun}$, $Z_{\sun}$)=(0.2705, 0.01510) and (0.2540, 0.01537), while the initial chemical composition of the solar calibrated model in the reference case is (0.2624, 0.01523). It is evident that the radiative opacity perturbation induces a change in the initial helium (3~percent) and metal content (1~percent) in the Sun. On the other hand the variation of \ml, due to the solar calibration, is very small, about 0.02--0.03 (KS66 BCs) or 0.03--0.04 (AHF11 BCs). Given such a situation, the change due to the different initial helium content (and metallicity) is much larger than that caused by the solar use of the calibrated \ml$_{\sun}$. However, keeping fixed the initial chemical composition (as discussed), we do not account for such an effect. As a consequence, the radius variation for the \zams{} or isochrones models obtained in the case of solar calibration of \ml{} on the perturbed models is the same obtained when no calibration is performed at all, even at masses close to 1~\msun. The same occurs if the opacity variation is caused by the adoption of a different mixture, i.e. the GS98, and the solar calibration does not change the results shown in Fig.~\ref{fig:kappa_gs98}.

\subsection{Equation of state}
\label{sec:eos}
\begin{figure}
	\centering
	\includegraphics[width=\columnwidth]{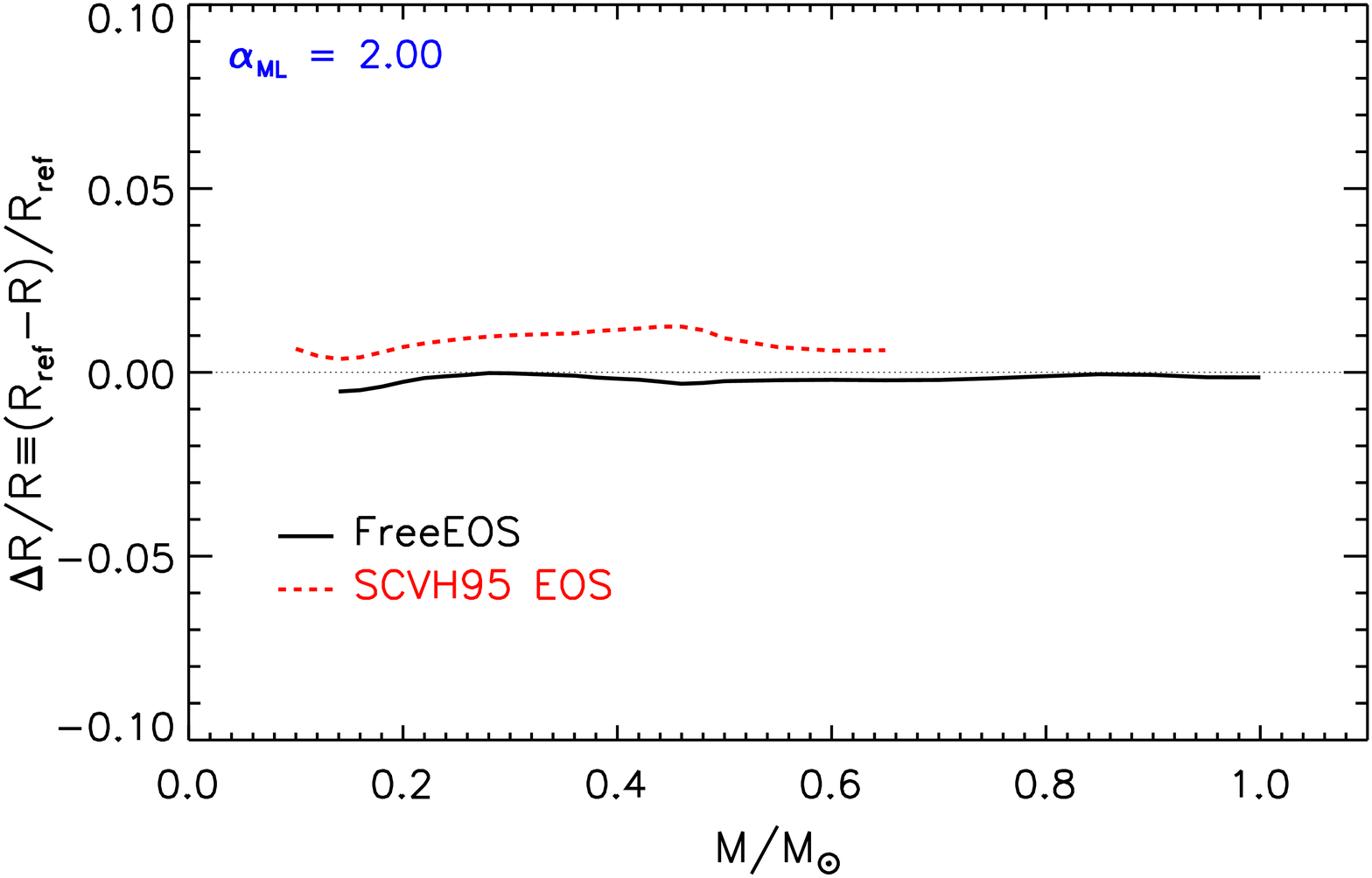}\\
	\includegraphics[width=\columnwidth]{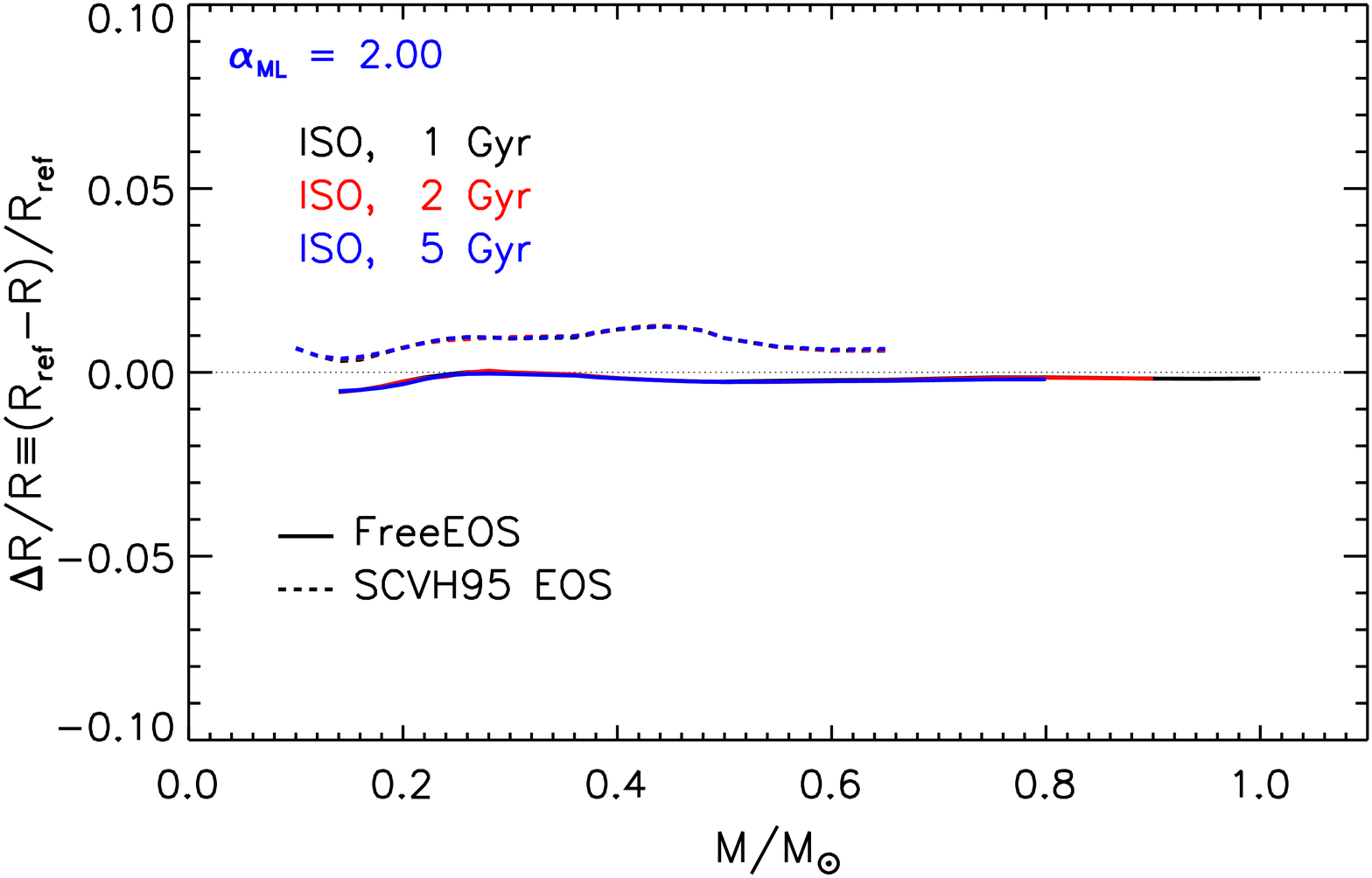}
	\caption{Relative radius variation as a function of the stellar mass due to the adoption of a different EOS (i.e. \textsc{scvh95} and \textsc{FreeEOS}) with respect to the reference one (\textsc{opal06}$+$\textsc{scvh95}). Top panel: models on the ZAMS. Bottom panel: models on the 1, 2, and 5 Gyr isochrones.}
	\label{fig:eos}
\end{figure}
As in the case of the radiative opacity, the current generation of equation of state (EOS) tables do not provide the uncertainties affecting the thermodynamical quantities. As such, a proper analysis of their propagation in stellar models cannot be performed. Even the simplified approach followed in the previous section for radiative opacity -- by rigidly increasing/decreasing the EOS values by a constant factor -- is not feasible since the different thermodynamical quantities of interest (i.e. pressure, density, adiabatic gradient, specific heat, etc.) are strictly correlated among each other. To have a first idea of the impact of the EOS uncertainty on stellar radius, we thus simply computed sets of models by changing the EOS tables, keeping fixed all the other input physics and parameters. We used two EOS largely adopted in the literature, namely the \textsc{scvh95} \citep{saumon95} and the \textsc{FreeEOS}\footnote{We used the \textsc{FreeEOS} in the EOS~1 configuration (as recommended by the author), which accounts for all the available ionizations states treated in detail. This configuration should give the best agreement with the \textsc{opal} and \textsc{scvh95} \eos.} \citep{irwin08}. This comparison is intended to give an estimation of the agreement/disagreement level between few \eos{} largely used in the regime of LM and VLM stars. Note that, less recent and out-dated \eos{} that neglects non-ideal effects or approximatively treat partial ionizations might strongly affect the characteristics of VLM and LM stars \citep[see e.g.][]{chabrier97,siess01,dicriscienzo10}.

The results of the computations are shown in Fig.~\ref{fig:eos}. For $M~\ga~0.65$~\msun{} the sole \textsc{scvh95 EOS} is not able to fully cover the temperature-pressure range of \zams{} models, thus the comparison is restricted on the mass range $M~\lt~0.65$~\msun. 

As previously mentioned, our reference \eos{} is the \textsc{opal} extended with the \textsc{scvh95} in the low temperature - high density regime. As such, in  $M\la 0.$2~\msun, both the \eos{} are used (but in different regions of the structure).

From Fig.~\ref{fig:eos} it is evident that the effect of the \eos{} variation on the radius is very small if the \textsc{FreeEOS} is adopted. This is expected as the \textsc{FreeEOS} has been developed to produce results similar to the \textsc{opal}. The largest differences between the predicted radii reaches about 1--1.5 percent if the sole \textsc{scvh95} is adopted.

It is not easy to clearly address the main cause of such differences. We checked that in the typical temperature-pressure regime of LM and VLM stars the \textsc{opal06} and the \textsc{FreeEOS} tables provides similar values of the thermodynamical quantities of interest (i.e. $C_\rmn{P}$ and $\nabla_\rmn{ad}$), usually within $\pm 1$--2~percent, with slightly larger values in the worst cases. On the other hand the \textsc{scvh95} is sensitively different with respect to the \textsc{opal}. For temperatures lower than about $10^5$~K, the relative differences of the adiabatic gradient (and specific heat) ranges between $\pm 5$ percent, but they can reach, in the the hydrogen and helium ionizations regions, differences up to about $\pm 10$--15~percent. 

The uncertainty due to the EOS is independent of both the stellar age (bottom panel of Fig.~\ref{fig:eos}) and the \ml{} value.

In this case we did not perform a comparison between solar calibrated \ml{} models because (1) the sole \textsc{scvh95} does not allow to compute a solar model, and (2) the \textsc{FreeEOS} is very similar to the \textsc{opal} at 1~\msun{} and the solar calibration is inconsequential.

\subsection{Nuclear cross-sections}
\label{sec:cross_section}
We analysed the effect of perturbing the proton-proton (pp)-chain reaction rates, namely, p(p,$e^+\nu_e$)d, p(d,$\gamma$)$^3$He, $^3$He($^3$He,2p)$^4$He, $^3$He($^4$He,$\gamma$)$^7$Be, p($^7$Be,$\gamma$)$^8$B and $^7$Be($e^-$,$\nu_e$)$^7$Li, within their current uncertainty. The perturbation of the selected burning channels does not affect the radius of models in \zams. We also verified that it does not change the radius of models on the 1, 2, and 5~Gyr isochrones. In the following we have neglected these uncertainty sources.

\subsection{Mixing length}
\label{sec:ml}
\begin{figure}
	\centering
	\includegraphics[width=\columnwidth]{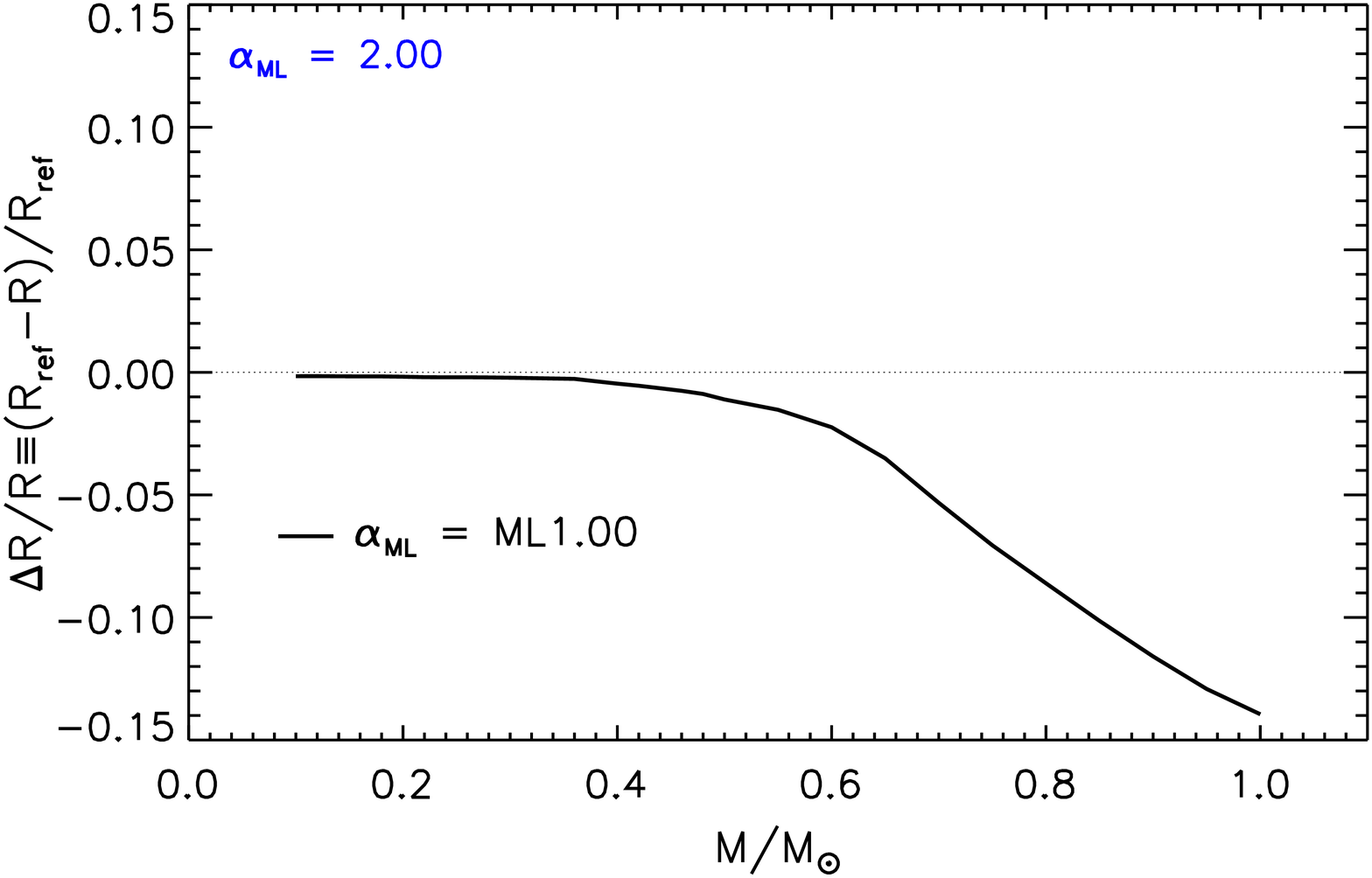}\\
	\includegraphics[width=\columnwidth]{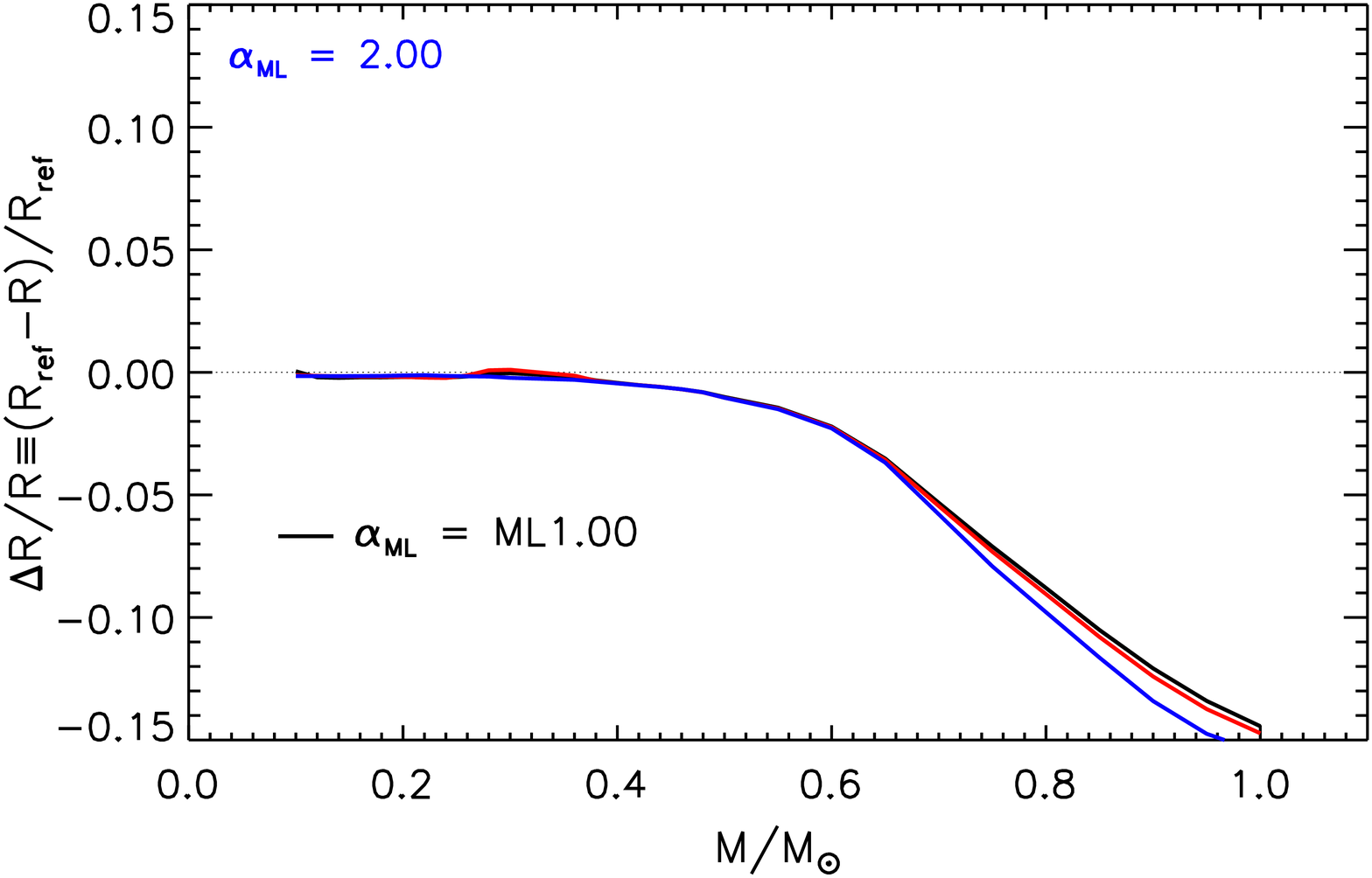}
	\caption{Relative radius variation as a function of the stellar mass due to the adoption of \ml=1.00 with respect to the reference one (\ml=2.00). Top panel: models on the ZAMS. Bottom panel: models on the 1, 2, and 5 Gyr isochrones.}
	\label{fig:ml}
\end{figure}
The radius of stars with a convective envelope depends on the efficiency of the convective transport and, following the mixing length formalism, on the mixing length parameter \ml. A decrease of \ml{} translates into a less efficient energy transport -- hence into a larger temperature gradient -- and into a larger radius. In order to quantify such an effect in the regime of LM and VLM stars, we computed two set of models, one with the reference solar calibrated \ml{} value (i.e. \ml~=~2.00) and the other with \ml~=~1.00. We recall that in the reference atmospheric table (AHF11) \ml{} is fixed to 2.00 and it cannot be modified. So, we can actually analyse the impact of the adopted \ml{} only the interior of the star. However, for VLM stars the value of \ml{} used in the atmosphere should be not crucial, because this objects are so dense (in ZAMS) to be almost adiabatic even in the layers at the bottom of the atmosphere, which mainly determine the ($T_\rmn{bc}$, $P_\rmn{bc}$) used to derive the outer BCs \citep[see e.g.][]{chabrier97,baraffe15}. For larger masses with partially-convective and super-adiabatic atmosphere the situation might be different, and the effect is not quantifiable without proper atmospheric models.

Top panel of Fig.~\ref{fig:ml} shows the radius relative differences between \zams{} model computed with \ml~=~2.00 and 1.00. The effect of changing \ml{} is negligible for $M\la 0.3$-0.4~\msun, while it progressively increases up to about 14~percent with the stellar mass. The reason of such a behaviour relies in the super-adiabaticity degree inside the star. In the VLM star tail of the explored range, stars are so compact that even the convective envelope is adiabatic and hence the temperature gradient is independent of \ml. On the other hand, for more massive stars the extension of the super adiabatic region in the convective envelope increases with stellar mass leading to a progressively larger sensitivity of the radius on the mixing length parameter value.

Bottom panel shows the effect of the mixing length on the radius of models along the isochrones. There is a slight dependence on the age for the larger masses. The maximum radius variation is achieved for the 1~\msun{} at 5~Gyr (i.e. 15--16~percent).

\section{Uncertainty in the adopted initial chemical composition}
\label{sec:chm}
To compute a stellar model, suitable initial chemical abundances have to be provided (i.e. initial helium $Y$, total metallicity $Z$, and heavy elements mixture). Unfortunately, helium abundance cannot be observed and in most of the cases only [Fe/H] is available from spectroscopy.

The initial helium abundance and the total metallicity can be obtained from [Fe/H] value by making some assumptions. Assuming a solar-scaled mixture and a linear relation between the initial helium and metallicity (both valid at least for Population I stars), $Y$ and $Z$ can be obtained using the following relations \cite[see e.g.][]{gennaro10}:
\begin{eqnarray}
Y&=& Y_\rmn{P} + \frac{\Delta Y}{\Delta Z}Z\label{eq:y}\\
Z&=& \frac{(1-Y_\rmn{P})(Z/X)_{\sun}}{10^{-[\rmn{Fe}/\rmn{H}]} + (1+\Delta Y / \Delta Z)(Z/X)_{\sun}}\label{eq:z}
\end{eqnarray} 
The derived $Y$ and $Z$ depend on some parameters, namely the helium-to-metals enrichment ratio $\Delta Y / \Delta Z$ \citep[we adopted 2 as reference, see][]{casagrande07}, the solar metals-to-hydrogen ratio $(Z/X)_{\sun}$ (we adopted 0.0181 as reference; see AS09), and the primordial helium abundance $Y_\rmn{P}$ \citep[we adopted 0.2485 as reference; see][]{cyburt04}. Such quantities are known within an uncertainty that eventually propagates into the final helium $\delta Y$ and total metallicity $\delta Z$ error. The uncertainties in [Fe/H], $\Delta Y / \Delta Z$, $(Z/X)_{\sun}$, and $Y_\mathrm{P}$ we use are, respectively, $\pm 0.1$~dex, $\pm 1$, $\pm 15$~percent and $0.0008$ \citep{tognelli15b}.

To estimate the effect of the initial chemical composition uncertainty on the stellar radius, we first analysed the effect of perturbing separately $Y$ and $Z$, using the error in [Fe/H] and $\Delta Y / \Delta Z$. To do this, first we fixed $Z$ to its reference value ($Z=0.013$) and vary $\Delta Y / \Delta Z$ to compute the approximated values for the perturbed $Y$. Secondly, we fixed $Y$ to its reference value ($Y=0.274$) and vary [Fe/H] to obtain the approximated perturbed $Z$ values.

However, $Y$ and $Z$ are not independent among each other, consequently, as a second step, we analysed the effect on the total radius of simultaneously varying the couple ($Y$, $Z$) as obtained directly from equations~(\ref{eq:y}) and (\ref{eq:z}) once [Fe/H], $\Delta Y / \Delta Z$, and $(Z/X)_{\sun}$ are perturbed within their maximum variability range.

Where not explicitly stated, we used $Y=0.274$, $Z=0.013$ as reference. We set the initial deuterium abundance to $X_\rmn{d} = 2\times10^5$ and the light elements abundances (inconsequential for this work) to the same values given in \citet{tognelli15b}. The heavy elements abundances are obtained assuming a solar-scaled metal distribution with the AS09 abundances.

\subsection{Independent variation of the initial helium abundance}
\label{sec:y}
\begin{figure}
	\centering
	\includegraphics[width=\columnwidth]{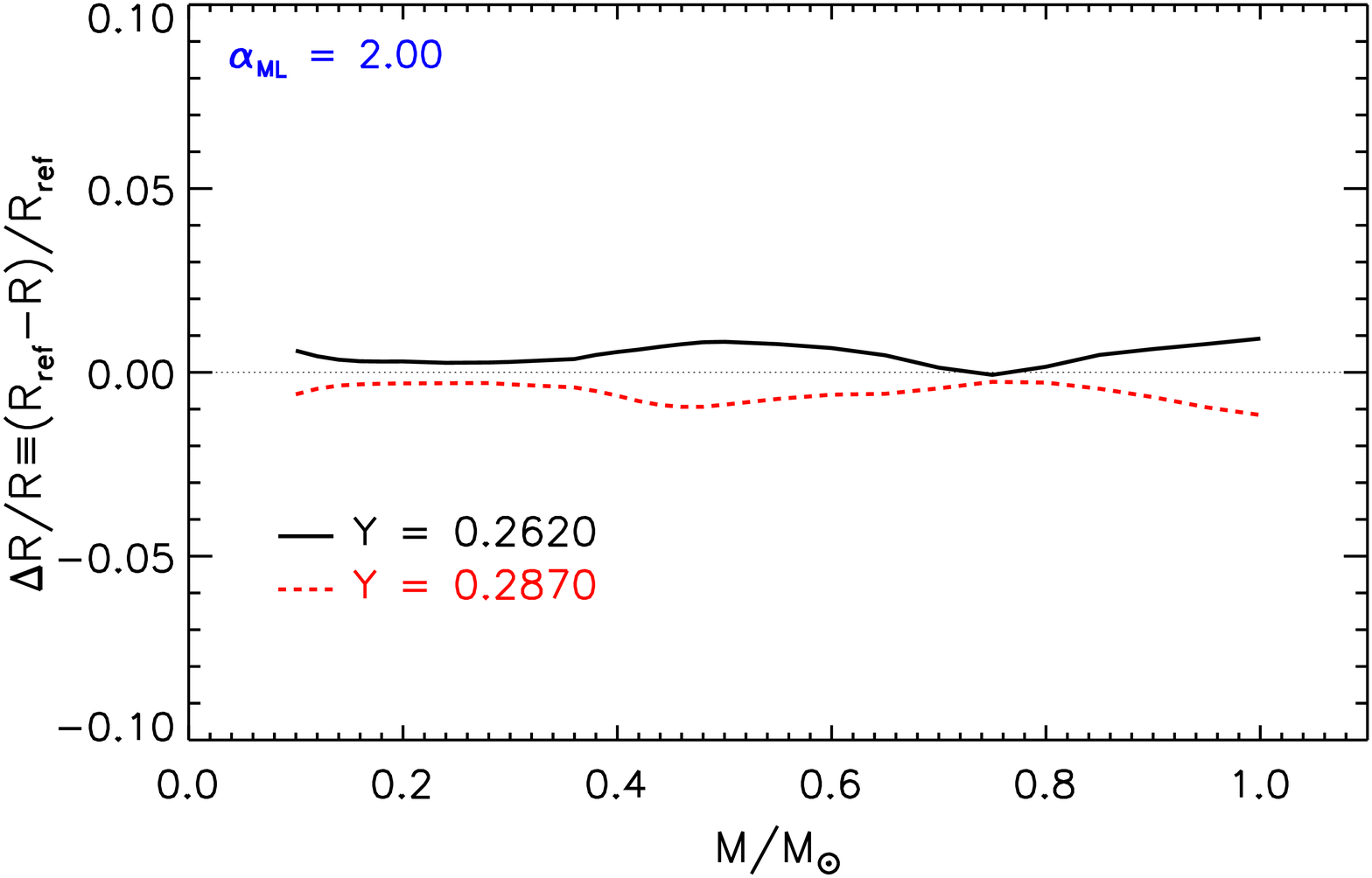}\\
	\includegraphics[width=\columnwidth]{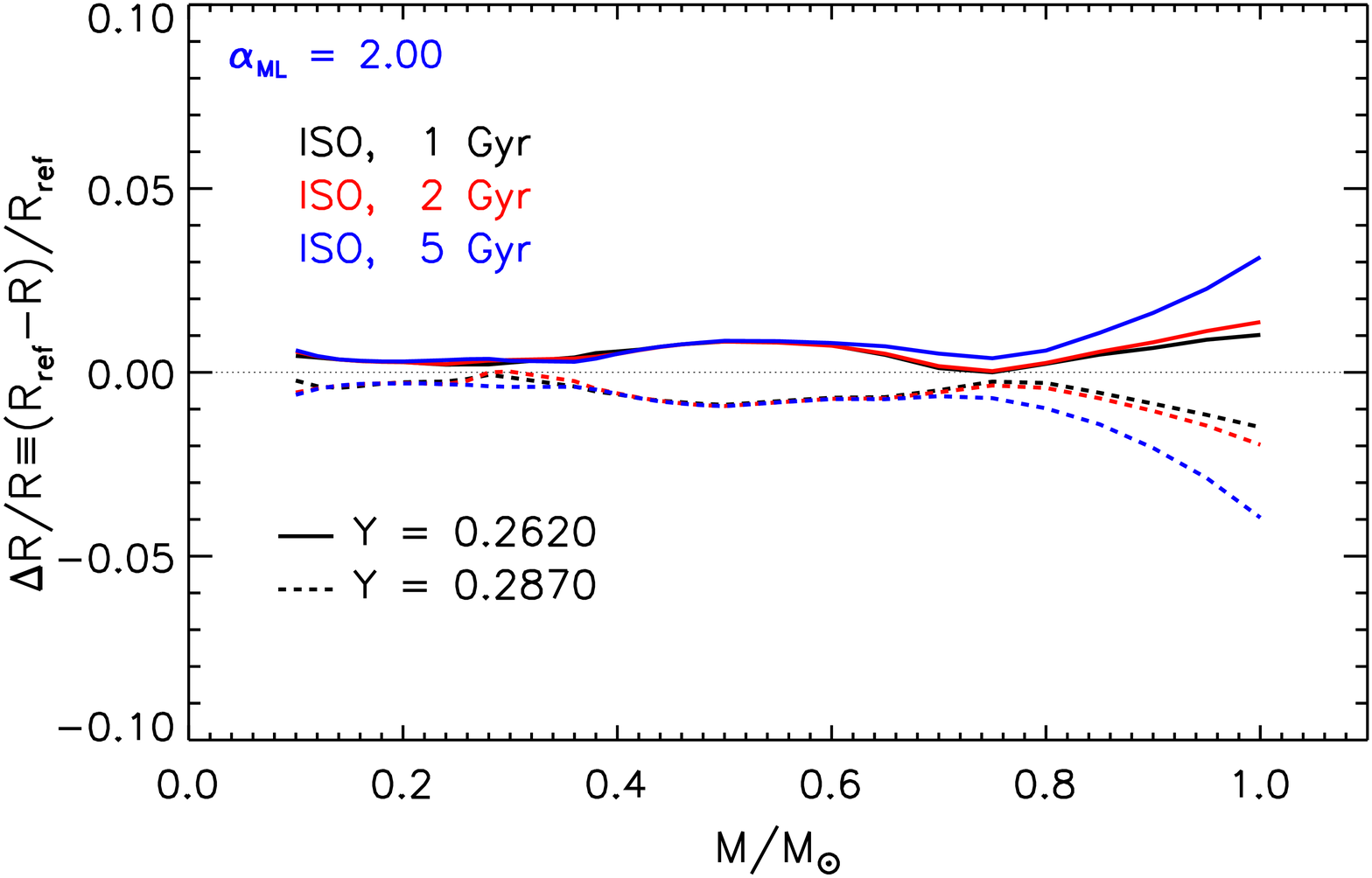}
	\caption{Relative radius variation as a function of the stellar mass due to the adoption of a different initial helium abundance (i.e. $Y=0.262$ and 0.287) with respect to the reference one ($Y=0.274$). Top panel: models on the ZAMS. Bottom panel: models on the 1, 2, and 5 Gyr isochrones.}
	\label{fig:y}
\end{figure}
The variation of the initial helium abundance $Y$ comes essentially from the uncertainty on $\Delta Y/\Delta Z$ and $Y_\rmn{p}$.  Regarding $Y_\rmn{P}$ we used the recent value $Y_\rmn{P} = 0.2485\pm 0.0008$ \citep{cyburt04}. Given its relatively small uncertainty, in the following, we will neglect the error of $Y_\rmn{p}$. 

Keeping fixed the total metallicity to its reference value $Z=0.013$ (and the other parameters) and perturbing only $\Delta Y/\Delta Z$ in equation~(\ref{eq:y}), we obtained the following extreme values for $Y$, $Y=0.262$ and $Y=0.287$, which correspond to a variation of $Y$ of about $\pm 4$~percent. 

Top panel of Fig.~\ref{fig:y} shows the relative radius difference between the models with the varied initial helium abundance and the reference one for the \zams{} models. An increase of the helium content produces, at a fixed mass, a star brighter and hotter in \zams. The total effect is to increase the stellar radius less than 1 percent for $M~\la~0.3$--0.4~\msun{} and of about 1 percent for larger masses. The opposite effect is achieved if the initial helium is reduced. The amount of the radius variation is symmetric because a symmetric perturbation on $Y$ is assumed.

Bottom panel of Fig.~\ref{fig:y} shows the effect of the helium variation on the radius for models on the isochrones. As in the other cases, only for masses above 0.7-0.8~\msun{} the radius variation depends on the age. The maximum radius change occurs at 5~Gyr, reaching about 4 percent.

Such differences are only marginally affected by the change of \ml{} from 2.00 to 1.00. Only stars in the mass range 0.70--0.85~\msun{} are affected. The waving disappears and the radius variation is essentially constant to $\pm 1$~percent. The same occurs for isochrones models.

\subsection{Independent variation of the initial metallicity}
\label{sec:z}
\begin{figure}
	\centering
	\includegraphics[width=\columnwidth]{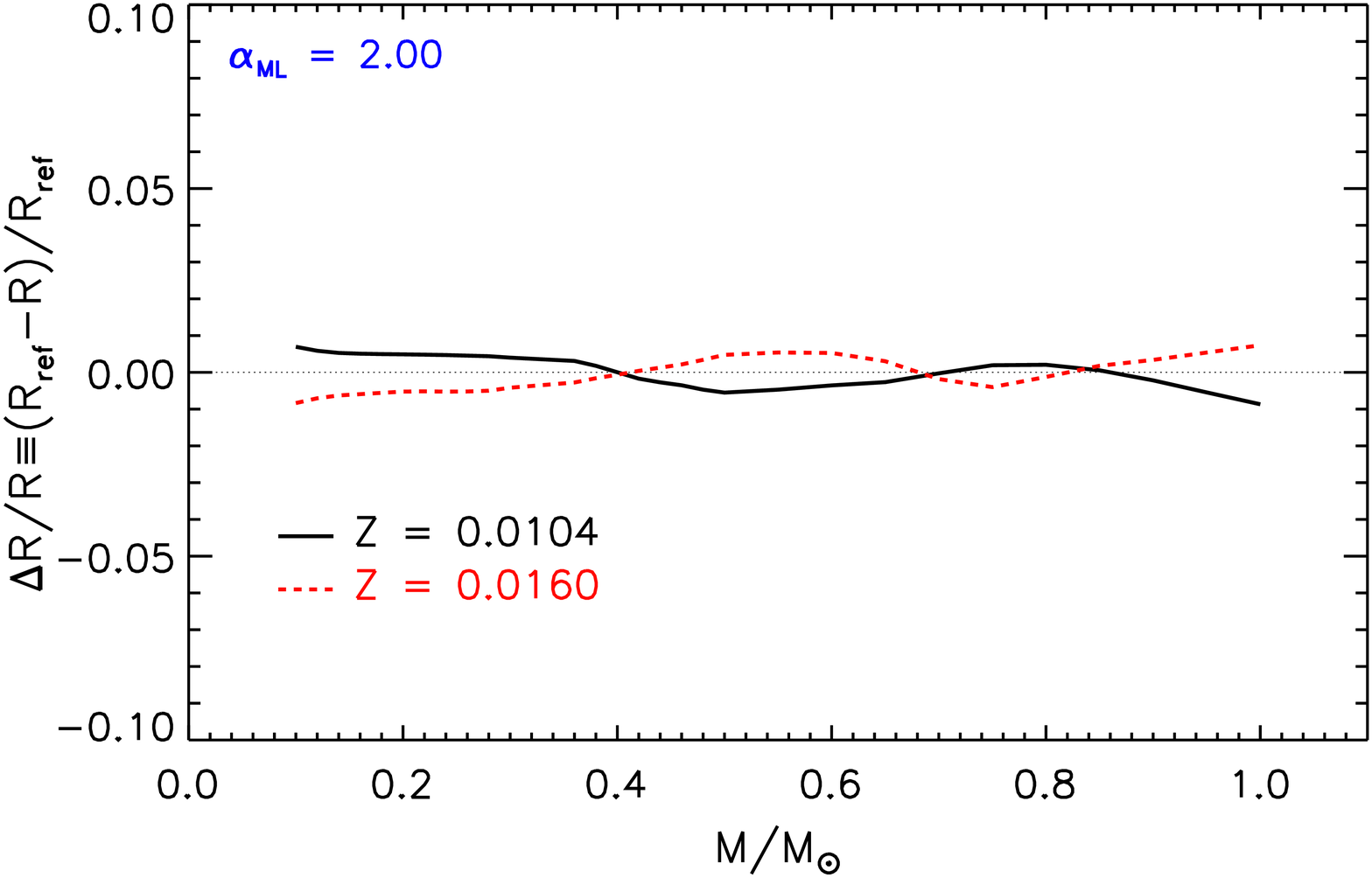}\\
	\includegraphics[width=\columnwidth]{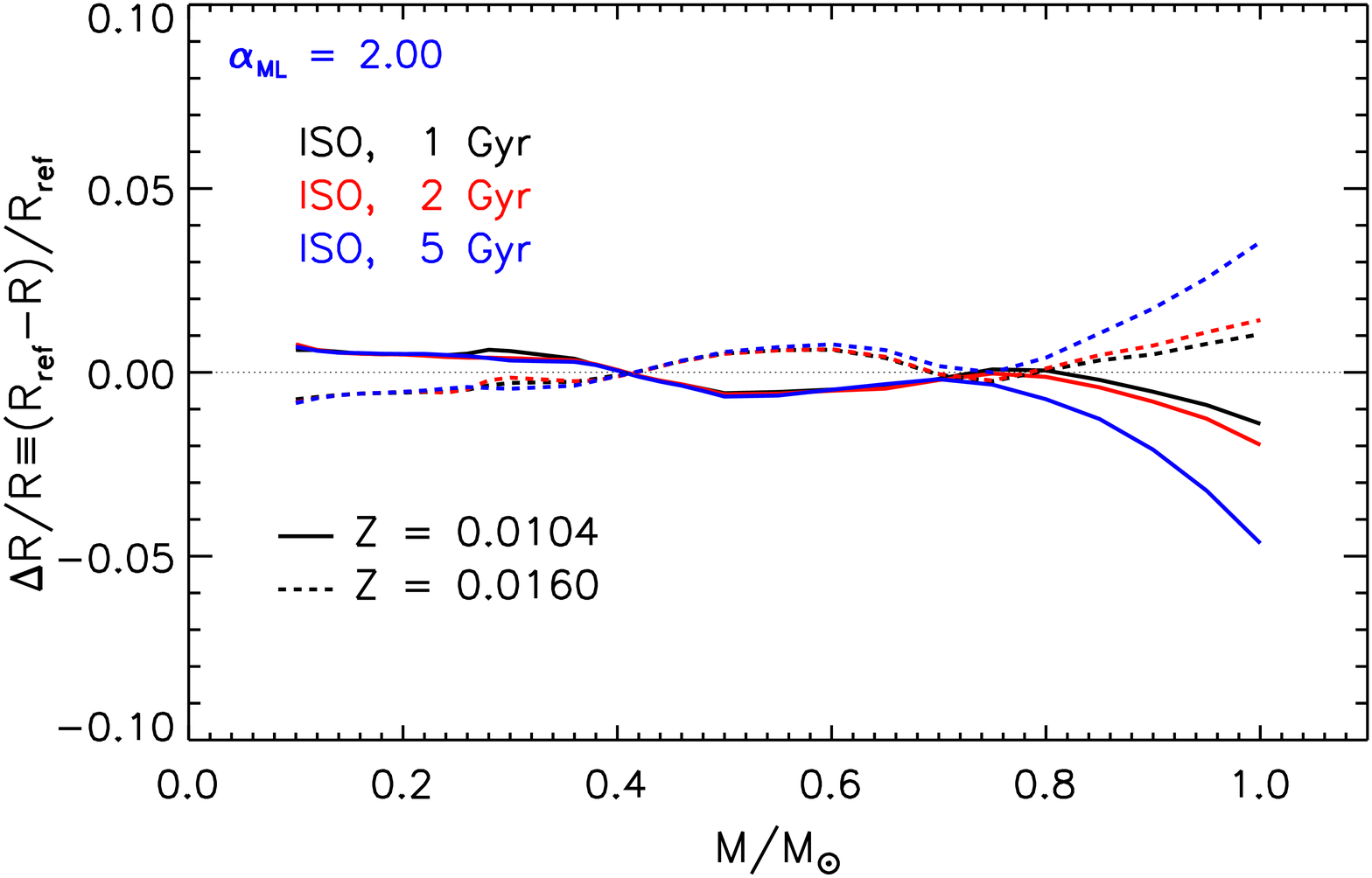}
	\caption{Relative radius variation as a function of the stellar mass due to the adoption of a different initial metallicity (i.e. $Z=0.0104$ and 0.0160) with respect to the reference one ($Z=0.013$). Top panel: models on the ZAMS. Bottom panel: models on the 1, 2, and 5 Gyr isochrones.}
	\label{fig:z}
\end{figure}
The metallicity variation $\delta Z$ depends mainly on the [Fe/H] error, which we set to a conservative value of $\pm0.1$ dex. Keeping the initial helium abundance fixed to its reference value $Y=0.274$ and perturbing only [Fe/H] we obtained $Z = 0.0104$ and $Z=0.0160$, hence a variation of about 20 percent with respect to the reference value.

Fig.~\ref{fig:z} shows the total radius relative differences between models computed with the perturbed initial metallicity and the reference ones. A variation of $Z$ affects the radiative opacity coefficients, thus producing an effect on both the interior and the atmosphere. As already discussed, depending on the stellar mass, the star reacts differently to an opacity variation in the atmosphere or in the interior. For fully convective and adiabatic \zams{} stars ($M \la 0.4$~\msun) the variation of $Z$ (i.e. the change of $\overline{\kappa}_\rmn{R}$) is inconsequential in the stellar interior ($\tau~\gt~\tau_\rmn{bc}$). For such models the change of $Z$ in the atmosphere produces a slight modification of the outer BCs that reflects in a small relative variation of the radius (less than 1~percent). For larger masses ($M~\ga~0.4$~\msun) the interior gets progressively more and more sensitive to the metallicity change through the opacity, as discussed. In this case the effect on the stellar interior dominates over the effect of $\delta Z$ on the atmosphere. As a result, an increase of $Z$ leads to a reduction of the radius (the opposite occurs if $Z$ decreases). However, even in this case, the total effect on the radius of the metallicity is small, being of the order of about 1~percent. 

The effect of the initial metallicity variation on the radius depends on the age for $M\ga 0.7$~\msun. The maximum variation of the surface radius is attained by the 1~\msun{} model at 5~Gyr (i.e. about 4-5 percent).

The radius change slightly depends on the \ml{} in the mass range 0.70--0.85~\msun, where the use of \ml=1.00 leads to an almost constant radius change of about 1~percent.

\subsection{Simultaneous variation of the initial helium and metallicity}
\label{sec:chm_dependent}
As discussed in the previous section, $Y$ and $Z$ are correlated and the value of the couple ($Y$, $Z$) is determined by specifying the values of a triplet defined by ([Fe/H], $\Delta Y / \Delta Z$ and $(Z/X)_{\sun}$), as a consequence the uncertainty on these last three quantities propagates into the final uncertainty on $Y$ and $Z$. In this section we discuss the variation of the stellar radius caused by the uncertainty on [Fe/H], $\Delta Y / \Delta Z$ and $(Z/X)_{\sun}$. 

\begin{figure}
	\centering
	\includegraphics[width=\columnwidth]{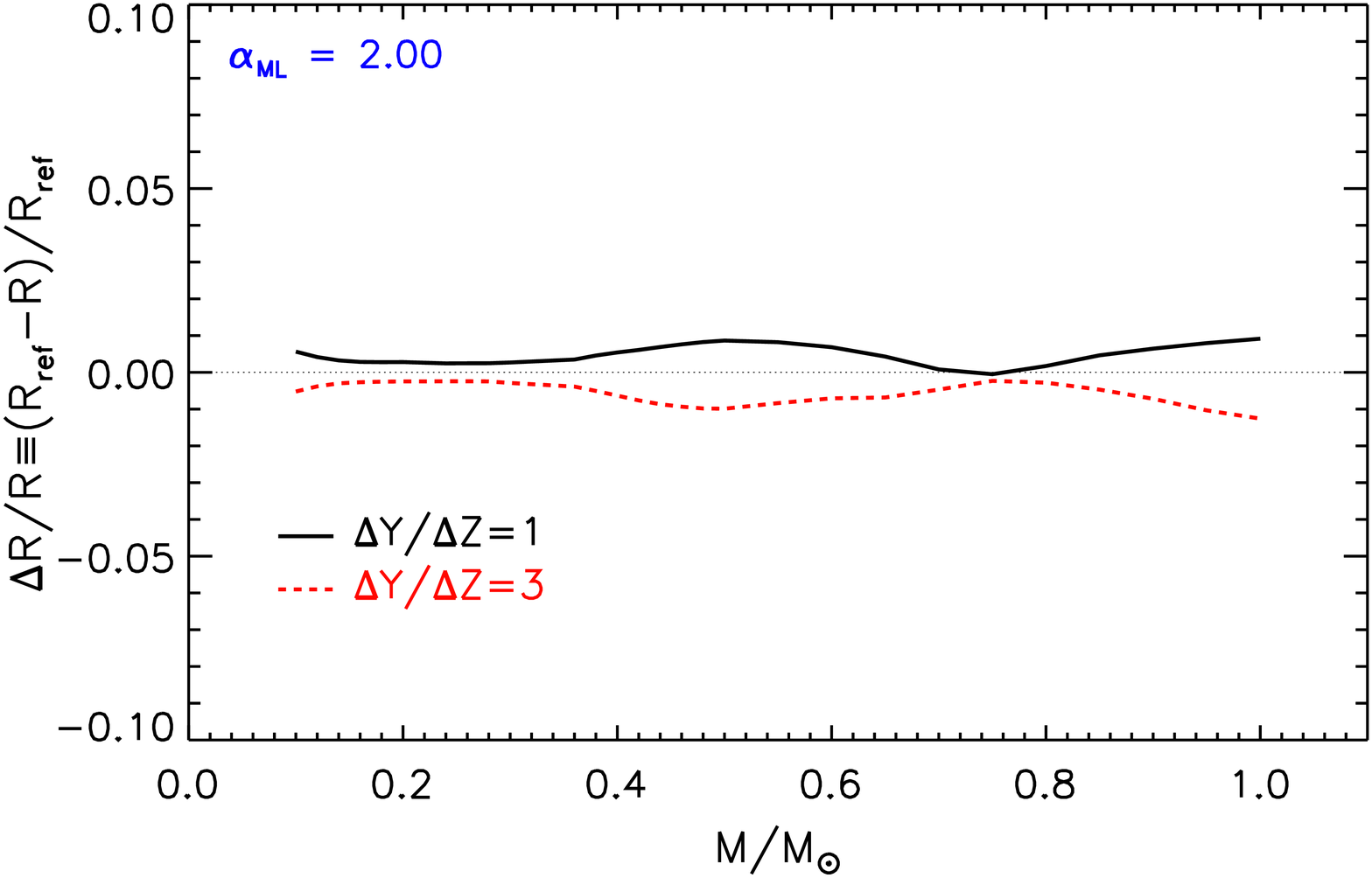}\\
	\includegraphics[width=\columnwidth]{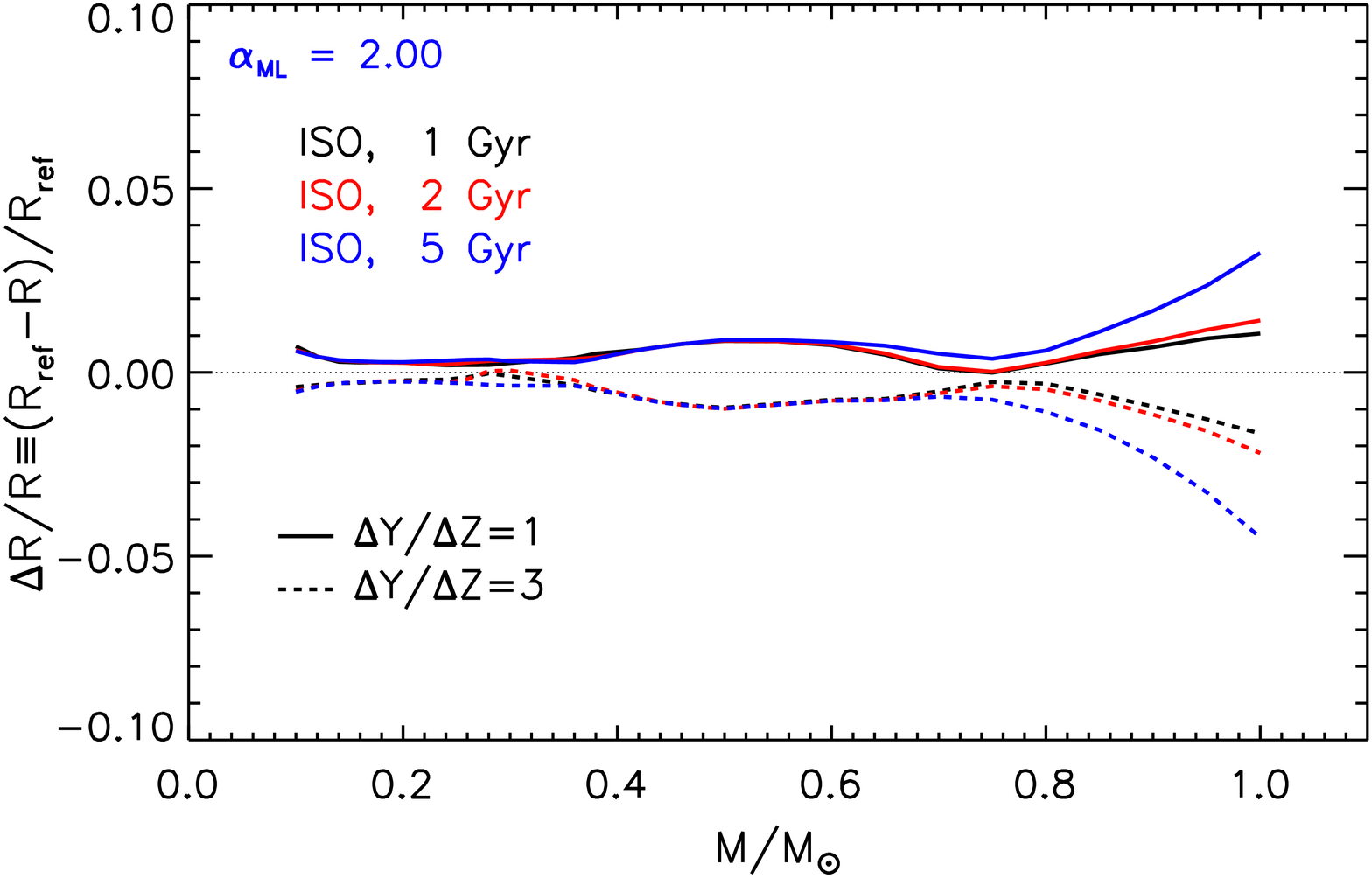}
	\caption{Relative radius variation as a function of the stellar mass due to the adoption of a different helium-to-metal enrichment ratio (i.e. $\Delta Y/\Delta Z = 1$ and 3) with respect to the reference one ($\Delta Y/\Delta Z = 2$). Top panel: models on the ZAMS. Bottom panel: models on the 1, 2, and 5 Gyr isochrones.}
	\label{fig:dydz}
\end{figure}
Fig.~\ref{fig:dydz} shows the total radius variation for the \zams{} and isochrone models due to the uncertainty in $\Delta Y / \Delta Z$. A variation of $\Delta Y / \Delta Z$ of $\pm 1$ affects primarily the initial helium abundance ($\delta Y/Y \sim 4$~percent) changing the metallicity at the level of $\delta Z/Z\sim$~1--2~percent. As the total radius variation is mainly due to the perturbation of $Y$, the net effect on the models is similar to that discussed in Section~\ref{sec:y}. 

\begin{figure}
	\centering
	\includegraphics[width=\columnwidth]{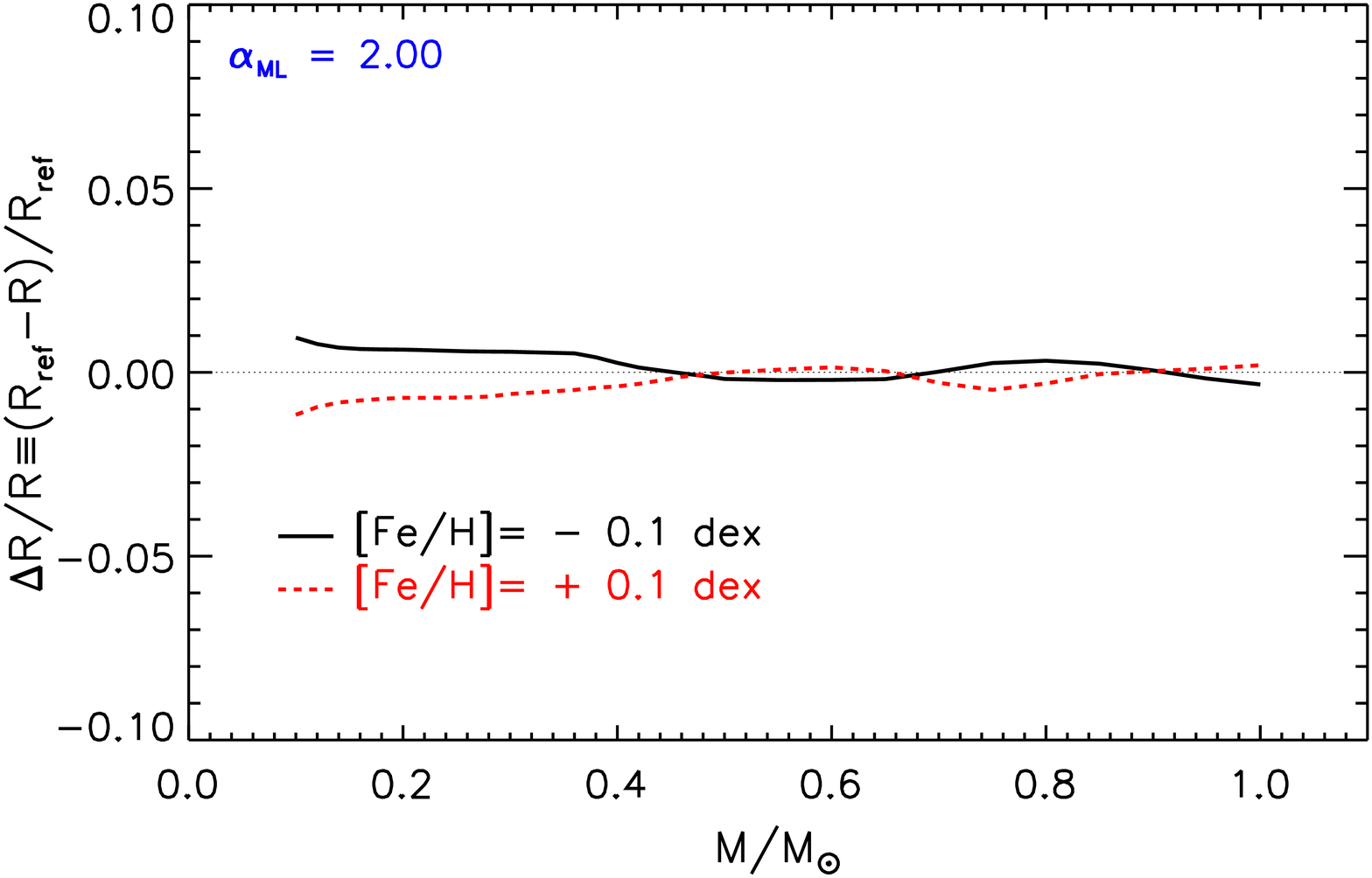}\\
	\includegraphics[width=\columnwidth]{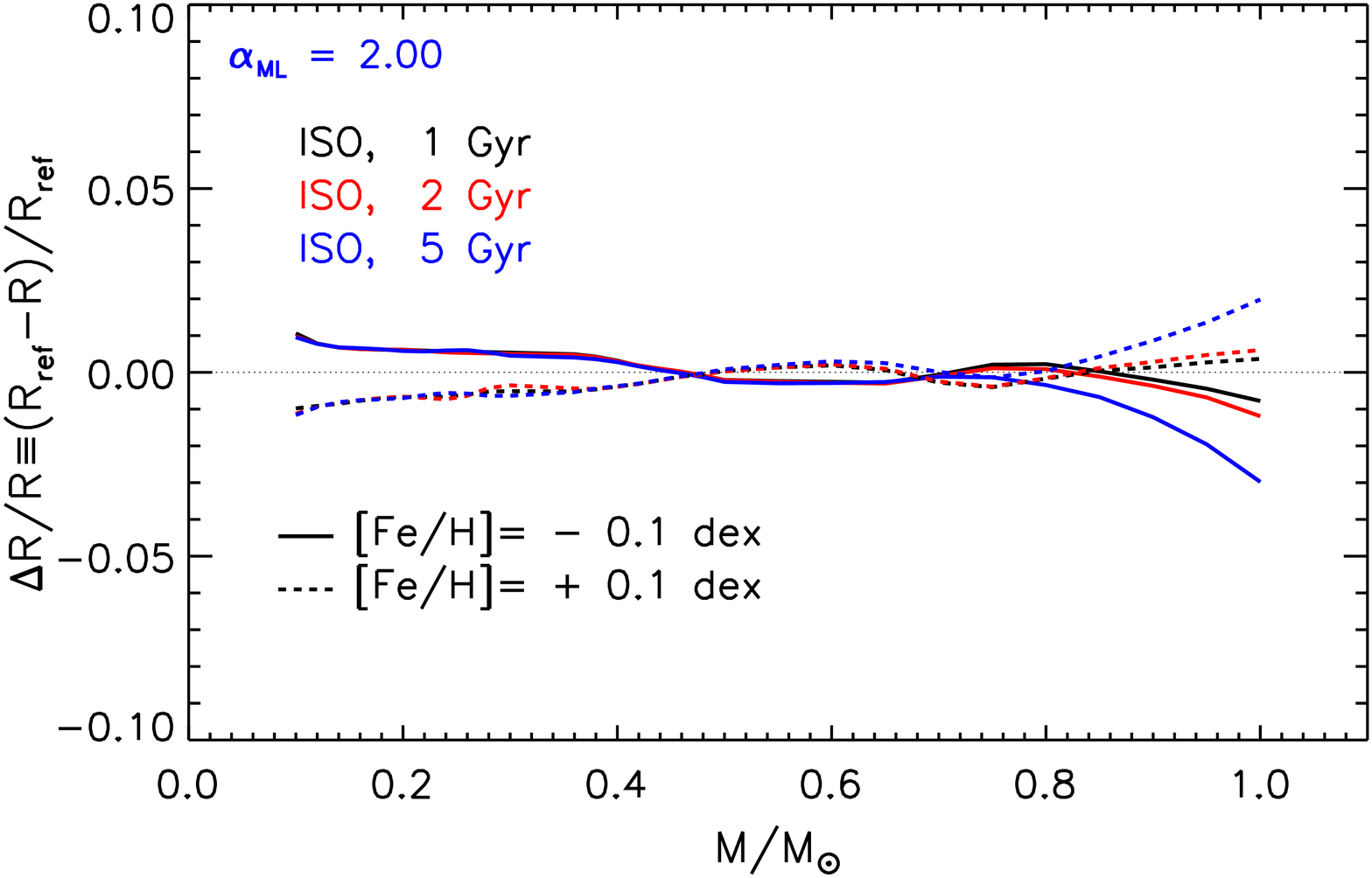}
	\caption{Relative radius variation as a function of the stellar mass due to the adoption of an uncertainty of $\pm 0.1$~dex on [Fe/H] (i.e. [Fe/H]=$-0.1$ and $+0.1$) with respect to the reference one ([Fe/H]=0). Top panel: models on the ZAMS. Bottom panel: models on the 1, 2 and 5 Gyr isochrones.}
	\label{fig:feh}
\end{figure} 

Top panel of Fig.~\ref{fig:feh} shows the radius relative difference between models computed with the reference and the perturbed [Fe/H] values for models along the \zams. A variation of [Fe/H] affects both $Y$ and $Z$, but the effect on $Z$ (about 20~percent) is larger than that on $Y$ (about 2~percent). Referring to Fig.~\ref{fig:y}, a decrease of initial helium content causes a radius reduction in all the selected mass range. On the other hand the effect of $Z$ is much complicated (Fig.~\ref{fig:z}), as for $M\la 0.4$~\msun{} the star shrinks (if $Z$ decreases) while for larger masses it slightly increases. As a consequence of such a behaviour, the radius change due to the variation of $Y$ almost counterbalances that caused by the metallicity perturbation for $M\ga 0.4$--0.5~\msun, while for lower values of the mass the two effects add up producing a total radius change by about $\pm 1$~percent.

Bottom panel of Fig.~\ref{fig:feh} shows the effect on the radius of the same variation of [Fe/H] for models on the isochrones. Only for $M~\ga~0.7$~\msun{} the radius variation gets progressively more and more sensitive to the stellar ages, reaching a maximum relative difference (with respect to the reference case) of about 3 percent for 1~\msun{} at 5~Gyr. 

The uncertainty on [Fe/H] produces a radius modification that is almost independent of the adopted \ml. The only effect is to further flatten the radius change in the 0.70--0.85~\msun{} region leaving unchanged the variation found in the other models.

\begin{figure}
	\centering
	\includegraphics[width=\columnwidth]{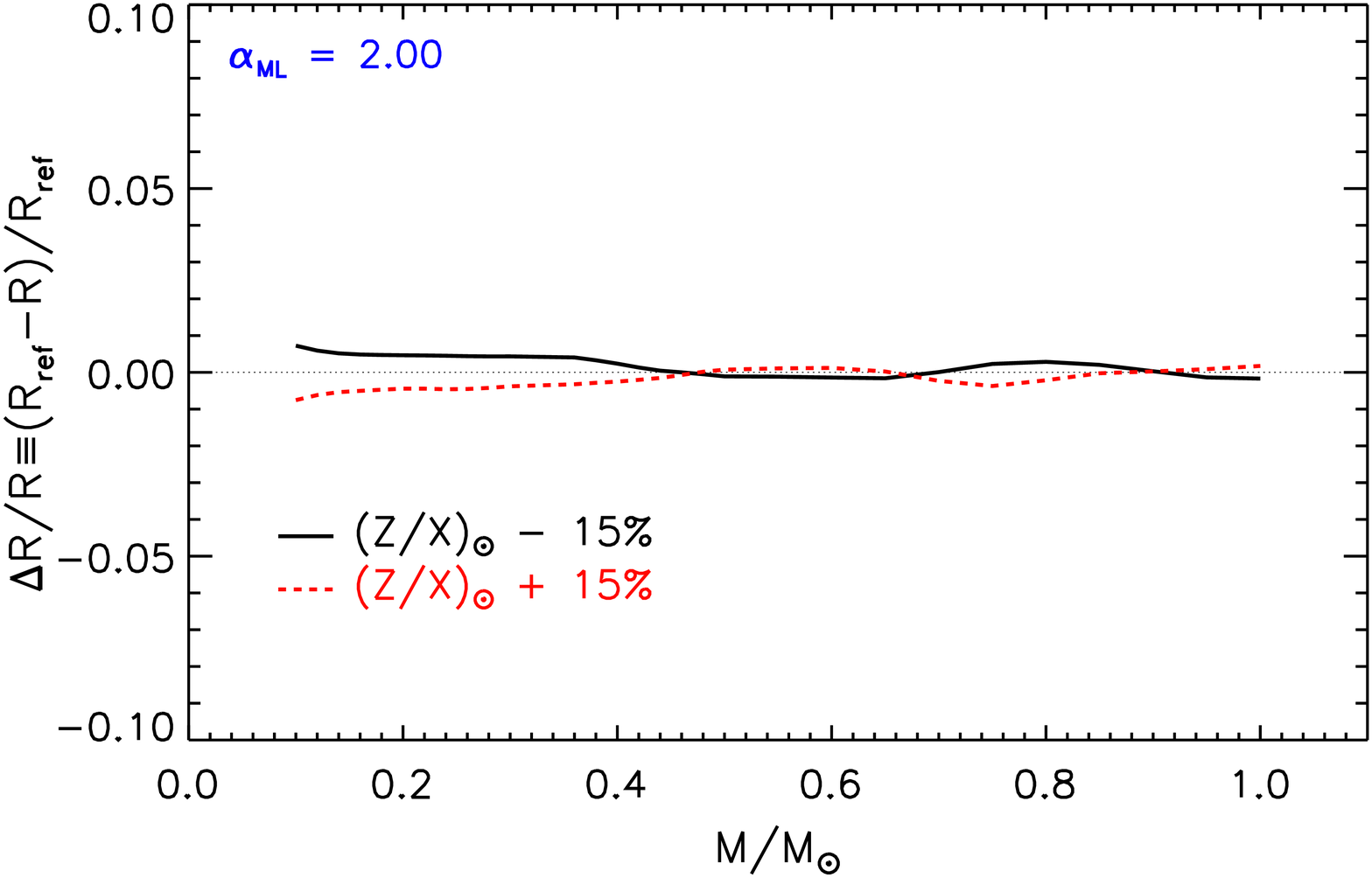}\\
	\includegraphics[width=\columnwidth]{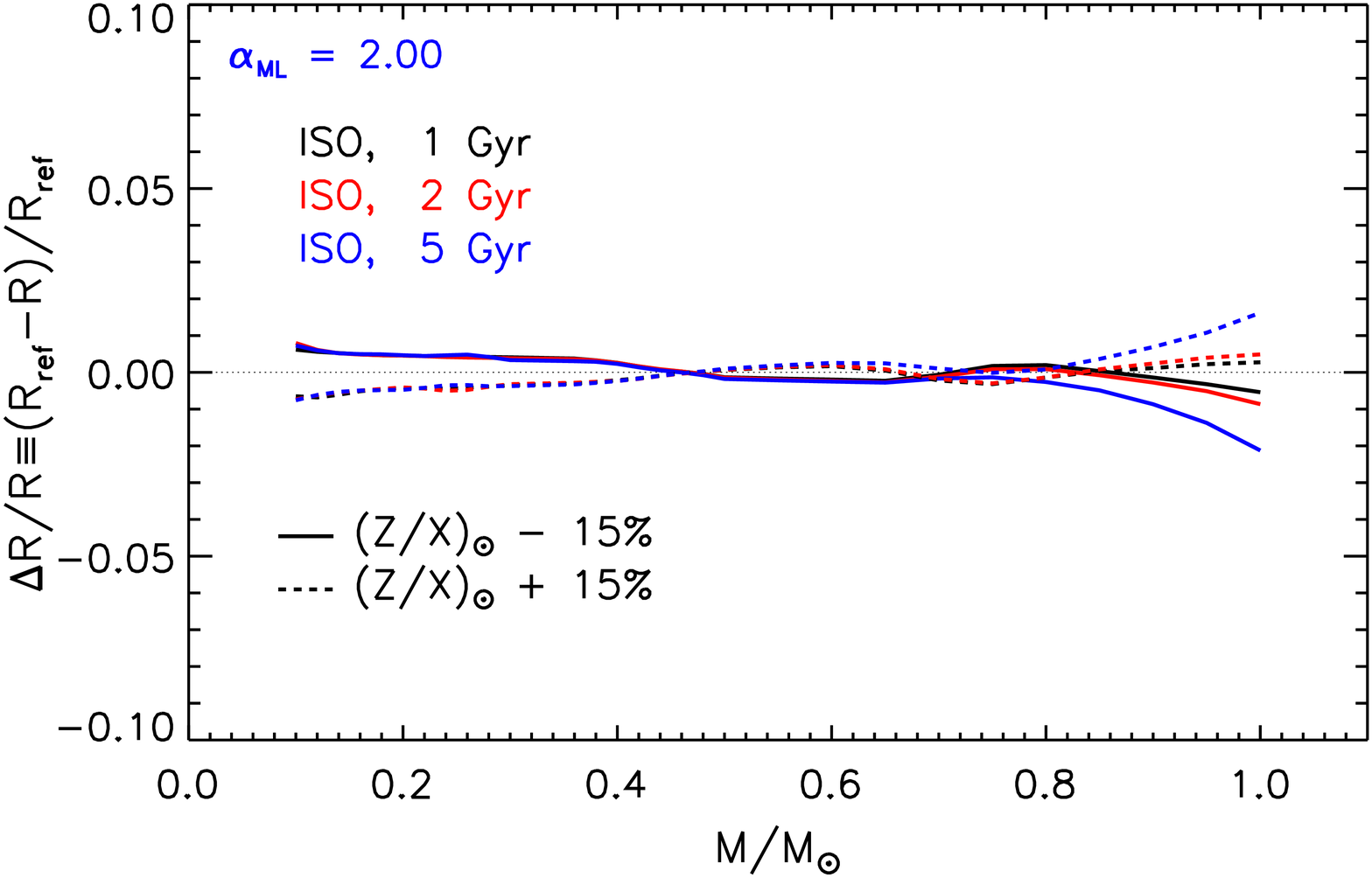}
	\caption{Relative radius variation as a function of the stellar mass due to the adoption of an uncertainty of $\pm 15\%$ on $(Z/X)_{\sun}$ with respect to the reference one \citep[$(Z/X)_{\sun}=0.0181$;][]{asplund09}. Top panel: models on the ZAMS. Bottom panel: models on the 1, 2, and 5 Gyr isochrones.}
	\label{fig:zx}
\end{figure}

Fig.~\ref{fig:zx} shows the effect on the radius of the variation of $\pm15$ percent on $(Z/X)_{\sun}$. We recall that the heavy elements mixture -- the relative abundances of the elements heavier than boron -- affects the stellar computations in two different ways: (1) through the opacity coefficients; and (2) through the total metallicity-over-hydrogen ratio $(Z/X)_{\sun}$, needed to compute $Z$ and $Y$ (see equation~(\ref{eq:z})). We already discussed the impact of the heavy element mixture on the opacity in Section~\ref{sec:opacity}, so here we limit to discuss its effect on the derived values of $Y$ and $Z$.

An uncertainty of $\pm 15$ percent on $(Z/X)_{\sun}$ results in a variation of about 1--2~percent in $Y$ and about 15~percent on $Z$, which is similar to that obtained by a perturbation of $\pm 0.1$~dex on [Fe/H]. Thus, the effect on the radius is essentially the same already discussed and shown in Fig.~\ref{fig:feh}. 

\section{Cumulative Uncertainty}
\label{sec:total}
\begin{table}
\centering
\caption{Quantities varied in the computation of perturbed stellar models and their corresponding uncertainty/range of variation. The quantity marked with the flag `Yes' has been taken into account in the calculation of the cumulative uncertainty on the stellar radius.}
\label{tab:err_fis_chm}
\begin{tabular}{lcc}
\hline
Quantity & Perturbation & Cumulative\\
\hline
\hline
\multicolumn{3}{c}{Input physics}\\
$\tau_\mathrm{ph}$ & $T(\tau_\rmn{bc})=T_\rmn{eff}$ & No \\
& $\tau_\rmn{bc} = 2/3$ & Yes \\
& $\tau_\rmn{bc} = 100$ & Yes \\
\bcs & BH05 & Yes\\
& AHF11+GN93 & No \\
& CK03 & No \\
& KS66  & No \\
$\overline{\kappa}_\mathrm{rad}$ & $\pm 5\%$, & Yes\\
Solar mixture & GS98 & Yes \\
\eos & \textsc{scvh95}, & Yes\\
& \textsc{FreeEOS} & Yes \\
\ml & 1.00, 2.00 & No\\
\hline
\multicolumn{3}{c}{Chemical composition}\\
$[$Fe/H$]$ & $\pm 0.1$dex & Yes\\
$\Delta Y/\Delta Z$ & $\pm 1$& Yes\\
$(Z/X)_\odot$ & $\pm 15\%$ & Yes\\
\hline
\end{tabular}
\medskip
\flushleft
\textit{Note. }AHF11-- \citet{allard11}; KS66 -- \citet{krishna66}; GS98 -- \citet{grevesse98}; \textsc{FreeEOS} -- \citet{irwin08}; \textsc{scvh95} -- \citet{saumon95}
\end{table}
\begin{figure*}
	\centering
	\includegraphics[width=\columnwidth]{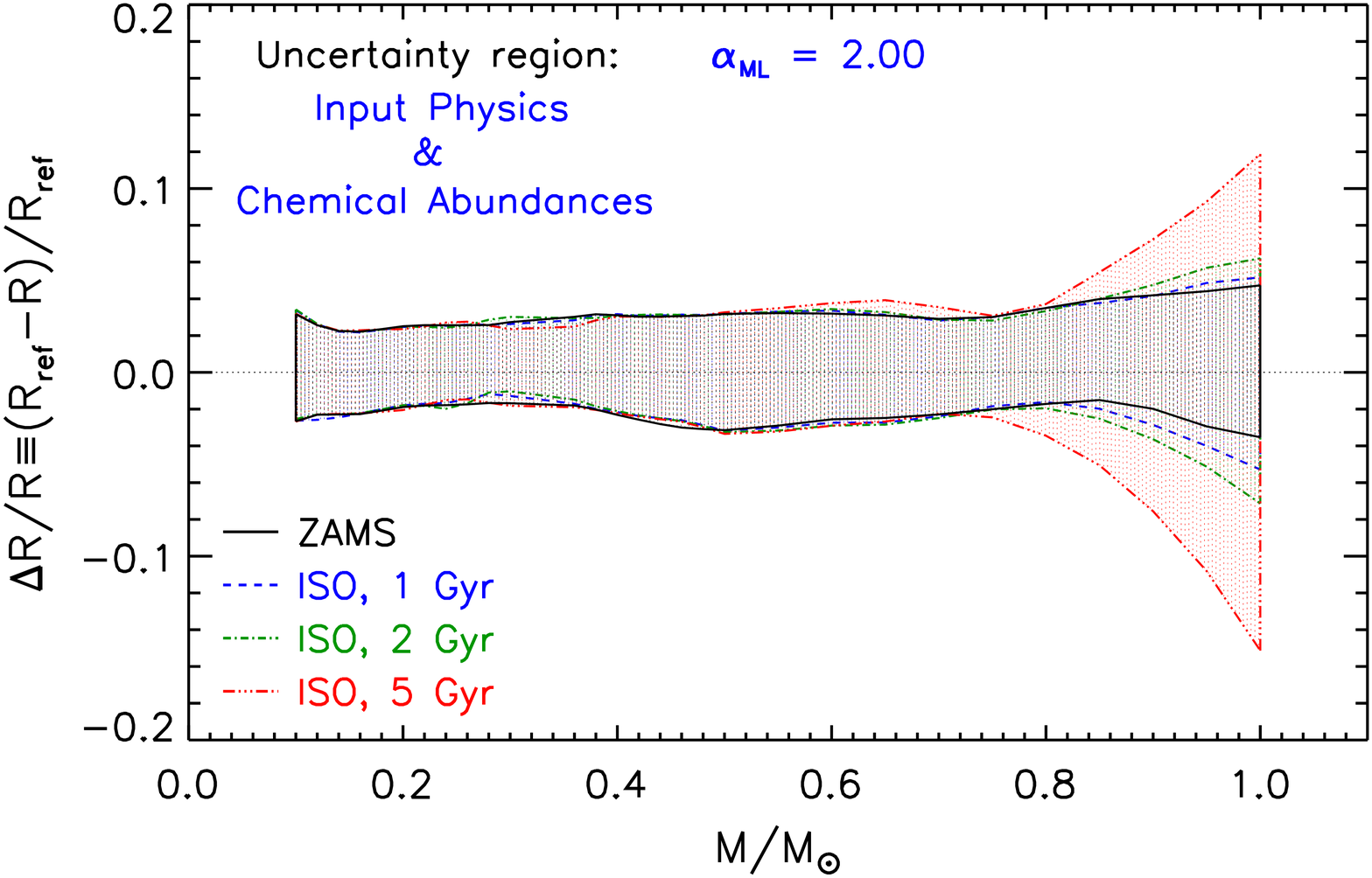}
	\includegraphics[width=\columnwidth]{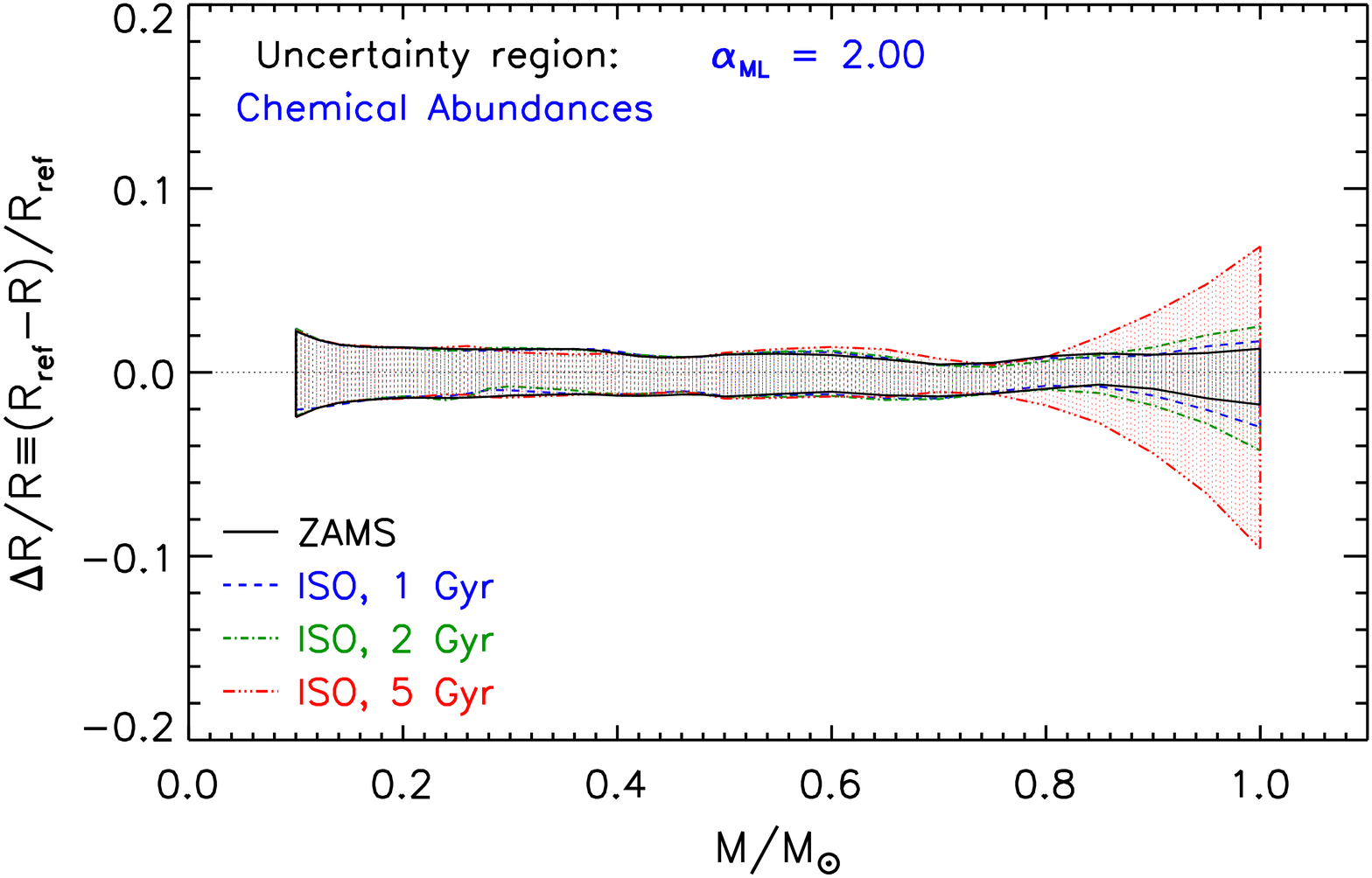}
	\includegraphics[width=\columnwidth]{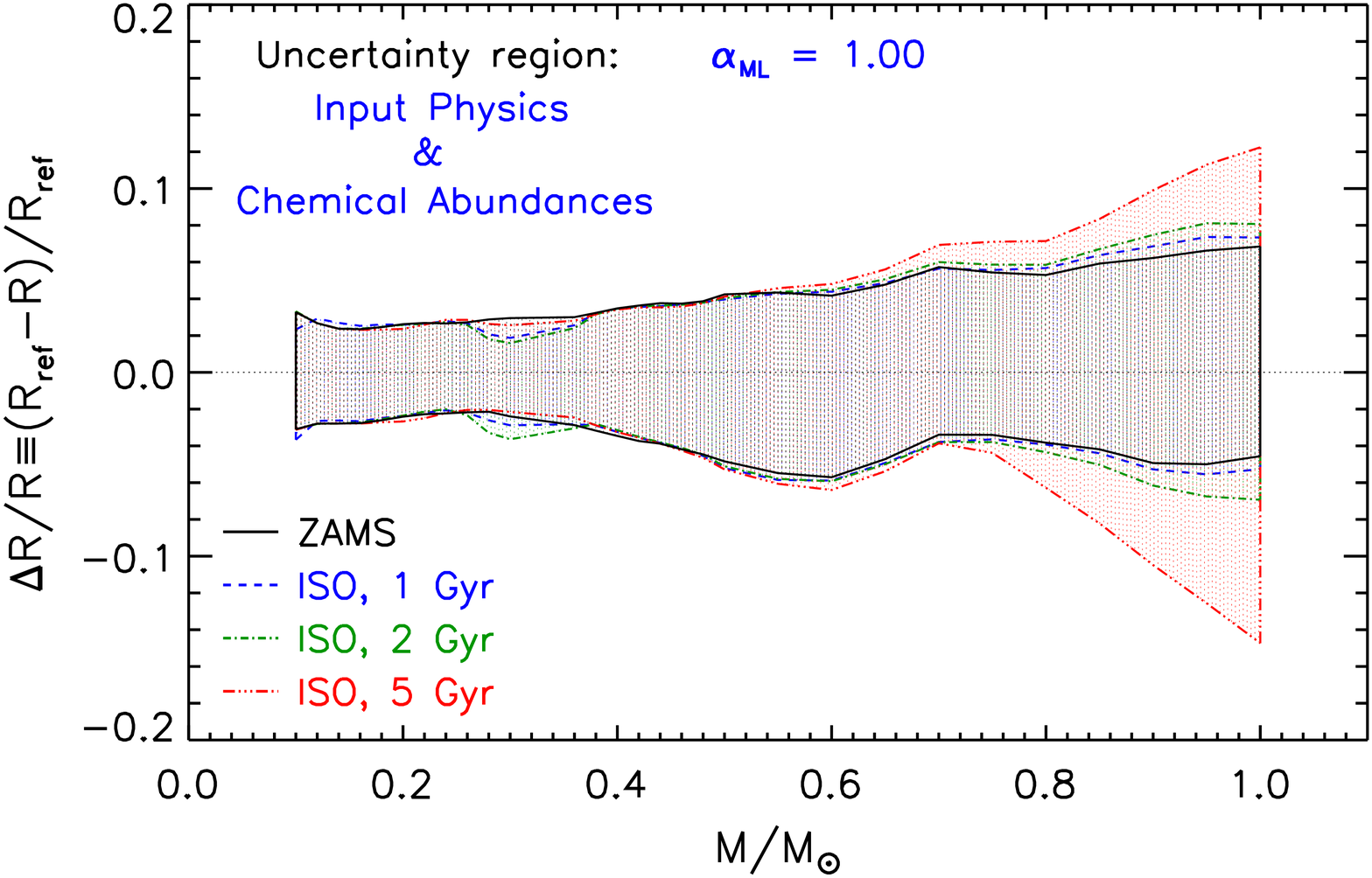}
	\includegraphics[width=\columnwidth]{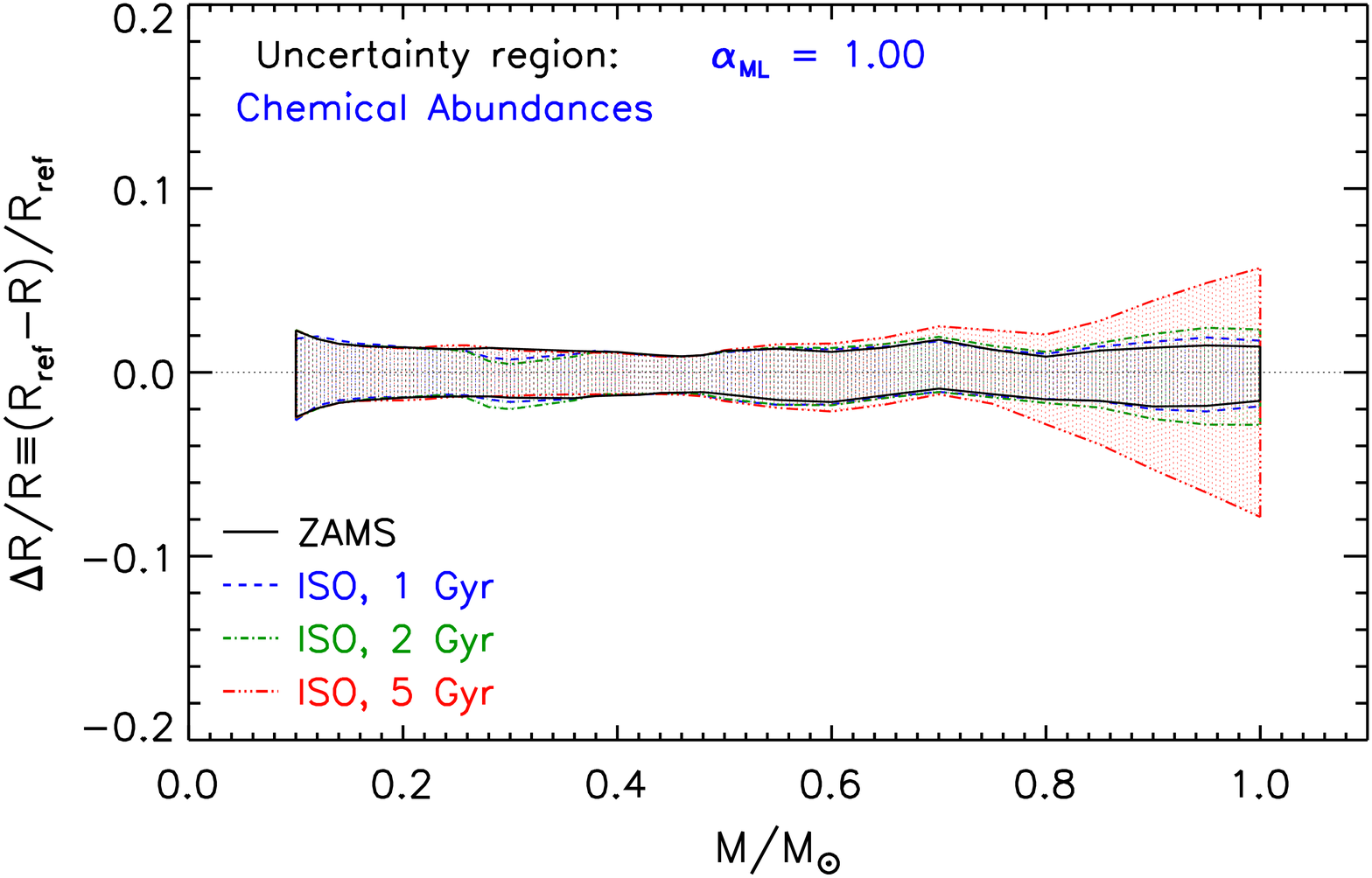}
	\caption{Cumulative relative radius difference for models on the ZAMS and on the 1, 2, and 5~Gyr isochrones, as a function of the stellar mass. Left-hand panels: total uncertainty due to both the errors on the adopted input physics and initial chemical composition. Right-hand panels: uncertainty due to only the errors on the initial chemical composition. Top panels: models with \ml=2.00. Bottom panels: models with \ml=1.00.}
	\label{fig:tot_fis_chm_iso}
\end{figure*}

In the previous sections we quantified the individual contribution to the uncertainty of the predicted radius of each quantity, keeping fixed all the others to their reference value. Here, we evaluate the cumulative uncertainty caused by simultaneously perturbing all the investigated input physics and parameters. To do this, we  started from the results obtained in previous papers, which showed that the models respond almost linearly to the perturbation/change of the quantities analysed here \citep{valle13a,valle13b,tognelli15b}. Thus, instead of computing a huge amount of models sets where several parameters are simultaneously perturbed, we preferred to follow a more simple -- but robust -- approach. The edges of the variability region (i.e. the cumulative error stripe) can be obtained by linearly adding the contribution of each individual perturbation. 

In the following we did not include the effect of changing \ml{} in the total error bar, but we preferred to treat it separately from the other uncertainty sources, analysing the results for \ml=2.00 and 1.00. 

The input physics and parameters that contribute to the cumulative error stripe are listed in Table~\ref{tab:err_fis_chm}. In many of the cases we were able to use symmetric perturbations, as in the case of the chemical composition and radiative opacity, thus obtaining upper and lower values for the radius change. In the other cases such as the \eos{} and BCs this was not possible, and we obtained only one edge of the error stripe. This will introduce an asymmetry in the cumulative error stripe. As for the impact of the atmospheric models on the radius we took as representative the differences between our reference (AHF11) and the BH05, neglecting both the out of date KS66 and the CK03 BCs, which is less suitable for LM stars and not available for VLM (for $M\la 0.36$~\msun). For the evaluation of the uncertainty due to the adopted $\tau_\rmn{bc}$ we used models with $\tau_\rmn{bc}=100$ and $\tau_\rmn{bc}=2/3$.

When analysing the impact on the models of the radiative opacity uncertainty, we computed models with both non-grey AHF11 and semi-empirical KS66 BCs to take into account the opacity change in the atmosphere too. The contribution of the opacity uncertainty to the cumulative error stripe has been computed considering both cases, i.e. adding the two (quite similar) contributions to the radius change and then dividing the resulting radius variation by 2. In other words we adopted the mean value of the two sets.

Left-hand panels of Fig.~\ref{fig:tot_fis_chm_iso} shows the resulting cumulative error stripe on the stellar radius that accounts for the uncertainty on both the adopted input physics and initial chemical composition in the case of ZAMS and isochrones models, for the reference mixing length parameter (i.e. \ml=2.00, upper panel) and for \ml=1.0 (bottom panel).  

In the case of \ml=2.00, the ZAMS cumulative radius error stripe is almost symmetric, ranging from about $\pm$2, $\pm$3~percent for VLM stars, to about about $\pm$4, $\pm$5~percent at larger masses. If \ml=1.0 is used, the uncertainty on VLM and LM stars (i.e. $M\la 0.4$--0.5~\msun) is similar to that for \ml$=2.00$ models, while at larger masses the stripe gets progressively broader (and less symmetric), in particular for masses between 0.5 and 0.8~\msun{}, where it reaches values of $\pm$6~percent.

Fig.~\ref{fig:tot_fis_chm_iso} shows also the resulting cumulative error stripe for models on the 1, 2, and 5~Gyr isochrones, compared to the ZAMS case. The cumulative error stripe on the radius is the same of that obtained for the ZAMS for $M\la 0.7$~\msun, while for larger masses the stripe gets progressively more and more sensitive to the chosen age, at least for ages larger than about 2~Gyr. The maximum value of the error stripe is achieved by the 1~\msun{} model at 5~Gyr, which reaches about $+12$($-15$)~percent, for \ml=2.00. The stripe for isochrones models is sensitive to the adopted mixing length parameter similarly to the ZAMS, getting broader if \ml=1.00 for $M \ga 0.5$~\msun.

We want also to clearly identify the total effect on the stellar radius due to the error on the sole initial chemical composition (i.e. helium and metal abundances). Right-hand panels of Fig.~\ref{fig:tot_fis_chm_iso} shows the error stripe when only the chemical composition errors are accounted for in the radius variation. The uncertainty on the initial helium and metal abundance produces a symmetric variation of the radius between $\pm 1$ and $\pm 2$~percent for ZAMS models. The isochrone models are similar to the ZAMS for $M\la 0.7$~\msun, while for larger masses the stripe depends on the age. In this case, the largest variation reaches about $+7$,$-10$~percent at 5~Gyr. The uncertainty due to the chemical composition on ZAMS/isochrone models is only slightly dependent on the adopted \ml{} (as discussed ins Section~\ref{sec:chm}).

The presented results showed the effect of the quoted uncertainties on the stellar radius at a fixed mass. However, it is worth to  provide also the cumulative error stripe as a function of the luminosity for a direct comparison with observations when the stellar mass is not available. The perturbation of the input physics and/or chemical composition might modify both the stellar radius and the luminosity. Consequently, at the same $\log L/$\lsun{} value, the reference and the perturbed models might have different masses. Thus, in this case, the uncertainty on the stellar radius is the result of the combined variation of the radius caused by the perturbed quantity at a given mass plus the variation of the radius due to the mass change to get the same luminosity.

\begin{figure}
	\centering
	\includegraphics[width=\columnwidth]{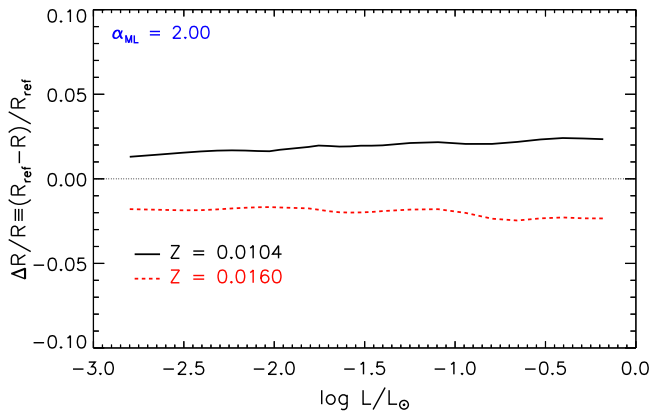}
	\caption{Relative radius variation caused by the perturbation of $Z$ at a fixed luminosity, between the reference ($Z=0.0130$) and the perturbed ($Z=0.0104$ and $Z=0.0160$) ZAMS models.}
	\label{fig:z_lum}
\end{figure}

We verified that only in the case of perturbed outer BCs (atmospheric structures and $\tau_\rmn{bc}$) and \ml{} the luminosity does not change appreciably, and the radius variation at a fixed luminosity is the same of that at a fixed mass. In the other cases, the luminosity depends on the adopted value of the perturbed quantity. It might happen that the luminosity variation at a fixed mass, due to a certain perturbed quantity,  shifts the radius-luminosity sequence in such a way to amplify the radius difference found at a fixed mass. This is exactly what happens, as an example, if $Z$ is perturbed. In this case, the luminosity variation (due to the opacity enhancement/decrease) overrides the intrinsic radius change at a constant mass, producing a larger radius variation if compared to that obtained at a fixed mass. We show this effect, as a representative example, in  Fig.~\ref{fig:z_lum}. To be noted that the radius change is essentially the same in the whole luminosity range.

\begin{figure*}
	\centering
	\includegraphics[width=\columnwidth]{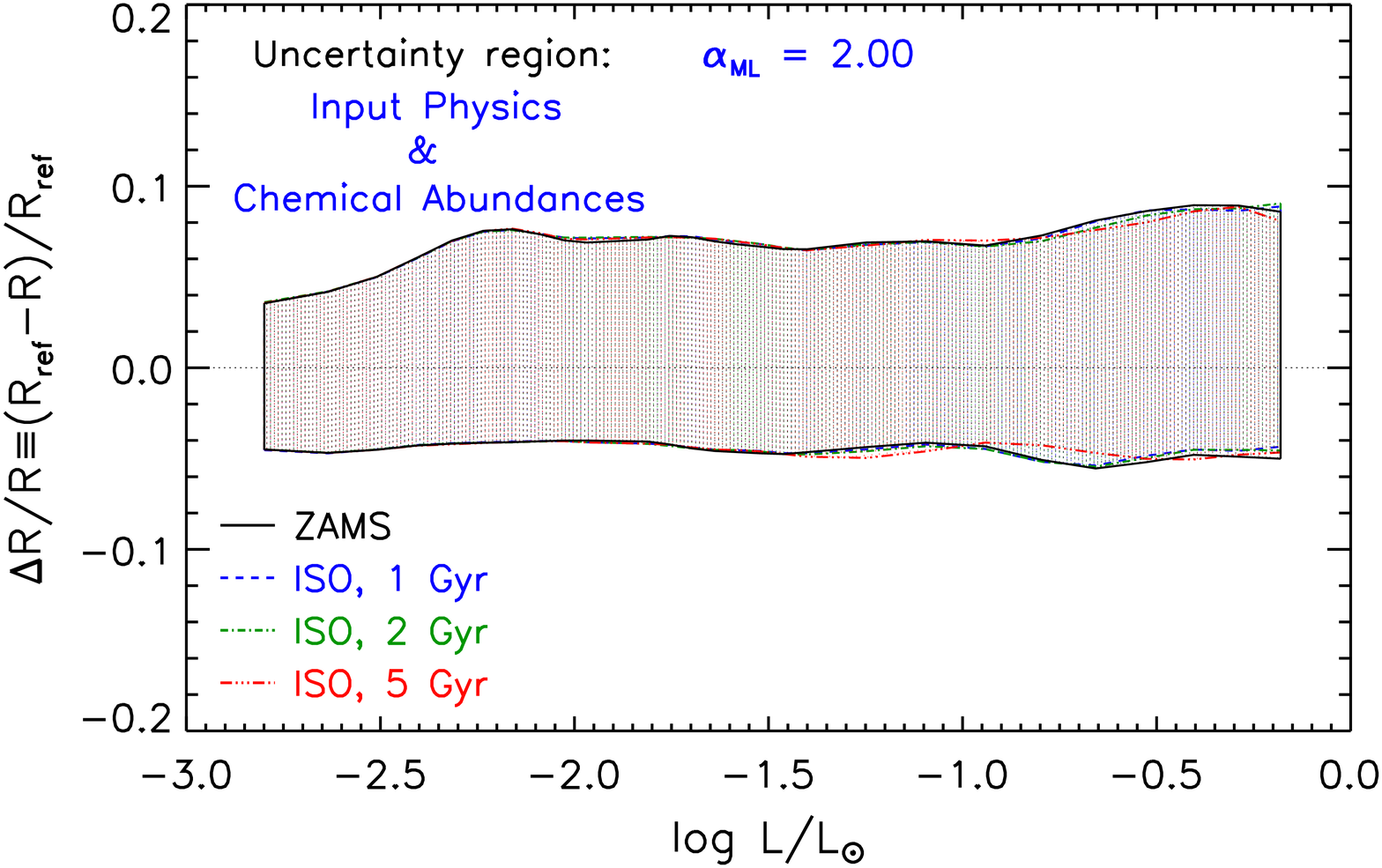}
	\includegraphics[width=\columnwidth]{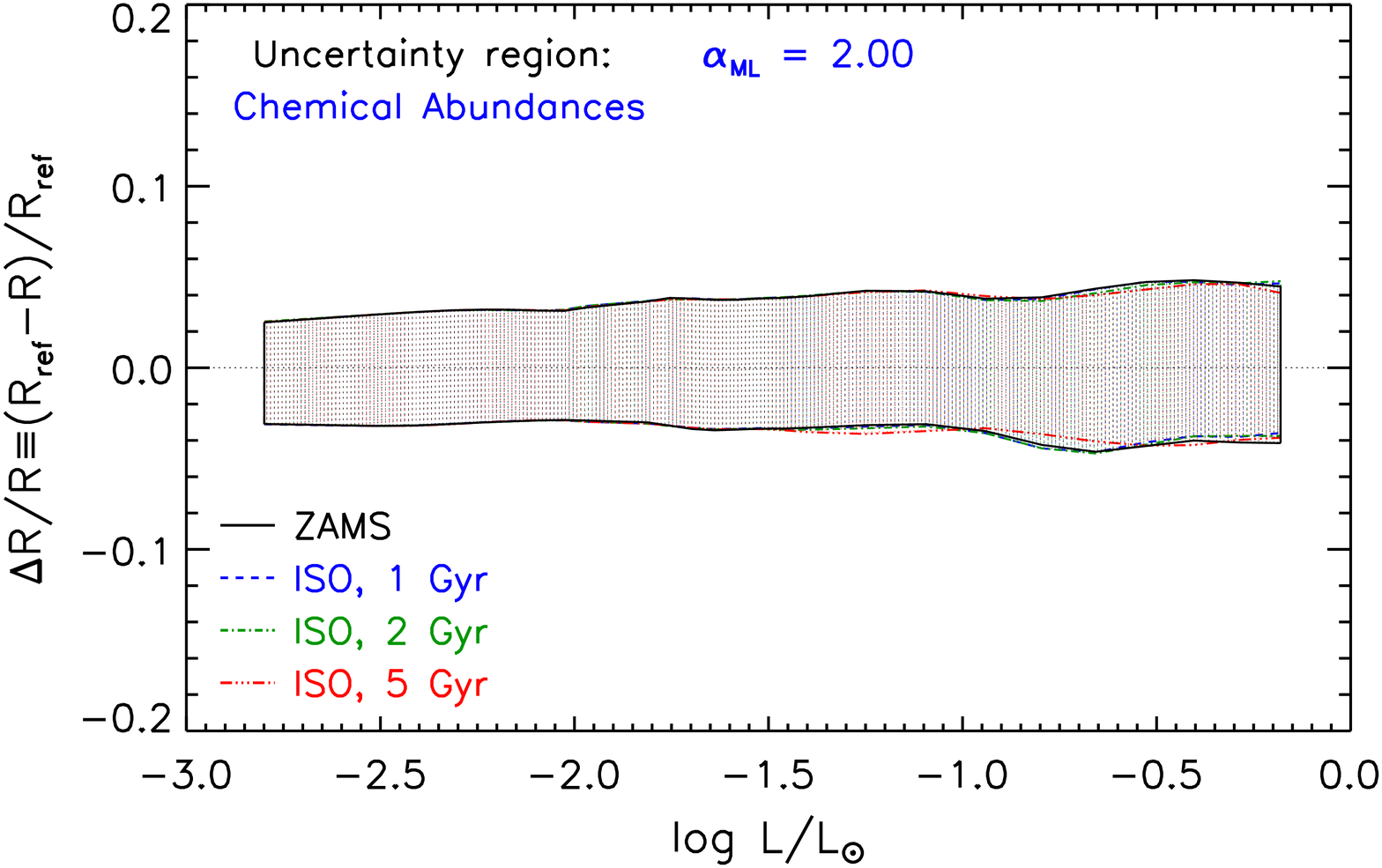}
	\includegraphics[width=\columnwidth]{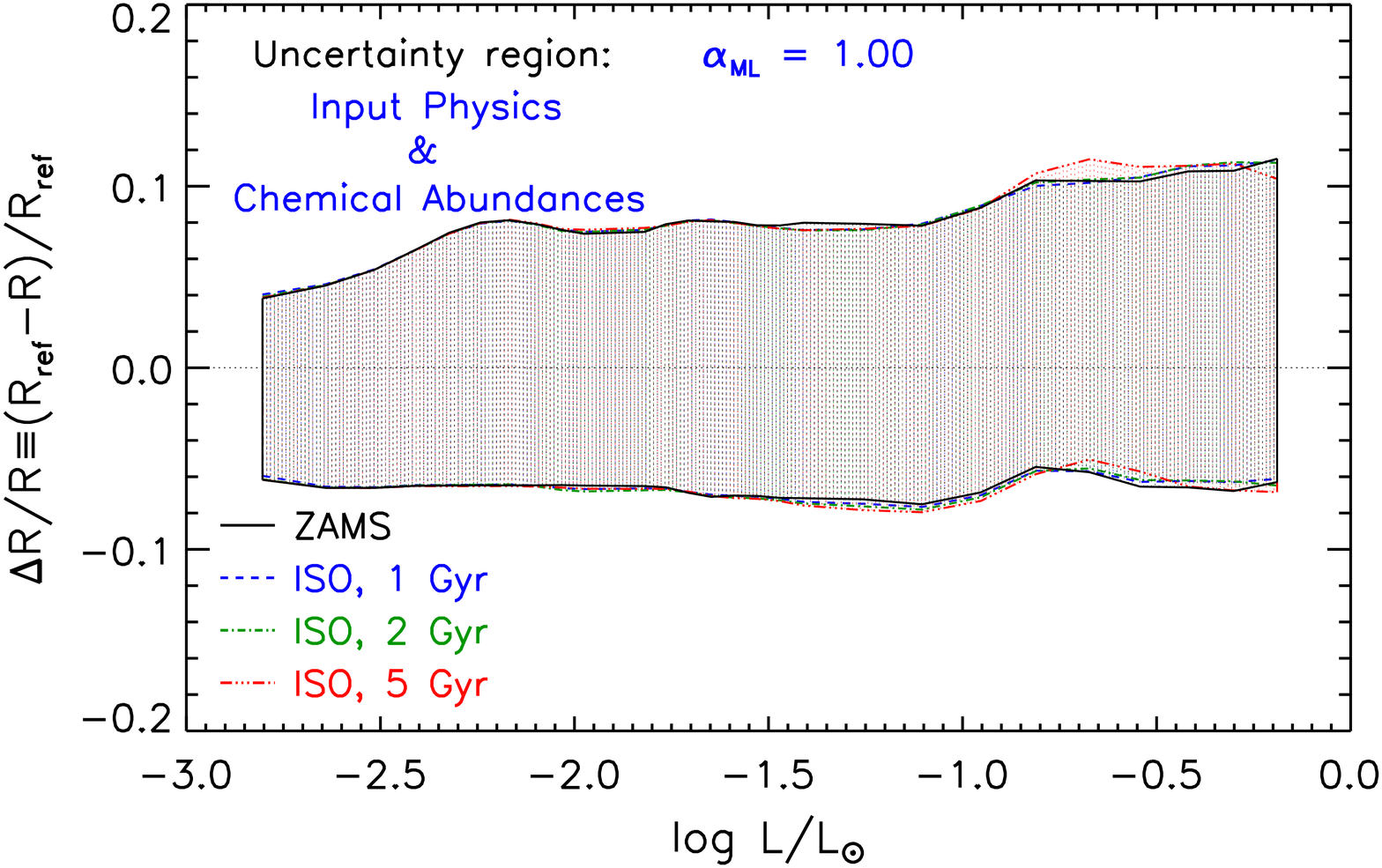}
	\includegraphics[width=\columnwidth]{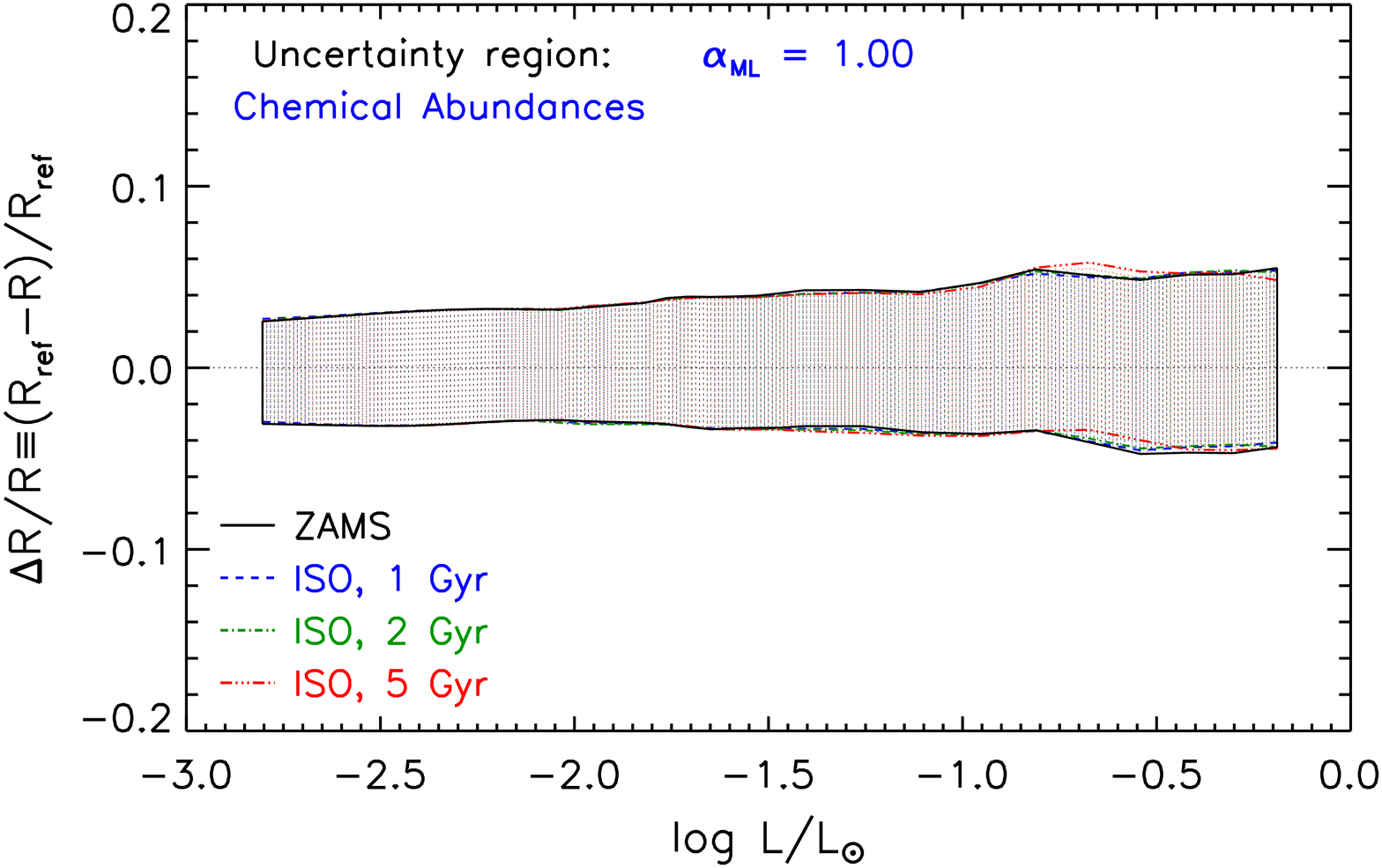}
	\caption{Cumulative relative radius difference between the perturbed and reference models as a function of the stellar luminosity, for models on the ZAMS and on the 1, 2, and 5~Gyr isochrones. Left-hand panels: total uncertainty due to both the errors on the adopted input physics and initial chemical composition. Right-hand panel: uncertainty due to only the errors on the initial chemical composition. Top panels: models with \ml=2.00. Bottom panels: models with \ml=1.00.}
	\label{fig:tot_fis_chm_iso_lum}
\end{figure*}

Fig.~\ref{fig:tot_fis_chm_iso_lum} shows the cumulative error stripe on the radius for models on the ZAMS and on the 1, 2, and 5~Gyr isochrones as a function of the luminosity. The error stripe for ZAMS or isochrone models is essentially the same. In the case of \ml=2.00, the cumulative error stripe is asymmetric and ranges from $\pm 4$~percent (for VLM) to about $+7$, $+9$($-5$)~percent in the selected luminosity range, with the largest uncertainties occurring at higher luminosities. The stripe has a slight dependence on the adopted \ml. If \ml=1.00 is considered, the cumulative error stripe ranges from $+4$($-6$) to $+8$, $+10$($-6$)~percent.

Right-hand panels of Fig.~\ref{fig:tot_fis_chm_iso_lum} show the impact of the sole chemical composition on the radius, which is between about $\pm 3$ and $\pm 5$~percent in the whole luminosity range considered (increasing with the luminosity). Such an effect is independent of the adopted \ml.

\begin{figure*}
	\centering
	\includegraphics[width=\columnwidth]{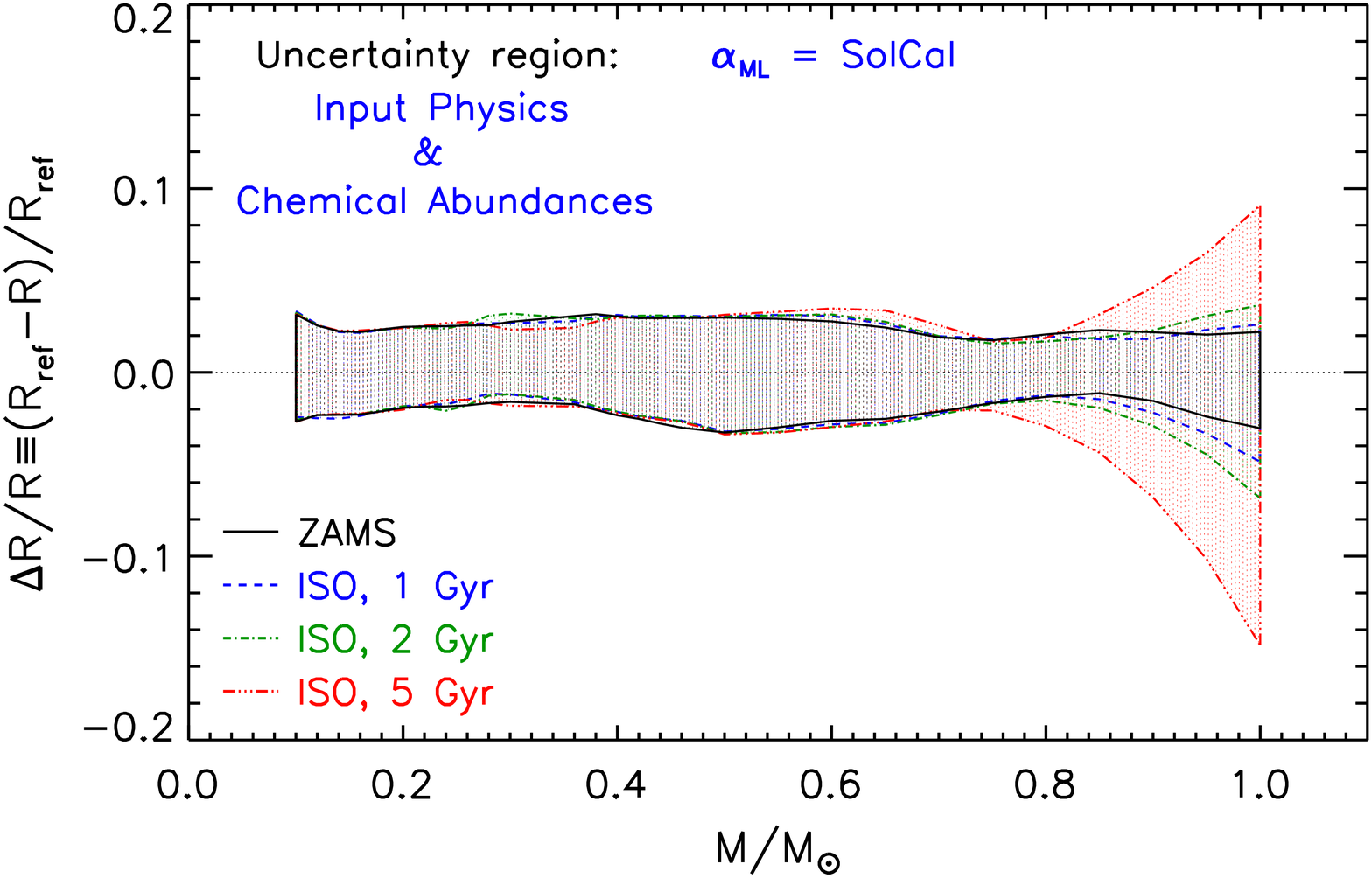}
	\includegraphics[width=\columnwidth]{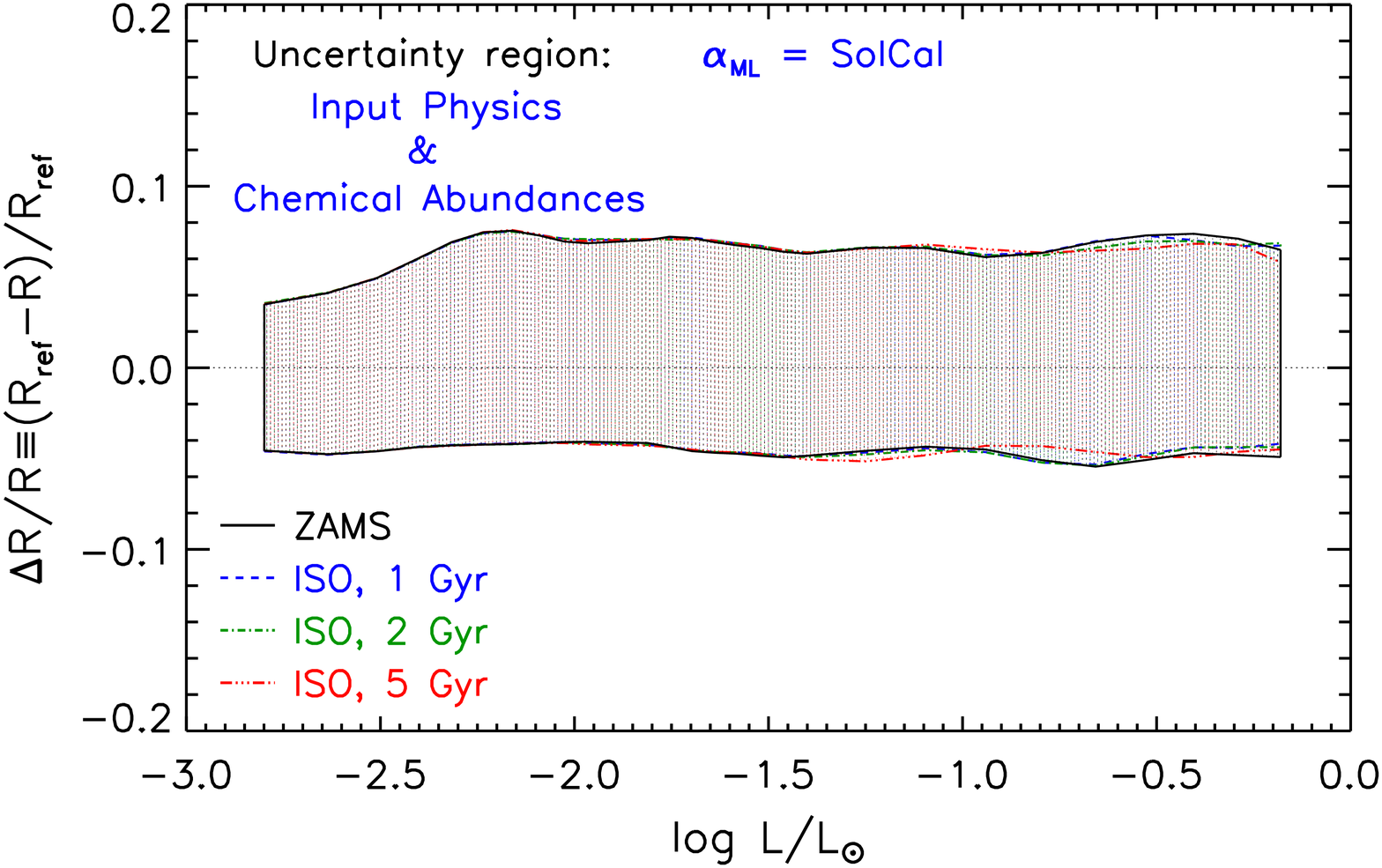}
	\caption{Cumulative error stripe for solar calibrated perturbed models. Left-hand panel: error stripe as a function of the stellar mass. Right-hand panel: stripe as a function of the luminosity.}
	\label{fig:tot_fis_chm_solcal}
\end{figure*}
Regarding the effect on the cumulative error stripe of the adoption of a solar calibrated \ml$_{\sun}$, we have showed in previous sections that the mixing length parameter calibration produces an effect only in the case of the BCs, reducing the radius change for $M\ga 0.7$~\msun. In all the other cases, the solar calibration is inconsequential, in particular when dealing with the chemical composition.

Fig.~\ref{fig:tot_fis_chm_solcal} shows the cumulative error stripe (input physics and chemical composition) for solar calibrated perturbed models as a function of the stellar mass (left-hand panel) and luminosity (right-hand panel). It is evident that the stripe is unaffected by the solar calibration for $M\le 0.7$~\msun, while it is slightly smaller than that shown in Figs.~\ref{fig:tot_fis_chm_iso} and \ref{fig:tot_fis_chm_iso_lum} for larger masses. In particular, at a fixed mass (left panel of Fig.~\ref{fig:tot_fis_chm_solcal}) for $M\ga 0.7$~\msun{} the radius variation in ZAMS is almost constant to $\pm 2$, $\pm 3$~percent, while in not solar calibrated cases it reaches 4--5~percent. A similar behaviour can be found in the isochrone models. The effect of solar calibration is visible also at a fixed luminosity, for $\log L/$\lsun$\ga -1$. The radius change is slightly smaller (about $+6,\,+7(-4)$~percent) than that obtained for not solar calibrated models (see Fig.~\ref{fig:tot_fis_chm_iso_lum}). However, comparing the results found for not-calibrated and calibrated \ml{} perturbed models, we can safely conclude that the effect of the solar calibration of \ml{} does not drastically affect the estimated error stripe.

\section{Conclusions}
\label{sec:conclusions}
We performed an analysis of the uncertainties affecting the radius predicted by stellar models of LM and VLM stars due to the errors in the adopted input physics and chemical composition. 

As a first step, we analysed the impact on the radius of each input physics quantity, initial helium and metal abundance by perturbing/varying only a single quantity and fixing the other to its reference value, at a fixed mass. This method allows us to clearly address the role of each analysed quantity in determining the stellar radius variation. 

Then, we computed the cumulative error stripe on the radius by adding each individual perturbation, for two different values of the super-adiabatic convection efficiency, namely \ml$ =2.0$ (our solar calibrated value) and \ml$ =1.0$ first at fixed mass and then at fixed luminosity.

The cumulative error stripe at a fixed mass for ZAMS models with \ml=2.00, ranges between $\pm 2$, $\pm 3$~percent for $M=0.1$~\msun{} to about $\pm 4$,$\pm 5$~percent at $M=1.0$~\msun. The relative error is less symmetric and broader if \ml$=1.0$ is adopted especially for masses in the range 0.5--0.8~\msun{} reaching about $\pm 6$~percent for $M=1.0$~\msun.

We evaluated the impact on the radius for models along the 1, 2, and 5~Gyr isochrones. For masses smaller than about 0.6-0.7~\msun, the cumulative error stripe is essentially that given for ZAMS models, while for larger masses  the radius uncertainty gets sensitive to the age, increasing with the isochrone age and reaching about $+12(-15)$~percent at 5 Gyr. Similarly to the ZAMS, the cumulative stripe for the isochrones gets slightly larger if \ml$=1.0$ is adopted.

We also showed the impact of the sole uncertainty on the adopted initial chemical composition on the radius, which is about $\pm 1$, $\pm 2$~percent in all the selected mass range for the ZAMS, while it is larger for the isochrones, reaching $+7$($-10)$~percent for the 5~Gyr isochrone.

We also analysed the error stripe at a fixed luminosity. In this case we found that the radius uncertainty is independent of the age and consequently the ZAMS and isochrones cumulative error stripes are coincident. The uncertainty ranges from about $\pm 4$~percent for the faintest stars to about $+7$, $+9$($-5$)~percent for the brightest stars, for \ml=2.0. The stripe gets larger if \ml$=1.0$ is considered ranging from about $+4$($-6$)~percent (faint stars) to $+8$, $+10$($-6$)~percent (bright stars). The contribution of the sole chemical composition produces a radius variation between about $\pm 3$~percent and $\pm 5$~percent in the selected luminosity range.

We also discussed the impact of a solar calibration on each of the perturbed model. We showed that only models with masses larger than about 0.7~\msun{} are actually affected by the solar calibration and that the error stripe slightly reduces in such a mass range.

\section*{Acknowledgements}
We thank Gregory Feiden for his useful comments that helped to improve the paper. ET acknowledges Osservatorio Astronomico di Teramo (\emph{Modelli di accrescimento per stelle di Pre Sequenza Principale e confronto teoria-osservazione con metodi bayesiani}, PI: S. Cassisi), INFN (Iniziativa specifica TAsP) and Universit\'a di Pisa (PRA 2016, \emph{Modellistica di stelle di piccola massa in fase di Pre-Sequenza Principale}, PI: S. Degl'Innocenti). 

\bibliographystyle{mn2e}
\bibliography{bibliography}
\label{lastpage}
\end{document}